\documentclass[fleqn,usenatbib]{mnras}

\usepackage{newtxtext,newtxmath}
\usepackage[T1]{fontenc}

\DeclareRobustCommand{\VAN}[3]{#2}
\let\VANthebibliography\thebibliography
\def\thebibliography{\DeclareRobustCommand{\VAN}[3]{##3}\VANthebibliography}

\usepackage{graphicx}	% Including figure files
\usepackage{amsmath}	% Advanced maths commands
\usepackage{placeins}
\usepackage{mathtools}
\usepackage{cuted}

\usepackage[dvipsnames]{xcolor}

\makeatletter
\newcommand*{\hyperlinkcite}[1]{\hyper@link{cite}{cite.#1}}%\hyperlinkcite takes 2 arguments: #1<- cite-key, #2<- link-text
\makeatother

\title[Viscous overstability in dense planetary rings]{Viscous overstability in dense planetary rings - Effect of vertical motions and dense packing}

\author[M. Lehmann $\&$ H. Salo]{
Marius Lehmann $^{1}$\thanks{E-mail: mlehmann@asiaa.sinica.edu.tw}
Heikki Salo,$^{2}$
\\
$^{1}$Institute of Astronomy and Astrophysics, Academia Sinica, Taipei 10617, Taiwan\\
$^{2}$Space Physics and Astronomy Research Unit, University of Oulu, Pentti Kaiteran katu 1, FI-90014, Finland\\
}

\date{Accepted XXX. Received YYY; in original form ZZZ}

\pubyear{2023}

\begin{document}
\label{firstpage}
\pagerange{\pageref{firstpage}--\pageref{lastpage}}
\maketitle

\begin{abstract}

We investigate the linear axisymmetric viscous overstability in dense planetary rings with typical values of the dynamical optical depth $\tau\gtrsim 0.5$. We develop a granular flow model which accounts for the particulate nature of a planetary ring subjected to dissipative particle collisions. The model captures the dynamical evolution of the disc's vertical thickness, temperature, and effects related to a finite volume filling factor of the ring fluid. We compute equilibrium states of self-gravitating and non-self-gravitating rings, which compare well with existing results from kinetic models and N-Body simulations.
Subsequently, we conduct a linear stability analysis of our model. We briefly discuss the different linear eigenmodes of the system and compare with existing literature by applying corresponding limiting approximations.
We then focus on the viscous overstability, analysing the effect of temperature variations, radial and vertical self-gravity, and for the first time the effects of vertical motions on the instability. In addition, we perform local N-body simulations incorporating radial and vertical self-gravity. Critical values for the optical depth and the filling factor for the onset of instability resulting from our N-body simulations compare well with our model predictions under the neglect of radial self-gravity. When radial self-gravity is included, agreement with N-body simulations can be achieved by adopting enhanced values of the bulk viscous stress.

\end{abstract}


%%%%%%%%%%%%%%%%%%%%%%%%%%%%%%%%%%%%%%%%%%%%%%%%%%

%%%%%%%%%%%%%%%%% BODY OF PAPER %%%%%%%%%%%%%%%%%%

\section{Introduction}\label{sec:intro}

The Voyager missions in the early 1980's, and particularly the recent Cassini (NASA/ESA) mission, revealed an intriguing amount of structure in Saturn's rings, including gaps, rings, and waves, on various length scales ranging from roughly 1000km down to about 100m.
This structural richness is indicative of an underlying dynamical richness \citep{burns2006,schmidt2009,cuzzi2018} which has been exposed in the rings. Due to their proximity and consequent accessibility by spacecraft, Saturn's rings can serve as a "local laboratory" to study processes, such as accretion and disk-satellite interactions, which are also at work in protoplanetary disks \citep{cuzzi2010,latter2018}. Furthermore, it may be expected that future missions, such as the planned ring skimmer mission \citep{skimmer} will provide observations of Saturn's rings with an even higher level of detail, which will help to further support the improvement of theories describing astrophysical disks.

The most appropriate theoretical framework to study the dynamics of dense planetary rings, such as Saturn's A, B and C rings is a kinetic description based on a Boltzmann or Enskog-Equation \citep{goldreich1978a,hameen1978,shukhman1984,stewart1984,Borderies83,shu1985c,araki1986,latter2006,latter2008,araki1988,araki1991}.
However, the equations resulting from this formalism bear a great mathematical complexity such that they have so far not been applied to dynamical problems beyond the level of linear stability calculations. The main alternative approaches constitute either N-body simulations, or the application of hydrodynamic models. The limiting factor of  N-body simulations is the entailing computational expense, especially when self-gravitational forces need to be included. Furthermore, it is often not straight forward to interpret the results from such simulations, and thereby to identify relevant physical processes at work. In order to get a handle on this, the development of simplified hydrodynamic models can be useful. Moreover, hydrodynamic simulations are typically less computationally expensive than N-body simulations, thus allowing to survey larger domains and timescales. On the other hand, hydrodynamic models neglect physical aspects that may or may not be important, depending on the problem at hand. 
In particular, hydrodynamic models typically assume a linear (Newtonian) relation between stress and strain in the fluid.
Collisions between ring particles may not be sufficiently frequent to neglect the dynamical evolution of the stress, such that stress and strain are actually not instantaneously coupled \citep{latter2006}, thus potentially violating the above assumption. Another issue is the presumed isotropy of the effective velocity dispersion tensor of the hydrodynamic model. When taking into account collisional contributions, the extreme thinness of Saturn's main rings implies that the vertical components of particles mutual impact velocities may be small, such that the vertical component of the collisional stress can be substantially reduced as compared with the planar components.
Furthermore, hydrodynamic models require, depending on the number of equations, a number of a priori unknown parameters which need to be supplemented.

That being said, hydrodynamic models have been found useful for at least a qualitative description of various structures in planetary rings, mainly those of Saturn.
Prominent examples are the propagation of satellite induced spiral density waves \citep{goldreich1978b,goldreich1978c,goldreich1979c,shu1984,shu1985a,shu1985b,borderies1985,bgt1986,schmidt2016,lehmann2016,lehmann2019}, the formation of moonlet induced gaps and 'propeller' structures \citep{goldreich1980,showalter1986,borderies1989,spahn1992a,hahn2009,hoffmann2015,spahn2018,graetz2019,seiss2019,seiler2019}
or the small-scale axisymmetric viscous overstability \citep{schmit1995,schmit1999,spahn2000a,schmidt2001b,salo2001,schmidt2003,latter2009,latter2010,lehmann2017,lehmann2019}. 
It is the latter to which this study is devoted.

One quantity that has been found to be crucial to understand how the viscous overstability operates is the volume filling factor of particles \citep{latter2008,mondino2023}.
A proper description of variations in this quantity
affords inclusion of variations in the disc thickness.
%
% We develop a hydrodynamic model of a dense planetary ring which, in contrast to most previous hydrodynamic models, includes the dynamical evolution of the disc scaleheight, as well as the temperature (i.e. particle velocity dispersion).
% %
% The inclusion of the evolution of the disc scaleheight enables us to describe variations in the volume filling factor (i.e. the amount of 'packing') of the fluid,
%
%
Furthermore, hydrodynamic studies of the viscous overstability usually incorporate a bulk viscosity in the formulation of the viscous stresses, so as to mimic stabilizing processes not captured by a fluid description. 
% That is, the bulk viscosity has been found to significantly affect the linear stability properties of the system \citep{schmit1995,schmidt2001b}. 
Since bulk viscosity couples to compressional motions, and since it is known from theoretical studies \citep{borderies1985} and N-body simulations \citep{salo2001} of dense rings (with appreciable volume filling factor) that radial compression of particles results in vertical expansion, the effect of a bulk viscosity is potentially flawed if vertical motions are neglected.
Another motivation for our new model is to move away from the rather vague and not wholly satisfactory viscosity prescriptions of previous work, which were usually in the form of power laws (\citealt{schmit1995,schmidt2001b,lehmann2017}, to name a few). Connecting such models with N-body simulations (and real planetary rings) is not straightforward, though notable attempts to do so were undertaken \citep{salo2001,schmidt2001b} . Ideally, one would want a hydrodynamic model that is securely anchored in kinetic parameters, such as particle size, filling factor, etc., allowing direct comparison and generalisation from N-body simulations. 

In this paper we make a first attempt to address the above points. To this end construct expressions for the pressure, the viscosity, the heat conductivity and the collisional cooling rate that are based on purely dimensional arguments as inspired by the study of \citet{hutter1995}, and which are framed in fundamental quantities such as the particle size, the particle mean free path, and the particle collision rate.
% While previous hydrodynamic models that included the temperature equation on the evolution of the viscous overstability (\citealt{schmidt2001b,lehmann2017}) were found to be adequate to some degree by means of comparison with N-body simulations, they required assumptions on the functional dependence of aforementioned expressions on fundamental parameters, such as the surface mass density and the temperature. These functional forms contain a number of parameters. Values of the latter have been retrieved from N-body simulations \citep{salo2001}, but contain substantial uncertainties. 
% In contrast, the functional dependencies adopted in our model are mainly determined by parameters (such as the particle size and their elasticity) that are accessible through observations and laboratory experiments. 
This formulation facilitates a comparison with N-body simulations and kinetic models, but retains the mathematical simplicity of hydrodynamic models. 
% \citet{lehmann2017} found that a good model for the equation of state is the most important aspect in attaining good agreement between hydrodynamic and N-body simulations of the viscous overstability.
% For this reason we expect that our model will result in an improved description of ring particles at high volume filling factors as compared with previous hydrodynamic models. 
To test predictions resulting from our model, we additionally perform local N-body simulations which include vertical and radial self-gravity forces. Overall we find good agreement between critical values of the volume filling factor and the optical depth for the onset of overstability resulting from both approaches. However, in order to obtain good agreement in the presence of radial self-gravity, we find it necessary to adopt values of the fluid's bulk viscosity that are significantly larger than those determined by \citet{salo2001}. 

% The aim is to apply our new model to the nonlinear evolution of the viscous overstability in Saturn's rings.
% To this end we develop a hydrodynamics code, which we utilize to conduct 2D shearing sheet simulations. These simulations allow for the first time a hydrodynamic study of the interaction between axisymmetric wavetrains instigated by the viscous overstability and non-axisymmetric self-gravitational (turbulent) wake structures. 

The paper is organised as follows.
In \S \ref{sec:dense} we briefly review the dynamics of dense planetary rings and aspects of the viscous overstability, as observed in Saturn's rings. In \S \ref{sec:hydromodel} we present our hydrodynamic model, along with the definition of the transport coefficients and all the parameters involved. In \S \ref{sec:equilibrium} we compute ring equilibrium states resulting from our model, which we compare to results form our N-body simulations. In \S \ref{sec:linpert} we perform a detailed axisymmetric linear analysis of the model. We consider various limiting cases and derive analytical expressions for the eigenmodes, which are compared to previous studies. In \S \ref{sec:OS} we focus on the viscous overstability. There we provide a detailed investigation of its driving and damping agents and present numerically obtained linear growth rates. In \S \ref{sec:nbody} we present our N-body simulations and compare critical quantities for the onset of viscous overstability between our model and simulations. In addition, we also compare linear growth rates of the viscous overstability following from both approaches.
Finally, in \S \ref{sec:summary} we summarise and discuss our main findings, as well as directions for future research.

\section{Dense Planetary Rings and the Viscous Overstability}\label{sec:dense}

While the kinematic equilibrium of a planetary ring is essentially the same as for any type of astrophysical disk, i.e. the balance between central gravity and centrifugal forces, the thermal equilibrium is rather distinct.
That is, the latter is determined by the balance between heating due to viscous dissipation and cooling due to inelastic mutual particle collisions.

\subsection{Viscosity in planetary rings}\label{sec:visc}

Viscosity (or angular momentum transport) in planetary rings essentially stems from either particle collisions or non-axisymmeric self-gravitational filamentary (wake) structures.

The contribution due to collisions is typically divided into two distinct parts, namely the `local' and `non-local' contributions $\nu_{\text{local}}$ and $\nu_{\text{nl}}$, respectively \citep{schmidt2009}. The local contribution is governed by random motions of particles in between mutual collisions, whereas the non-local contribution describes the angular momentum transferred between particles at contact during collisions. The latter becomes important as soon as the particle mean free path becomes comparable to the particle size \citep{shukhman1984,araki1986}.
An expression for the local (kinematic) viscosity was derived by \citet{goldreich1978a}:  
\begin{equation}\label{eq:nuloc_gt}
  \nu_{\text{loc}} \sim \omega_{c} \frac{c^2}{\omega_{c}^2 + \Omega^2},
\end{equation}
where $\widehat{\nu}_{\text{local}}$ is a dimensionless factor of order unity.
On the other hand, the non-local (kinematic) viscosity scales as \citep{shukhman1984,araki1986} 
\begin{equation}\label{eq:nu_nonlocal}
    \nu_{nl}\sim \omega_{c} d^2.
\end{equation}
In these expressions $\Omega$ denotes the orbital frequency, $c$ is the particle random velocity ( velocity dispersion), $d$ the particle diameter, and $\omega_{c}$ the particle collision frequency. These expressions can also be derived from the basic expression for viscosity $\nu \sim \omega_{c} s^2$, where $s$ denotes the particle mean free path. In the case of the local viscosity $s\sim c/\omega_{c}$ if multiple collisions occur during one orbital period, whereas $s \sim c/\Omega$ if collisions are rare, since ring particles are forced to evolve on epicycles. On the other hand, for the non-local mode $s\sim d$.
Thus, their ratio is given by
\begin{equation}
\frac{\nu_{\text{nl}}}{\nu_{\text{loc}}}\sim \text{max} \left[\frac{\omega_{c}^2 }{\Omega^2 } ,1 \right] R^2, 
\end{equation}
depending on the value of $\omega_{c}/\Omega$. Here we defined
\begin{equation}
    R=\frac{ \Omega d}{c},
\end{equation}
known as the Savage and Jeffrey parameter \citep{savage1981,araki1991}, which in Saturn's rings is assumed to be of order unity (e.g. \citealt{salo2018}). Furthermore, in dense optically thick rings such as Saturn's A and B rings one has $\omega_{c}\gg \Omega$, such that the non-local angular momentum transport dominates the local contribution. However, even in optically thin regions of Saturn's rings, such as the C ring or the Cassini Division, where one expects $\omega_{c}\sim \Omega$, non-local transport effects are important.

Self-gravitational transport of angular momentum in Saturn's rings is realized through the emergence of self-gravity wakes, which are non-axisymmetric trailing density enhancements that form and dissolve on the orbital time scale \citep{salo1992a,daisaka2001}. This mode of angular momentum transport is expected to dominate throughout the A ring and possibly in some parts of the outer B ring.

\subsection{Viscous overstability}

It is common belief that some of the shortest scale structures which are detectable in Saturn's rings are most likely a result of the short-scale axisymmetric viscous overstability \citep{thomson2007,collwell2007,hedman2014a}. 
% The high resolution required to detect these structures has been realized in Cassini Radio Science Subsystem observations \citep{thomson2007} as well as stellar occultations that are passing a turnaround point in the ring plane \citep{collwell2007,hedman2014a}.
%
The axisymmetric viscous overstability is an intrinsic, oscillatory instability of the ring flow which draws energy from the Keplerian background shear and amplifies the collective epicyclic motion of the ring particles (i.e. a density wave) through a coupling of the viscous stress-oscillations to the background orbital shear.
In a hydrodynamic description this mechanism requires that the viscous stress is a steeply increasing function of the surface mass density. 
\citet{schmit1995} studied the conditions for the onset of overstability in Saturn's B-ring in terms of an isothermal hydrodynamic model, assuming a simple powerlaw parameterization for the dynamic shear viscosity
\begin{equation}\label{eq:beta}
\eta=\eta_{0} \left(\frac{\sigma}{\sigma_{0}}\right)^{\beta+1}.
\end{equation}
They adopted a value $\beta=1.26$ which was based on results from N-body simulations by \citet{wisdom1988}. They found that the condition for the viscous overstability can be formulated as $\beta>\beta_{c}$ with
\begin{equation}\label{eq:betc}
\beta_{c}=\frac{1}{3}\left(\frac{\xi}{\eta}-\frac{2}{3}\right),
\end{equation}
with $\xi$ denoting the dynamic bulk viscosity.
Their model assumed $\xi= \eta$ so that $\beta_{c}=1/9$. On the other hand, non-gravitating N-body simulations, as well as N-body simulations including vertical self-gravity \citep{salo2001} revealed a substantially larger critical value $\beta_{c} \approx 1$ for the emergence of
axisymmetric overstable oscillations. In terms of hydrodynamic modeling, this can be explained by including thermal effects \citep{spahn2000a,schmidt2001b}, as well as larger values of the bulk viscosity $\xi \sim (2-4) \,\eta$ \citep{salo2001}.
In non-self-gravitating N-body simulations \citep{salo2001} of meter-sized particles overstability was found for optical depths $\tau \gtrsim 4$. However, recent time-extended simulations by \citet{mondino2023} (\hyperlinkcite{mondino2023}{MS23} henceforth) revealed the onset of overstability already for $\tau\gtrsim 2.5$, which agrees well with the dense ring kinetic model of \citet{latter2008} (\hyperlinkcite{latter2008}{LO08} henceforth). On the other hand, in simulations by \citet{salo2001} where vertical self-gravity was taken into account in terms of an enhanced vertical frequency
of the particles $\Omega_{z} / \Omega=3.6$ (the same value was originally used by \citealt{wisdom1988}) overstability occurred for $\tau \gtrsim 1$. The reason for this difference is that the onset of the viscous overstability has been found to require a critical value of the volume filling factor
\begin{equation}\label{eq:ff_true}
    \mbox{FF}= \frac{\rho}{\rho_{b}}
\end{equation}
 of the particle configuration (typically its value at the ring midplane) to be exceeded (\hyperlinkcite{latter2008}{LO08};~\hyperlinkcite{mondino2023}{MS23}). In the above expression $\rho$ and $\rho_{b}$ denote the volume mass density and particle bulk density, respectively.
In simulations of uni-sized particles which neglect vertical self-gravity, the particles tend to spread in vertical direction, thereby reducing the filling factor. This vertical extension of the ring is significantly reduced if vertical self-gravity is included.

The N-body simulations of \citet{salo2001}, and recently \hyperlinkcite{mondino2023}{MS23}, including particle-particle self-gravity revealed that axisymmetric overstable oscillations can co-exist with non-axisymmetric self-gravity wakes. But if the self-gravitational perturbations are too strong (resulting in too large density contrasts), overstability is suppressed. \hyperlinkcite{mondino2023}{MS23} performed an extensive survey of gravitating N-body simulations and formulated a criterion for the onset of viscous overstability in terms of a critical filling factor, as well as a critical ratio between the viscous stress due to wakes \citep{daisaka2001} and the non-local viscosity.
% The viscosity (\ref{eq:nugrav}) associated with wakes would imply $\beta=2$, so that one could naively expect overstability to occur, since, at least in the absence of wakes $\beta_{c} \approx 1$. \mlc{However, since wakes vary on length and time scales comparable to those which characterize overstable oscillations, the concept of a viscous stress is not expected to be applicable.}
%At this point it should be noted again that the contribution of self-gravitational wakes to the stress acting on a $100 \mathrm{~m}$ scale in the presence of nonlinear overstable waves is less clear than their contribution to the angular momentum flux in an unperturbed ring. 

\citet{latter2006} have shown that the viscous overstability is sensitive to any non-locality of the viscous stress in time. Thus, the instability mechanism requires that the stress oscillations are sufficiently in phase with the oscillations of the (epicyclic) velocity perturbations. This affords that the relaxation time of the stress (the collisional time scale) is appreciably shorter than the orbital time scale. 
Similarly, \citet{lehmann2019} found that also the perturbation due to a nearby Lindblad resonance can destroy the phase relation of the viscous stress perturbation with the epicyclic oscillation in an overstable wave, offering an explanation for the mitigation of overstable waves by large scale density waves seen in their hydrodynamic simulations.
%

% Furthermore, \citet{ballouz2017} studied the effect of surface irregularities of ring particles on the co-existence of axisymmetric ovestable waves and self-gravity wakes in terms of N-body simulations where particles in contact were subject to static and rolling friction. They found that with increasing inter-particle friction overstable oscillations become more pronounced, whereas self-gravity wakes tend to be suppressed. They also showed that with increasing friction the non-local viscosity increases, which is likely to be crucial in promoting overstability.

The long-term nonlinear saturation of the viscous overstability in Saturn's rings has, thus far, mainly been investigated in terms of isothermal hydrodynamic models neglecting self-gravity \citep{schmidt2003,latter2009,latter2010}. An exception are the non-self-gravitating N-body simulations by \citep{latter2013} and the hydrodynamic simulations of \citep{schmit1999} that took into account radial self-gravity.
The result of these studies is that the viscous overstability typically saturates in form of traveling waves with wavelengths  $\sim 100 \mathrm{m}-500 \mathrm{m}$. The papers by \citet{latter2009}, \citet{latter2010} and \citet{latter2013} showed that the saturated state in general most likely consists of counter-propagating nonlinear wave trains, separated by defect structures (so called sink and source structures). 
%which are reminiscent of solutions of the coupled complex Ginzburg-Landau equations \citep{aranson2002}. 
% describing (weakly) nonlinearly interacting, counter-propagating waves (e.g. van \citealt{vanhecke1999}). 
% On the other hand, \citet{schmit1999} argued that radial self-gravity limits the saturation wavelength of the overstable pattern to a finite value, as their simulations without self-gravity did not show a halt in the increase of the wavelength within the time scale accessible by their simulations. This finding contradicts in some sense the results by \citet{latter2009,latter2010}, who showed that there is a well defined wavelength which the nonlinear saturated waves can attain in the absence of self-gravity. As explained in \citet{latter2010}, this difference is related to the implementation of the radial boundary conditions in simulations.
%   We note that the studies by \citet{salo2001}, \citet{latter2008}, \citet{ballouz2017} and reently \hyperlinkcite{mondino2023}{MS23} investigated under which conditions viscous overstability is expected to occur in a given setup, rather than how it eventually saturates.
More recently, \citet{lehmann2017} performed N-body and non-isothermal hydrodynamic simulations that incorporated radial self-gravity as well as a constant vertical self-gravity (see \S \ref{sec:sg}). The key result of this study is that radial self-gravity determines the wavelength of nonlinear saturated wave trains. That is,  under the influence of radial self-gravity the wave pattern tends to evolve into a state of minimum oscillation frequency, such that the pattern's wavelength is governed by the frequency minimum of the nonlinear dispersion relation of overstable waves. 
% This implies that the saturation wavelength decreases with increasing strength of the radial self-gravity force, which in the models considered in \hyperlinkcite{lehmann2017}{LSS17} is governed by the ground state surface mass density $\sigma_{0}$.}

\section{Hydrodynamic Model}\label{sec:hydromodel}

\subsection{Governing equations}\label{sec:equations}
We apply a local fluid model to describe a small patch of a planetary ring which orbits a planet of mass $M_{P}$ at constant distance $r_{0}$, using the shearing sheet approximation (\citealt{goldreich1965}). We adopt Cartesian coordinates $(x,y,z)$, centered at radius $r_{0}$, and rotating around the planet with the Keplerian frequency
\begin{equation}\label{eq:kepler}
\Omega_{0}\equiv\sqrt{\frac{G M_{P}}{r_{0}^3}},
\end{equation}
such that $x$ points radially outward, $y$ in the direction of orbital motion, and $z$ is parallel to the angular velocity vector given by $\vec{\Omega} \equiv \Omega_{0} \Vec{e}_{z}$.   
The flow of ring particles (a granular flow) is governed by the set of vertically averaged nonlinear fluid equations
% \begin{equation}
% D_{t} \sigma =  - \sigma \left( \partial_{x} u_{x} +\partial_{y} u_{y} \right),\label{eq:contsig}
% \end{equation}
% \vskip -0.5cm
% \begin{equation}
%  D_{t} u_{x}    = F_{\text{sg},x}+ 2 \Omega_{0} u_{y}  - \frac{1}{\sigma} \left[\partial_{x} \overline{P_{xx}} +\partial_{y} \overline{{P}_{yx}} + 
% \partial_{x} f\left(u_{z}\right) \right],\label{eq:contux}  
% \end{equation}
% \vskip -0.5cm
% \begin{equation} 
% D_{t} u_{y}  =  F_{\text{sg},y}  - 2 \Omega_{0} u_{x}  - \frac{1}{\sigma} \left[ \partial_{x}\overline{P_{xy}} + \partial_{y}\overline{{P}_{yy}}  +\partial_{y} f\left(u_{z}\right)  \right],
% \label{eq:contuy}
% \end{equation}
% \vskip -0.5cm
% \begin{equation} 
% \begin{split}
% D_{t} T & =  -\frac{2}{3 \sigma} \left[\hat{P}:\mathbf{\nabla} \mathbf{u} + \mathbf{\nabla} \cdot \mathbf{F}_{\kappa}  + \Gamma + \overline{P_{zz}}\,  \frac{K}{H} \right. \\%[0.1cm]
% &  \quad  \left. + f(u_{z}) \left( \partial_{x}u_{x} +\partial_{y}u_{y} \right) -\frac{1}{3} H^2 \eta \left( \left[\partial_{x} \frac{K}{H}\right]^2 +\left[\partial_{y} \frac{K}{H}\right]^2    \right) \right], \label{eq:conttemp} 
% \end{split}
% \end{equation}
% \vskip -0.5cm
% \begin{equation}
% \begin{split}
%   D_{t} K & = -\mu^2 H  - 2 \pi G \sigma+ \frac{3 \overline{P_{zz}} }{\delta H \sigma}\\
% &  \quad +\frac{H}{\sigma} \left[ \partial_{x} \left(\eta \, \partial_{x} \frac{K}{H} \right) +\partial_{y} \big(\eta \, \partial_{y} \frac{K}{H} \big)\right]\label{eq:conth2}  ,
% \end{split}
% \end{equation}
\begin{equation}
D_{t} \sigma =  - \sigma \left( \partial_{x} u_{x} +\partial_{y} u_{y} \right),\label{eq:contsig}
\end{equation}
\vskip -0.5cm
\begin{equation}
 D_{t} u_{x}    = F_{\text{sg},x}+ 2 \Omega_{0} u_{y} - 3\Omega_{0}^2 x - \frac{1}{\sigma} \left[\overline{\partial_{x} P_{xx}} +\overline{\partial_{y} {P}_{yx}} + 
\overline{\partial_{x} f\left(u_{z}\right)} \right],\label{eq:contux}  
\end{equation}
\vskip -0.5cm
\begin{equation} 
D_{t} u_{y}  =  F_{\text{sg},y}  - 2 \Omega_{0} u_{x}  - \frac{1}{\sigma} \left[ \overline{\partial_{x}P_{xy}} + \overline{\partial_{y}{P}_{yy}}  +\overline{\partial_{y} f\left(u_{z}\right)}  \right],
\label{eq:contuy}
\end{equation}
\vskip -0.5cm
\begin{equation} 
\begin{split}
D_{t} T & =  -\frac{2}{3 \sigma} \Bigg[ \sum\limits_{i,j=x,y}^{} \overline{P_{ij}}\partial_{i}u_{j} + \overline{P_{zz}\,  \partial_{z} u_{z}}  +  \overline{f(u_{z})} \left( \partial_{x}u_{x} +\partial_{y}u_{y} \right)  \\%[0.1cm]
&  \quad   + \overline{\mathbf{\nabla} \cdot \mathbf{q}}  + \overline{\Gamma} -\overline{\eta \left(\partial_{x} u_{z}\right)^2}  -\overline{\eta \left(\partial_{y} u_{z}\right)^2} \Bigg], \label{eq:conttemp} 
\end{split}
\end{equation}
\vskip -0.5cm
\begin{equation}
\begin{split}
  D_{t} \dot{H} & = -\Omega_{0}^2 H + \frac{\overline{\rho z F_{\text{sg},z}}}{\gamma H \sigma} \\
  \quad & +\frac{1}{\gamma H \sigma}\left[\overline{P_{zz}} - \overline{\partial_{x} \left(z P_{xz}\right)} -\overline{\partial_{y} \left(z P_{yz}\right)}\right]\label{eq:conth2}  ,
\end{split}
\end{equation}
and
\begin{equation}
D_{t} H  =   \dot{H}\label{eq:conth}. 
\end{equation}
In these equations $D_{t}=\partial_{t} + u_{x}\partial_{x} + u_{y}\partial_{y}$ and  $u_{x},u_{y}$ are the radial and azimuthal components of the total velocity, such that
 the azimuthal component $u_{y}$ includes the linear Keplerian shear $-\frac{3}{2} \Omega_{0}x $. 
The central planet is assumed to be spherical so that we have equality between the orbital frequency $\Omega(r)$, the epicyclic frequency and
the vertical frequency of test particles. The balance between central gravity and centrifugal force in equilibrium has been subtracted from the radial momentum equation to leading order in the small parameter $x/r_{0}$.
 A bar on a quantity $X$ indicates vertical integration:
 \begin{equation}
     \overline{X} \equiv \int \limits_{-\infty}^{+\infty} X \mathrm{d}z.
 \end{equation}
  As such, the surface mass density $\sigma$ is related to the volumetric mass density $\rho=\rho(x,y,z,t)$ (\S \ref{sec:vstruct}) via $\sigma = \overline{\rho}$.
  Furthermore, $T$ is the temperature, and $H$ is the vertical thickness parameter of the disc.   The temperature $T$ is related to the root mean squared velocity deviations $c$ of particles from the mean flow velocity field
 by $T\equiv c^2$ (e.g. \citealt{shu1985c}). The remaining quantities are specified below.
 
Eqs. (\ref{eq:contsig})---(\ref{eq:conth}) can be obtained from vertical integration of the three-dimensional Navier-Stokes equations and the heat flow equation, thus describing the balance laws of mass, momentum, and temperature of the ring fluid. This is shown in detail in Appendix \ref{app:zint}.
The quantities $\mathbf{q}=\left\{q_{x},q_{y}\right\}$ and $P$ denote the \emph{planar} heat flux and the pressure tensor (see \S \ref{sec:constit}). 
% Furthermore,
% \begin{equation}\label{eq:du}
% \begin{array}{@{}*{22}{l@{}}}
% \mathbf{\nabla} \mathbf{u} = \begin{pmatrix}   \partial_{x}u_x   \hspace{0.2 cm} &  \partial_{x}u_y  \\[0.15 cm] 
%  \partial_{y}u_x   \hspace{0.2 cm}  & \partial_{y}u_y \end{pmatrix} 
% \end{array}
% \end{equation}
% is the rate of strain tensor.
The symbols
$\eta$, $\xi$ and $\Gamma$ stand for the dynamic shear viscosity, dynamic bulk viscosity and cooling rate (\S \ref{sec:transportcoef}). The latter describes the loss of random kinetic energy due to inelastic particle collisions. We assume from the outset that the planar components of the velocity $u_{x}$, $u_{y}$ and the  self-gravity force $F_{\text{sg,x}}$,$F_{\text{sg,y}}$, as well as the temperature $T$ are independent of $z$ (see \S \ref{sec:sg} and Appendix \ref{app:zint}).

% Furthermore, the self-gravity potential $\phi^{\text{sg}}$ is understood to be evaluated at $z=0$ (see below), which is a good approximation for a thin disk.
Equations (\ref{eq:conth2})-(\ref{eq:conth}) govern the dynamical evolution of the disc thickness parameter $H$, where $\dot{H}$ describes the speed of vertical `disc breathing'. For the vertical component of velocity we assume \citep{papaloizou1988}
\begin{equation}\label{eq:velz}
 u_{z}=\frac{z}{H}D_{t} H = z \frac{\dot{H}}{H},
\end{equation}
such that the velocity $u_{z}$ is the vertical velocity of a  Lagrangian fluid element following the planar flow $\left\{u_{x},u_{z}\right\}$, and \emph{motions are assumed to be symmetrical with respect to the midplane $z=0$} (see also \citealt{stehle1999}). The latter restriction should be acceptable for our study of the viscous overstability in dense planetary rings. The terms on the right hand side of Eq. (\ref{eq:conth2}) in turn describe effects of vertical planetary gravity, vertical self-gravity and  effects related to vertical stress, respectively. Furthermore, $\gamma$ is a dimensionless scaling factor defined by
\begin{equation}\label{eq:gamma}
    \gamma \equiv \frac{1}{H^2 \sigma}\int \limits_{-\infty}^{+\infty} \rho z^2 \mathrm{d}z.
\end{equation}
Our approach to include vertical disc motions into an otherwise two-dimensional hydrodynamic model is very similar to the one of \citet{stehle1999}.  Recently a more generalised approach, which allows for more complex vertical motions, has been devised by \citet{ogilvie2018}. Furthermore, the kinetic model of \citet{shu1985c} allows for disc bending (or corrugation) motions, but assumes the disc to be in vertical hydrostatic equilibrium. Eq. (\ref{eq:conth2}) is also closely related to Equation (19) in \hyperlinkcite{latter2008}{LO08}. The difference is that we allow the disc to  depart from vertical hydrostatic equilibrium, in contrast to \hyperlinkcite{latter2008}{LO08}, who assumed $\dot{H}=0$. 
As a consequence, the planar momentum equations (\ref{eq:contux})-(\ref{eq:contuy}) and the heat flow equation (\ref{eq:conttemp}) contain terms arising from vertical disc motions,  resulting from viscous coupling and work done by pressure.
The function $f(u_{z})$ appearing in the equations for $u_{x}$, $u_{y}$ and $T$ describes viscous coupling of vertical 
and planar motions. This quantity, and also the vertical stress $P_{zz}$ are given in \S \ref{sec:constit}.

\subsection{Vertical density  stratification}\label{sec:vstruct}

In order to evaluate the various terms appearing in Eqs. (\ref{eq:contsig})---(\ref{eq:conth}) we need to make assumptions on the vertical structure of the planetary ring.
For the purpose of illustration we will consider two limiting cases.
In this work we will mainly apply a Gaussian distribution
\begin{equation}\label{eq:rho_gauss}
   \rho_{\text{G}}(x,y,z,t) \equiv \frac{\sigma}{\sqrt{2 \pi} H} \exp \left(-\frac{z^2}{2 H^2} \right),
\end{equation}
where the dependence on $x$, $y$ and $t$ is implicit via $\sigma$ and $H$, which evolve according to (\ref{eq:contsig}) and (\ref{eq:conth2}), respectively. 
Strictly, this distribution applies in the dilute limit \citep{goldreich1978a,shu1985c}, characterized by a large values of  $c/(\Omega d)$ (cf. \S \ref{sec:dense}). 
Using (\ref{eq:rho_gauss}) the midplane volume filling factor (\ref{eq:ff_true}) yields
\begin{equation}\label{eq:ff_gauss}
 \mbox{FF}_{G}\left(0\right) \equiv \frac{\sigma}{\sqrt{2} \pi H\rho_{b}}.
\end{equation}
Furthermore, for the Gaussian distribution we find $\gamma=1$.

In addition, we consider a vertically unstratified (or uniform) slab of constant volumetric mass density
\begin{equation}\label{eq:rho_slab}
 \rho_{\text{S}}(x,y,z,t) \equiv \frac{\sigma}{2 H} \hspace{0.5cm}\text{for} \hspace{0.2cm}-H\leq z \leq H,
\end{equation}
By using (\ref{eq:rho_slab}) the (midplane) volume filling factor becomes
\begin{equation}\label{eq:ff_slab}
 \mbox{FF}_{S}\left(0\right) \equiv \frac{\sigma}{2 H\rho_{b}}.
\end{equation}
Furthermore, for this distribution we find $\gamma=1/3$.
A convenient aspect of the uniform distribution (\ref{eq:rho_slab}) is that all vertical integrals can be trivially solved, allowing to provide analytical expressions for all quantities appearing in our model. 
A uniform profile (\ref{eq:rho_slab}) is expected to apply to a cool, densely packed ring in which equal-sized particles have settled to a thin layer surrounding the midplane \citep{shu1984,borderies1985}.

Note that the actual thickness of a particle configuration measured by $\sqrt{\overline{z^2}}$ yields $H$ for the Gaussian, and $H/\sqrt{3} $ for the uniform distribution. 

Indeed, N-body simulations show (\S \ref{sec:nbody}) that the vertical distribution of a planetary ring transitions from a Gaussian toward a more or less uniform central distribution for increasing values of the optical depth (and hence $\mbox{ff}$).
% Nevertheless, in our computations presented below we will see that results only mildly vary between the two choices (\ref{eq:rho_slab}) and (\ref{eq:rho_gauss}).

% In the remainder of this paper, unless stated otherwise, quantities correspond to slab model (\ref{eq:rho_slab}).

\subsection{Self-gravity}\label{sec:sg}
Since we study self-gravitating planetary rings, we additionally need to consider the Poisson equation 
\begin{equation}\label{eq:poissoneq}
 (\partial_{x}^2 + \partial_{y}^2 +\partial_{z}^2) \phi_{\text{sg}}  = 4 \pi G\rho,
 \end{equation}
with the self-gravitational potential $\phi_{\text{sg}}$.
From this equation one can derive the \emph{planar} components of the self-gravity force  $F_{\text{sg},x/y}=-\partial_{x/y}\phi_{\text{sg}}$ 
as a function of the surface density, provided that one neglects any variations of $F_{\text{sg},x/y}$ with the vertical coordinate $z$ (\citealt{shu1984}, see also \S \ref{sec:linpert}). This approximation is valid as long as radial variations in the density occur on much longer length scales than the vertical disc thickness. Nevertheless, corrections to the so obtained planar self-gravity forces can be applied to account for a finite vertical disc thickness (\S \ref{sec:linpert}).

The vertical component of self-gravity in Eq. (\ref{eq:conth}) in general depends on the choice of the vertical disc structure and reads 
\begin{align}
  \overline{\rho z F_{\text{sg},z}} & =  - \frac{2}{3} \pi G  H \sigma^2 \hspace{0.55cm}   \text{if}   \, \rho=\rho_{\text{S}},\label{eq:fsgz_slab}\\
   \overline{\rho z F_{\text{sg},z}} & = - 2 \sqrt{\pi} G  H \sigma^2  \hspace{0.355cm} \text{if} \, \rho=\rho_{\text{G}}\label{eq:fsgz_Gauss},
\end{align}
for the slab and Gaussian distribution, which follow directly from Eqs. (\ref{eq:fsgz_slab_z})- (\ref{eq:fsgz_Gauss_z}), respectively. In order to arrive at these expressions we assumed that the self-gravitational potential $\phi_{\text{sg}}$ varies much faster in vertical than in planar directions, as well as symmetry of the ring around the mid-plane.

In addition, from Equation (\ref{eq:conth2}) we can define an \emph{effective} dimensionless frequency of vertical motions
\begin{equation}\label{eq:omz}
\frac{\Omega_{z}}{\Omega_{0}}=\sqrt{1  - \frac{\overline{\rho z F_{\text{sg},z}}}{\gamma \Omega_{0}^2 H^2 \sigma} },
\end{equation}
which combines vertical planetary gravity and vertical self-gravity, neglecting collisions (contributions from the pressure tensor).
This quantity which is often referred to as the `Wisdom-Tremaine' factor features prominently in studies of self-gravitating planetary rings.
A \emph{constant} value of $\Omega_{z}=3.6 \Omega_{0}$ was originally proposed by \citet{wisdom1988} to account for the effect of vertical self-gravity in N-body simulations of Saturn's B ring. The same value was also used in N-body simulations by \citet{salo1995}, \citet{salo2001}, \citet{latter2013} and \citet{lehmann2017}.
% Furthermore, \citet{salo2001} derived sets of hydrodynamic transport coefficients from N-body simulations with different values for the ground state optical depth and $\Omega_{z}$, which were subsequently used in hydrodynamic model calculations \citep{schmidt2001b,latter2009,latter2010,lehmann2016,lehmann2017,lehmann2019}. 
Recently \hyperlinkcite{mondino2023}{MS23} found that the actual vertical forces in self-gravitating simulations are closer to $\Omega_{z} \approx 2 \Omega_{0}$. Moreover, they argued that an artificially increased value of $\Omega_{z}$ can to some extent mimmick the inclusion of particle-particle surface (or frictional) forces which can promote viscous overstability in favor of self-gravity wakes \citep{ballouz2017}.

%Furthermore, we will consider models where the vertical self-gravity is described through an artificially increased vertical frequency denoted by 
%$\Omega_{z}$, such that
%\begin{equation}\label{eq:keqwtf}
%  D_{t} K = -\Omega_{z}^2 H   + \frac{\overline{P_{zz}} }{\delta H \sigma}
%\end{equation}
%with $\Omega_{z}>\Omega_{0}$.

% Although the quantity $\delta$ is fixed as as soon as $\zeta(z)$ is specified, we use it as a free parameter, taking the value obtained with 
% (\ref{eq:rho_slab}) only as a guide.

\subsection{Constitutive relations}\label{sec:constit}

The hydrodynamic equations (\ref{eq:contsig})-(\ref{eq:conth}) need to be supplemented with constitutive relations for the viscous stress tensor and the heat flux 
$\mathbf{q}$. 

We assume a vertically isothermal planetary ring, such that $T$ is independent of $z$.
Similar to previous studies  (\citealt{spahn2000a,schmidt2001b};~\hyperlinkcite{lehmann2017}{LSS2017}) we parameterize the heat flux as
\begin{equation}\label{eq:heatflux}
\mathbf{q} =- \kappa \, \left\{\partial_{x} T,\partial_{y} T\right\}
\end{equation}
with the dynamic heat conductivity $\kappa$ (see \S \ref{sec:transcoefs}).

For the pressure tensor we adopt the usual Newtonian form
\begin{equation}\label{eq:pten}
    P_{i j} =p \delta_{i j}-\eta\left(\partial_{i} u_{j}+\partial_{j} u_{i}\right)+\left(\frac{2}{3} \eta-\xi\right) \delta_{i j} \sum \limits_{k=x,y,z}^{}\partial_{k}u_{k}
\end{equation}
with the scalar pressure $p$, the dynamic shear viscosity $\eta$ (\S \ref{sec:transcoefs}) and the bulk viscosity $\xi$ (see below).
% The components of the vertically integrated planar pressure tensor are given by (Appendix \ref{app:zint})
% \begin{equation}\label{eq:pten}
% \overline{P_{ij}} = \overline{p} \delta_{ij} - \overline{\eta} \left( \partial_{i} u_{j} + \partial_{j}u_{i}\right) + \left(\frac{2}{3}\overline{\eta} -\overline{\xi}\right)\delta_{ij} \nabla \cdot \mathbf{u},
% \end{equation}
% where $i,j=\{x,y\}$,
% and is thus completely described by the planar velocities $u_x$, $u_y$, as well as the dynamic shear and bulk viscosities $\overline{\eta}$, $\overline{\xi}$, the heat conductivity $\overline{\kappa}$, and the total isotropic pressure $\overline{p}$ 
%
%
% Specifically, the (viscous) stress terms in (\ref{eq:contux})---(\ref{eq:conttemp}) resulting from coupling of vertical and planar disc motions using (\ref{eq:velz})
% read
Furthermore,
\begin{equation}\label{eq:fz}
f\left(u_{z}\right)  =  \left(\frac{2}{3} \eta-\xi\right) \partial_{z} u_{z} = \left(\frac{2}{3} \eta-\xi\right) \frac{\dot{H}}{H},
% P_{zz}& = p +  \left(\frac{2}{3}\eta-\xi \right) \left( \partial_{x} u_{x} + \partial_{y} u_{y} \right) - \left(\frac{4}{3}\eta + \xi \right) \frac{\dot{H}}{H} ,
\end{equation}
describes coupling of vertical and planar disc motions.
% as well as (assuming the slab model (\ref{eq:rho_slab}))
% \begin{align}
%   \overline{\eta \left(\partial_{i} u_{z}\right)^2}  & = \frac{1}{3} H^2 \overline{\eta} \left(\partial_{i} \frac{K}{H}\right)^2,\label{eq:nuh1}\\ 

% \subsection{Bulk viscosity and heat conductivity}\label{sec:linpars}
% % \subsection{Additional parameters}\label{sec:add_pars}

Furthermore, we define the bulk viscosity as 
\begin{equation}
\overline{\xi} \equiv  \widetilde{\xi}\left(\omega_{c}\right) \, \overline{\eta},
\end{equation}
with the dimensionless parameter $\widetilde{\xi}\left(\omega_{c}\right)$, which we formally let depend on the collision frequency (defined in Eq. (\ref{eq:omc}) below).
In our calculations below we find the bulk viscosity to have a significant impact on the stability boundary for viscous overstability. That is, too small values of the bulk viscosity result in viscous overstability for very  small filling factors $\mbox{FF}\ll 1$. This is not surprising, as our hydrodynamic model does not account for the non-Newtonian behavior of the viscous stress which prevails in dilute particulate rings with $\mbox{FF}\to 0$ \citep{latter2006}. The bulk viscosity in our model is used to mimic this, and  other possible stabilizing processes that may occur in N-body simulations of planetary rings, but which are not explicitly captured in hydrodynamics. 
% In addition from the aforementioned non-locality in time of the viscous stress, these could for instance be related to departures from the actual vertical velocity from our simple expression (\ref{eq:velz}).
%
%
We find that a constant value of $\widetilde{\xi}_{0}=4$ (the equilibrium value)
 leads to a reasonable overall agreement between our linear calculations and results from N-body simulations presented below. 
 A slightly better agreement is achieved when we use the relation of \citet{salo2001} (their Figure 17, upper right panel).
 % , where we identify the collision frequency $\omega_{c}$ with our expression (\ref{eq:omc}). 
 We then
 have 
 \begin{equation}\label{eq:nub_omc}
 4\lesssim \widetilde{\xi}_{0} \lesssim 2 \hspace{0.3 cm} \text{for} \hspace{0.1 cm} 1\lesssim\omega_{c,0}<\infty,
 \end{equation}
   such that $\widetilde{\xi}_{0}$ drops from a value of $\widetilde{\xi}_{0}\approx 4 $ with increasing collision frequency, and levels of to $\widetilde{\xi}_{0}\approx 2$. 
  The latter asymptotic value 
is similar to that derived from the Enskog kinetic
theory of dense systems, which is $\widetilde{\xi}\approx 1.3$ \citep{chapman1970}. A possible explanation for the rise of $\widetilde{\xi}_{0}$ at smaller collision frequencies may be the increasing non-locality in time of the viscous stress in simulations, as mentioned above.
The above relation is used in all calculations below that involve the bulk viscosity. However, for very small collision frequency $\omega_{c,0} \lesssim 0.2$ we find it necessary to apply a strongly increased value of $\widetilde{\xi}_{0}\gg 1$ to suppress spurious viscous overstability.

%
%  
%  
%   \partial_{i} \left(\overline{z P_{iz}}\right) & = 
%   -\frac{1}{H \sigma}  \partial_{i} \left(\eta H^2\, \partial_{i} \frac{K}{H} \right)\label{eq:nuh2}\\
% %   
%  \end{align}
% where $i\in \{x,y\}$. For a Gaussian vertical stratification (\ref{eq:rho_gauss}) the expressions (\ref{eq:nuh1})-(\ref{eq:nuh2}) need to be computed numerically.

\subsection{Transport coefficients and model parameters}\label{sec:transportcoef}
\subsubsection{Particle collision model}

We assume a planetary ring consisting of identical, non-spinning, spherical, indestructable particles with diameter $d$ (radius $r_{p}=d/2$) and bulk density $\rho_{b}$. The particles undergo partially inelastic mutual binary collisions and $s$ is their mean separation distance. While the most accurate theoretical description of such a system is likely one based on a dense gas kinetic theory (see \S \ref{sec:intro}), in this paper we  apply a granular flow model. That is, we describe the planetary ring through the equations of hydrodynamics (\ref{eq:contsig})---(\ref{eq:conth}), but with expressions for pressure $\overline{p}$, shear and bulk viscosity $\overline{\eta}$ and $\overline{\xi}$, the heat conductivity $\overline{\kappa}$, and the cooling rate $\overline{\Gamma}$, which to some degree take into account the particulate and collisional nature of the system.
 \citet{borderies1985} applied a granular flow model to Saturn's dense rings, based on an incompressible hydrodynamic model using transport coefficients derived from a simple kinetic model of binary particle collisions derived by \citet{haff1983b}. Using this model they were able to predict the onset of viscous overstability in a sufficiently dense planetary ring with optical depth $\tau \gtrsim 1$. This was subsequently confirmed in N-body simulations by \citet{mosqueira1996}. Later, \citet{hutter1995} revised the model by \citet{haff1983b} and generalised it to account for two additional physical aspects of binary particle collisions which the original model omitted, and which may be of general importance to granular flows.
For one, they allowed for variations of the volumetric mass density $\rho$ of the particle ensemble, which requires the particle mean separation distance $s$ to vary. Furthermore, they accounted for a finite contact time at collisions, related the particle material properties.

Since we adopt a compressible  hydrodynamic model we follow \citet{hutter1995}, and allow for variations in the volume mass density. However, we neglect the finite contact time at particle collisions. We start by noting that the volumetric mass density may be written as
\begin{equation}\label{eq:dens2}
\rho = \left(\frac{d}{d+s}\right)^3 \rho_{p},
\end{equation}
where  where $\rho_{p}$ denotes the \emph{maximal} density of the particle configuration under consideration. We assume
\begin{equation}\label{eq:rhop}
\rho_{p}\equiv 0.61 \rho_{b},
\end{equation}
which is approximately the maximal volume density of uni-sized randomly packed spherical particles. Based on  Eq. (\ref{eq:dens2}) we
define the \emph{effective} filling factor
\begin{equation}\label{eq:ff}
\mbox{ff}\equiv \frac{\rho}{\rho_{p}}  ,
\end{equation}
in addition to the actual filling factor given by (\ref{eq:ff_true}).
% 
%A larger filling factor leads to a larger collision frequency between ring particles, which in turn results in an increased non-local viscosity and an increased value of $\beta$. 
%
The motivation is that the transport coefficients defined below should diverge if the densest possible packing is achieved, i.e. if $\rho\to \rho_{p}<\rho_{b}$, rather than if $\rho\to \rho_{b}$. Thus, the effective filling factor can take values $0<\mbox{ff}<1$.

If we assume $s\ll d$ (as was done in \citealt{haff1983b}) we have $\rho=\rho_{p}$, consistent with an incompressible model.
From (\ref{eq:dens2}) we directly obtain the particle \emph{mean separation distance}
\begin{equation}\label{eq:mfp}
s = d \left( \mbox{ff}^{-\frac{1}{3}}-1\right),
\end{equation}
where we used (\ref{eq:ff}). Thus, $s$ depends on the local state of `packing' of the system. 
More precisely, (\ref{eq:mfp}) describes the average distance particles traverse between two successive direct\footnote{by a `direct' collision we mean a collision where particles get in contact with each other.} collisions with other particles.
However, even in the absence of direct collisions, the particle \emph{mean free path} in a planetary ring is limited from above by the epicyclic excursion length $\sim c/\Omega_{0}$, since particles are constrained to undergo epicycles rather than stream indefinitely \citep{goldreich1978a}.
% Thus, Eq. (\ref{eq:mfp}) requires a modification to take this circumstance into account. 
We formally write the particle mean free path as
\begin{equation}\label{eq:mfpt}
    \widetilde{s} = d \left(\widetilde{\mbox{ff}}^{-\frac{1}{3}}-1\right),
\end{equation}
where we defined the modified filling factor
% \begin{align}\label{eq:fft}
%   \widetilde{\mbox{ff}} & = \frac{\rho+\rho_{l}}{\rho_{p}+\rho_{l}} = \frac{\mbox{ff} + \alpha}{1+ \alpha},
% \end{align}
\begin{align}\label{eq:fft}
   \widetilde{\mbox{ff}} & = \frac{\mbox{ff} + \alpha}{1+ \alpha},
\end{align}
with the dimensionless quantity $\alpha>0$,
% \begin{equation}\label{eq:alpha}
%     \alpha \equiv \frac{\rho_{l}}{\rho_{p}},
% \end{equation}
which ensures that the mean free path $\widetilde{s}$, unlike $s$, does not diverge with vanishing density $\rho\to 0$. On the other hand, for $\rho \to \rho_{p}$ we find   $\widetilde{s}\to s\to 0$. Thus, 
Eq. (\ref{eq:mfpt}) interpolates smoothly between these limiting cases, in a similar fashion as Eq. (\ref{eq:nuloc_gt}) for the local viscosity.
The quantity $\alpha$ follows from solving
\begin{equation}\label{eq:epic}
  \widetilde{s}(\rho\to 0) =  \frac{c(\rho\to 0)}{\delta \Omega_{0}} ,
\end{equation}
yielding
\begin{equation}\label{eq:rhol}
    \alpha = \left[\left(1+\frac{\widetilde{s}(\rho\to 0)}{ d}\right)^3-1\right]^{-1},
\end{equation}
where we defined the parameter $\delta\sim 1$, which will be stipulated below. 

For definition of the transport coefficients below we first need to define the collision frequency of ring particles.
From the time between two collisions we can define \citep{hutter1995} 
\begin{equation}\label{eq:omcol_dens}
    \omega_{c}=\frac{c}{s}=\frac{c}{d\left(\mbox{ff}^{-1/3}-1\right)},
\end{equation}
such that $\omega_{c}\to 0$ for $\mbox{ff}\to 0$ and $\omega_{c}\to \infty$ for $\mbox{ff}\to 1$. However, it is to be expected that this expression holds only for a sufficiently dense ring, since at small filling factor particle motions will be affected by epicyclic oscillations, as discussed above. Indeed, for $\mbox{ff}\to 0$ Eq. (\ref{eq:omcol_dens}) does not converge to the collision frequency of a dilute ring, which is given by \citep{shu1985c}:
\begin{equation}\label{eq:omcol_dil}
   \omega_{c}(\mbox{ff}\to 0) =\frac{24 \,c}{\pi^{3/2} d}\mbox{FF}.
\end{equation}
Recall that $\mbox{FF}= 0.61\, \mbox{ff}$ by Eq. (\ref{eq:rhop}).
We note that simply replacing $s$ in (\ref{eq:omcol_dens}) by $\widetilde{s}$ does not resolve this discrepancy, as then the collision frequency does not vanish for $\mbox{ff}\to 0$. Thus, we seek an expression for the collision frequency which converges to (\ref{eq:omcol_dens}) and (\ref{eq:omcol_dil}) for sufficiently large and small values of $\mbox{ff}$, respectively.
We find that such an expression is given by
\begin{equation}\label{eq:omc}
   \omega_{c}= \frac{3}{2}\frac{24 c}{\pi^{3/2} d}  \left(\frac{1}{5}\frac{1}{\mbox{ff}^{-1/3}-1}+1\right) \tanh \mbox{ff},
\end{equation}
which matches reasonably well the results from N-body simulations for small and large $\mbox{ff}$, as will be shown below.
Note that for $\mbox{ff}\to 0$ we have $\tanh \mbox{ff}\to \mbox{ff}$, and the bracket yields unity, such that $\omega_{c}(\mbox{ff}\to 0)$ becomes proportional to (\ref{eq:omcol_dil}). For $\mbox{ff}\lesssim 1$ the $\tanh$-function varies slowly and $\omega_{c}$ becomes essentially equal to (\ref{eq:omcol_dens}).

Furthermore, we define the dynamical optical depth of the planetary ring as 
\begin{equation}\label{eq:tau}
    \tau= \pi r_{p}^2 \frac{\sigma}{m_{p}} = \frac{3\sigma}{2 d \rho_{b}},
\end{equation}
$m_{p}$ denoting the particle mass, as well as
\begin{equation}\label{eq:rh}
    r_{h} = 0.82 \left(\frac{\rho_{b}}{900 \text{kg} \text{m}^{-3}}\right)^{\frac{1}{3}}\left( \frac{r}{10^{5} \text{km}}\right),
\end{equation}
which is the scaled mutual Hill radius for a pair of particles.
The latter is a convenient parameter to quantify the effect of self-gravity (\hyperlinkcite{salo2018}{S18};~\hyperlinkcite{mondino2023}{MS23}).
An increasing value of $r_{h}$ implies an increased self-gravity force compared as with tidal forces.

\subsubsection{Pressure, viscosity and  cooling rate}\label{sec:transcoefs}

Following \citet{hutter1995} we now define pressure via
\begin{equation}
\begin{split}\label{eq:pressz}
    p  =\frac{\text{"momentum"}}{\text{"area"}\times \text{"time"}}  \equiv \widehat{p}_{0} \frac{\rho\left(d+\widetilde{s}\right)^3 c}{\left(d+\widetilde{s}\right)^2 (\widetilde{s}/c)} =\widehat{p}_{0} \frac{\rho c^2}{1-\widetilde{\mbox{ff}}^{\frac{1}{3}}} 
    \end{split}
\end{equation}
where $\widehat{p}_{0}$ is a dimensionless constant, formally of order unity, which will be specified below. 
In addition to the \emph{total} pressure (\ref{eq:pressz}) we define its \emph{local} contribution (cf. \S \ref{sec:dense}) as:
\begin{align}
     p_{\text{loc}} & \equiv \rho c^2,\label{eq:press_locz} 
\end{align}
which directly follows from (\ref{eq:pressz}) in the limit $\widetilde{s}\gg d$. The latter is formally expected to apply if $\mbox{ff} \to 0$, such that (\ref{eq:pressz}) becomes  (\ref{eq:press_locz}) for $\mbox{ff} \to 0$ if the constant $\widehat{p}_{0}$ is fixed properly.
Thus, we expect our expression for the pressure (\ref{eq:pressz})  to be adequate throughout the dilute and dense regimes, i.e. for  $0< \mbox{ff}\lesssim 1$ (or $0< \mbox{FF} \lesssim 0.61$).
Based on (\ref{eq:pressz}) we can define an \emph{effective}  velocity dispersion (cf. \citealt{schmidt2001b})
\begin{equation}\label{eq:c0eff}
    c_{\text{eff}} =\sqrt{ \frac{\widehat{p}_{0}  }{1-\widetilde{\mbox{ff}}^{\frac{1}{3}}}}c,
\end{equation}
which effectively includes local as well non-local contributions when used in the local pressure formula (\ref{eq:press_locz}). Thus, the quantity $c_{\text{eff}}$ generally takes larger values than the actual velocity dispersion $c$.

By following again \citet{hutter1995} we can define the dynamic shear viscosity via
\begin{equation}\label{eq:viscz_dens}
    \begin{split}
        \eta = \frac{\text{"mass density"}\times \text{"area"}}{\text{"time"}}  = \widehat{\nu}_{0}\frac{\rho \left(d+\widetilde{s}\right)^2}{(s/c)}=\widehat{\nu}_{0} \rho  d^2 \omega_{c}  \widetilde{\mbox{ff}}^{-\frac{2}{3}},
    \end{split}
\end{equation} 
with the dimensionless constant $\widehat{\nu}_{0}$ specified below. Note that we here consider the time between direct particle collisions $s/c$, as the latter are  required to produce a viscosity, in contrast to pressure defined above. 
However, it turns out that Eq. (\ref{eq:viscz_dens}) is only adequate for sufficiently large values of $\mbox{ff}$, similar to (\ref{eq:omcol_dens}). 
In order to correctly describe the viscosity for $\mbox{ff}\to 0$ we therefore combine (\ref{eq:viscz_dens}) with the classical formula (\ref{eq:nuloc_gt}) for a dilute ring, such that
\begin{equation}\label{eq:viscz}
    \eta \equiv \rho d^2 \omega_{c} \left[ \widehat{\nu}_{\text{0}}  \widetilde{\mbox{ff}}^{-\frac{2}{3}}  + \widehat{\nu}_{\text{1}}  \frac{\left(\frac{c}{\Omega d}\right)^2}{1 + \left(\frac{\omega_{c}}{\Omega}\right)^2} \right] ,
\end{equation}
where we introduced the dimensionless coefficient $\widehat{\nu}_{1}$.
Note that (\ref{eq:viscz_dens}) differs from the classical formula for non-local viscosity (\ref{eq:nu_nonlocal}) only by a factor of $\widetilde{\mbox{ff}}^{-\frac{2}{3}}$, which mildly decreases with increasing $\mbox{ff}$. 
% In fact, the actual behavior of the non-local viscosity as found in N-body simulations is more complicated than the simple expression (\ref{eq:nu_nonlocal}), and shows a drop with increasing velocity dispersion $c$, not captured by (\ref{eq:nu_nonlocal}). On the other hand, an increased value of $c$ implies an increased value of the ring thickness $H$, and thus a reduced filling factor $\mbox{ff}$, resulting in a mild decrease of (\ref{eq:viscz_dens}).
Note that for $\mbox{ff}\to 0$ we have $\omega_{c}\to 0$ and $\widetilde{\mbox{ff}}\to const.$. If in addition $c$ approaches a constant value (which will be confirmed below) then (\ref{eq:viscz}) converges to (\ref{eq:nuloc_gt}) for small $\mbox{ff}$, as it should. At this point it should be noted that neither (\ref{eq:nu_nonlocal}) nor (\ref{eq:viscz_dens}) are rigorously derived expressions for the non-local viscosity at large filling factor. Nevertheless, based on the above, we expect our expression for the viscosity (\ref{eq:viscz}) to behave at least qualitatively correct across the dilute and dense regimes, which will be confirmed below.
% In the dense limit $\rho \to \rho_{p}$ the two time scales coincide. However, in the limit of small density $\rho\to 0$, the latter remains finite, whereas the former diverges, since then direct collisions are absent. 

Since heat conductivity and viscosity have the same dimensions, we simply define the dynamic heat conductivity via
\begin{equation}
\kappa \equiv \widehat{\kappa}_{0}\eta,
    % \kappa = \widehat{\kappa}_{0}\rho c d \frac{\widetilde{\mbox{ff}}^{-\frac{2}{3}} \mbox{ff}^{\frac{1}{3}} }{1-\mbox{ff}^{\frac{1}{3}}},
\end{equation}
where $\widehat{\kappa}_{0}$ is again a dimensionless constant of order unity.
  Typical values  $\widehat{\kappa}_{0}\sim 1-5$ were derived from N-body simulations \citep{salo2001b} applied to Saturn's B ring. In our stability analysis below we do not find any notable influence of heat conductivity on the linear viscous overstability. However, the inclusion of a finite value  $\widehat{\kappa}_{0}\gtrsim 0.2$ is found to be necessary to avoid spurious secular instability on length scales $\lambda \gtrsim H_{0}$. Therefore, we simply set
\begin{equation}
\overline{\kappa_{0}}=\overline{\eta_{0}}
\end{equation}
in all calculations presented below which involve the heat flow equation (\ref{eq:linT}).

Finally, the cooling rate (or energy annihilation rate) is given by \citep{goldreich1978c,hutter1995}
\begin{equation}
\begin{split}\label{eq:gamz}
    \Gamma = \frac{\text{"dissipated specific kinetic energy"}}{\text{"time"}} & =  \frac{\frac{1}{2} \rho c^2 \left(1-\epsilon^2\right)}{(s/c)} \\
    \quad & =  \frac{1}{2 } \rho c^2 \left(1-\epsilon^2\right) \omega_{c},
    \end{split}
\end{equation}
which describes the amount of specific kinetic energy lost via collisions, characterised by the normal coefficient of restitution $\epsilon$ (see \S \ref{sec:epsilon}).
Note that we again use the time between inter-particle collisions, which are required to dissipate energy. 

% In addition to the \emph{total} pressure (\ref{eq:pressz}) and viscosity (\ref{eq:viscz}), we define their \emph{local} contributions (cf. \S \ref{sec:dense}) as:
% \begin{align}
%      p_{\text{local}} & \equiv \rho c^2,\label{eq:press_locz}\\
%     \eta_{\text{local}} & \equiv  \widehat{\nu}_{0,\text{local}} \,\rho  d^2 \big(\widetilde{\mbox{ff}}^{-\frac{1}{3}}-1\big)^2 \omega_{c}\label{eq:visc_locz}, 
% \end{align}
% which directly follow from (\ref{eq:pressz}) and (\ref{eq:viscz}) in the limit $\widetilde{s}\gg d$. The latter is expected to apply if $\rho \to 0$, such that (\ref{eq:pressz}) and (\ref{eq:viscz}) should approach (\ref{eq:press_locz}) and (\ref{eq:visc_locz}) for $\rho \to 0$, respectively. 

%
%
If we assume the uniform density distribution (\ref{eq:rho_slab})
vertical integration of above transport coefficients yields
\begin{align}
   \overline{p} & =\widehat{p}_{0} \frac{\sigma c^2}{1-\widetilde{\mbox{ff}}^{\frac{1}{3}}}\label{eq:press} = \sigma c_{\text{eff}}^2,\\
   \overline{\eta} & = \sigma  d^2 \omega_{c}\left[ \widehat{\nu}_{0}\widetilde{\mbox{ff}}^{-\frac{2}{3}} +  \widehat{\nu}_{1}\frac{\left(\frac{c}{\Omega_{0} d}\right)^2}{1+\left(\frac{\omega_{c}}{\Omega_{0}}\right)^2}\right] \label{eq:visc},\\
   \overline{\kappa} & = \widehat{\kappa}_{0}\overline{\eta},\\
   \overline{\Gamma} & =  \frac{1}{2} \sigma c^2 \left(1-\epsilon^2\right) \omega_{c}\label{eq:gammaf},
\end{align}
as well as
\begin{align}
    \overline{p_{\text{loc}}} & = \sigma c^2\label{eq:press_loc}.
    % \overline{\eta_{\text{loc}}} & = \widehat{\nu}_{\text{loc}} \sigma  \omega_{c}\frac{c^2}{\Omega_{0}^2 + \omega_{c}^2}\label{eq:visc_loc}.
\end{align}
If we apply the Gaussian vertical distribution (\ref{eq:rho_gauss}) the vertical integrals need to be computed numerically.
% In what follows, for clarity the bars will be dropped with the understanding that all quantities are vertically integrated.

% In addition it is useful to define the (dimensionless) sound speed, given by
% \begin{equation}\label{eq:soundspeed}
% c_{s} = \sqrt{\overline{p_{\sigma}} +  \frac{2}{3} \overline{p} \, \overline{p_{T}} },
% \end{equation}
% where $\overline{p_{\sigma}}=\overline{\partial p /\partial \sigma}$ and $\overline{p_{T}}=\overline{\partial p / \partial T}$.
% In the limit of a dilute gas ($c_{\text{eff}} \to c$) the sound speed becomes $c_{s}=\sqrt{5/3} c$, which is the sound speed of an ideal mono-atomic gas (with adiabatic index 5/3).
% With (\ref{eq:soundspeed}) we can define the hydrodynamic Toomre parameter
% \begin{equation}\label{eq:toomreH}
%     Q_{\text{hydro}} = \frac{\Omega_{0} c_{s} }{\pi G \sigma},
% \end{equation}
% where the subscript "hydro" is used to distinguish (\ref{eq:toomreH}) from the actual Toomre parameter (\ref{eq:toomreK}).

% We note that unlike \citet{hutter1995}, we assume that particle collisions are instantaneous and do not consider a finite contact time at collision. This would result in an additional term in the denominators of (\ref{eq:press})-(\ref{eq:visc_loc}), that depends on the  particle radii and material sound speeds. However, this extension and its consequences may be considered in future studies.

\subsubsection{Coefficient of restitution}\label{sec:epsilon}

Finally, $\epsilon$ is the normal coefficient of restitution, describing the irreversible loss of random kinetic energy of particles subsequent to a collision, as described by (\ref{eq:gamz}). In Saturn's main rings, collisions are expected to be rather dissipative \citep{porco2008}, characterised by small values of $\epsilon$. For simplicity, and in order to facilitate comparison with the recent N-body simulations of \hyperlinkcite{mondino2023}{MS23} we will mainly work with a constant $\epsilon=0.1-0.5$.
Complementary, we also consider a velocity-dependent coefficient of restitution using the piece-wise powerlaw prescription \citep{bridges1984}
\begin{equation}\label{eq:bridges}
\epsilon(v_{n})=\left\{\begin{array}{cl}
\left(\frac{v_{n}}{v_{c}}\right)^{-0.234} & \text { if } v_{n}>v_{c} \\
1 & \text { if } v_{n} \leq v_{c}
\end{array}\right.
\end{equation}
which has been used in most kinetic studies and N-body simulations of Saturn's dense rings.
Here we identify $v_{n} \equiv c$ as an averaged value of the mutual normal velocity between particles, and the scale parameter $v_{c}$, with nominal value $v_{c}\equiv v_{b}=0.077 \mathrm{mms}^{-1}$. For illustration purposes below we will also consider the case $v_{c}=10 v_{b}$, which results in a significantly hotter ground state.

\subsubsection{Remarks}

Our expresions for the pressure (\ref{eq:pressz}) and the viscosity (\ref{eq:viscz_dens}) are analogous to Eqs.
(46) and (47) in \citet{borderies1985}, except that we (as \citealt{hutter1995}) do not assume $\widetilde{s}\ll d$, and we use the (corrected) mean
free path expression $\widetilde{s}$, rather than the mean separation distance $s$.
Furthermore, using (\ref{eq:mfpt}) the requirement $\widetilde{s}\gg d$, used to derive the local pressure (\ref{eq:press_locz}), translates to the condition $\alpha \ll 1$. As will turn out in our computations presented below, this is realized only if $\delta \lesssim 1$. Physically, the fulfilment of the condition $\alpha \ll 1$ means that the epicyclic departure length of ring particles (Eq. \ref{eq:epic}) should be much larger than the particle diameter $d$. However, as outlined in \S \ref{sec:dense}, in Saturn's rings this condition is not expected to be fulfilled on account of their small velocity dispersion. In our calculations below we use $\delta=2$, but generally values $1<\delta<4$ result in reasonable agreement with results from N-body simulations.

\begin{table}
 \caption{Important symbols and definitions.}
 \label{tab:anysymbols}
 \begin{tabular*}{\columnwidth}{@{}l@{\hspace*{50pt}}l@{\hspace*{50pt}}l@{}}
  \hline
  Symbols  & Definitions\\
  \hline
  $\omega$  & complex eigenvalue of perturbation\\[2pt]
     $k$  & radial wavenumber of perturbation\\[2pt]
   %  $d$  & particle diameter\\[2pt]
     $s$  & particle mean separation length\\[2pt]
       $\tau$  & dynamical optical depth\\[2pt]
    $\widetilde{s}$  & particle mean free path\\[2pt]
    $\mbox{FF}$  &  volume filling factor\\[2pt]
    $\mbox{ff}$  & effective volume filling factor\\[2pt]
     $\widetilde{\mbox{ff}}$  & modified volume filling factor\\[2pt]
  $\beta$  & viscous slope\\[2pt]
  $\gamma$  & vertical stratification factor\\[2pt]
   $p$ & total pressure \\[2pt]
  $\eta$  & total dynamic shear viscosity \\[2pt]
     $p_{\text{loc}}$ & local pressure \\[2pt]
  $\eta_{\text{loc}}$  & local dynamic shear viscosity \\[2pt]
    $\xi$  & dynamic bulk viscosity \\[2pt]
     $\widetilde{\xi}$  & ratio of $z$-averaged dynamic bulk and shear viscosity \\[2pt]
    $\kappa$  &  dynamic heat conductivity\\[2pt]
      $\Gamma$  &  cooling rate\\[2pt]
  $\Omega_{0}$  & orbital reference frequency\\[2pt]
    $\Omega_{z}$  & effective vertical frequency \\[2pt]
  %$r_{0}$  & reference disc radius\\[2pt]
  % $\epsilon$  & coefficient of restitution\\[2pt]
    $\rho$  &  volume mass density \\[2pt]
        $\rho_{p}$  &  maximal volume density of granular fluid \\[2pt]
   $\rho_{b}$  &  particle bulk density  \\[2pt]
      $\sigma$  &  surface mass density\\[2pt]
        $u_{x/y}$  &   Eulerian radial/azimuthal velocities \\[2pt]
      $u_{z}$  &   Lagrangian vertical velocity \\[2pt]
  $H$  &  disc thickness parameter \\[2pt]
    $\dot{H}$  &  speed of vertical disc breathing \\[2pt]
  $T$  &  temperature\\[2pt]
   $c$  &  velocity dispersion\\[2pt]
     $g$  &  self-gravity parameter \\[2pt]
          $\zeta$  &  thickness correction factor for radial self-gravity \\[2pt]
        % $Q_{\text{hydro}}$  &  hydrodynamic Toomre parameter \\[2pt]
     $r_{h}$  &  scaled mutual Hill radius of particle pair\\[2pt]
  \hline
 \end{tabular*}
\end{table}

%
%

% \begin{table}
%  \caption{Model parameter values.}
%  \label{tab:anysymbols}
%  \begin{tabular*}{0.7\columnwidth}{@{}l@{\hspace*{20pt}}l@{\hspace*{20pt}}l@{}}
%   \hline
%   Model parameter  & Value\\
%   \hline
%   $\delta$  & 2\\[2pt]
%     $\epsilon$  & 0.1-0.5 or velocity dependent\\[2pt]
%      $\widehat{\nu}_{0}$  &  0.053\\[2pt]
%       $\widehat{\nu}_{1}$  &  0.053\\[2pt]
%      $\widetilde{\xi}_{0}$   & 2-4\\[2pt]
%      $\widehat{\kappa}_{0}$ & 1\\[2pt]
%      $d$  &  2m\\[2pt]
%        $r_{0}$  & 100,000km\\[2pt]
%   \hline
%  \end{tabular*}
% \end{table}

%

\section{Equilibrium state}\label{sec:equilibrium}

In this section we derive the equilibrium state of a planetary ring corresponding to stationary solutions of Eqs. (\ref{eq:contsig})---(\ref{eq:conth2}) using the constitutive relations and transport coefficients described in the previous sections.

\subsection{Numerical procedure}

The ring equilibrium is characterised by $\sigma_{0}  = \text{const.}, u_{x0}  =0, u_{y0}  =-3/2\Omega_{0} x$ and $\dot{H}_{0}  =0$,
where $\sigma_{0}$ is a free parameter.
Note that since $\sigma_{0}$ is a constant, radial self-gravity forces vanish at equilibrium.
Furthermore, values for the particle bulk density $\rho_{b}$, particle diameter $d$ and the dimensionless viscosity coefficients $\widehat{\nu}_{0}$ and $\widehat{\nu}_{1}$ need to be specified. A list of important parameters and scales can be found in Table \ref{tab:anysymbols}.
In what follows we will work with dimensionless quantities, such that
time is scaled with $1/\Omega_{0}$, length is scaled with $d$, and surface density is scaled with $\sigma_{0}$.
Throughout the remainder of this paper all values of $\rho_{b}$ and $\sigma_{0}$ are understood to be in units of kg/m$^3$ and kg/m$^2$, respectively.

\subsubsection{Energy balance - equilibrium velocity dispersion}

From Equation (\ref{eq:conttemp}) at equilibrium we directly find the thermal equilibrium condition
\begin{equation}\label{eq:ebalance}
\overline{\Gamma_{0}} =  \frac{9}{4} \overline{\eta_{0}},
\end{equation}
which describes the balance between viscous heating and collisional cooling and where $\overline{\Gamma_{0}}$ and $\overline{\eta_{0}}$ are given by (\ref{eq:visc}) and (\ref{eq:gammaf}) at equilibrium, respectively. For the slab profile (\ref{eq:rho_slab}) this yields a quartic equation for the velocity dispersion, whose solutions can be obtained analytically, but are unwieldy. Instead, we consider some limiting cases. In particular, if the ring is hot and dilute with $c_{0}\gg 1$ and $\mbox{ff}\ll 1$, the viscosity (\ref{eq:visc}) becomes equal to (\ref{eq:nuloc_gt}), and we find
\begin{equation}\label{eq:eps_tau_relation}
    \left(1-\epsilon^2\right)\left(1+\omega_{c}^2\right) = \frac{9}{2}\widehat{\nu}_{1}.
\end{equation}
For $\mbox{ff}\ll 1$ we have $\omega_{c}\sim \tau$ (as shown below, see also \citealt{shu1985c}) and the above equation becomes the classical \citeauthor{goldreich1978c} $\epsilon_{\text{cr}}-\tau$ relation, where the unique solution $\epsilon_{\text{cr}}$ separates thermally stable and unstable equilibria.
On the other hand, if we consider a dense ring ($c_{0} \sim 1$) in the limit of large collision frequency $\omega_{c}\gg 1$ Eq. (\ref{eq:viscz}) becomes equal to (\ref{eq:viscz_dens}), and the energy balance (\ref{eq:ebalance}) for the slab distribution readily yields
 \begin{equation}\label{eq:veldisp}
     c_{0,\text{S}(\omega_{c}\gg 1)} = \sqrt{\frac{9 \widehat{\nu}_{0} }{2\left(1-\epsilon^2\right) \widetilde{\mbox{ff}}_{0}^{\frac{2}{3}}}},
 \end{equation}
 such that now the parameter $\widehat{\nu}_{0}$ controls the amount of (non-local) viscous heating, which is balanced by collisional cooling, quantified by $\epsilon$.
% Note that the velocity dispersion (\ref{eq:veldisp}) remains finite for $\rho_{0}\to 0$, in agreement with kinetic models and N-body simulations.
Note that the equilibrium temperature $T_{0}\equiv c_{0}^2$.

 \subsubsection{Vertical hydrostatic balance - equilibrium scaleheight}
Next, from (\ref{eq:conth2}) at equilibrium we readily find the equilibrium disc scale height, which for the slab profile reads
\begin{equation}\label{eq:scaleheight}
    H_{0,\text{S}}   \equiv g  \left(\sqrt{\frac{3 \overline{p_{0}}}{ g^2} +1} -1\right),
\end{equation}
where we defined the dimensionless self-gravity parameter
\begin{equation}
    g = \frac{\pi G \sigma_{0}}{\Omega_{0}^2 d}
\end{equation}
and $\overline{p_{0}}$ denotes the pressure (\ref{eq:press}) at equilibrium.
Thus, the equilibrium scaleheight is determined by the competition between vertical planetary gravity and self-gravity acting against pressure. For a Gaussian profile (\ref{eq:rho_gauss}) we find
\begin{equation}\label{eq:scaleheight_gauss}
    H_{0,\text{G}}   \equiv \frac{g}{\sqrt{\pi}}  \left( \sqrt{\pi \frac{ \overline{p_{0}}}{ g^2} +1} -1\right).
\end{equation}
Moreover, if we model vertical self-gravity through a constant value of $\Omega_{z}$ (cf. (\ref{eq:omz})), we find 
\begin{equation}\label{eq:scaleheight_omz}
    H_{0,\Omega_{z}}  \equiv \frac{1}{\Omega_{z}}\sqrt{\frac{\overline{p_{0}}}{\gamma}},
\end{equation}
with $\gamma$ given by (\ref{eq:gamma}). In absence of vertical self-gravity ($g\to 0$, or equivalently, $\Omega_{z}\to 1$), the equilibrium scaleheight becomes $H_{0}= \sqrt{ \overline{p_{0}}/\gamma }$.

Furthermore, as stated in \S \ref{sec:transportcoef}, in the limit $\sigma_{0}\to 0$ we require
\begin{align}\label{eq:ptoploc}
 \overline{p_{0}}\to \overline{p_{0,\text{loc}}}.
  % \overline{\eta_{0}}\to \overline{\eta_{0,\text{loc}}}\label{eq:etatoetaloc}.
\end{align}
Using (\ref{eq:fft})---(\ref{eq:rhol}) this yields the relation
\begin{align}
    \widehat{p}_{0} & =\frac{c_{0}(\rho\to 0)}{\delta+c_{0}(\rho\to 0)}\label{eq:p0h}.
      % \widehat{\nu}_{0,\text{loc}} & = \widehat{\nu}_{1} + \widehat{\nu}_{0} \left(\frac{1}{c_{0}(\rho\to 0)} +\frac{1}{\delta}\right)^2\label{eq:nu0hloc} 
\end{align}

The ring equilibrium is then computed as follows.
We first specify values for $r_{0}$, $\rho_{b}$, $\epsilon$,  $d$ , $\widehat{\nu}_{0}$ and $\widehat{\nu}_{1}$ which are then kept fixed.
Then, for sufficiently small $\sigma_{0}\to 0$ we assume initial values for $H_{0}$ and $c_{0}$. From these we readily obtain $\alpha$ via (\ref{eq:rhol}) and (\ref{eq:p0h}). Using (\ref{eq:ff}) and (\ref{eq:fft}) we can evaluate the vertical integrals (\ref{eq:press})---(\ref{eq:press_loc}) and apply a multi-dimensional Newton-Raphson method to solve (\ref{eq:ebalance}) and (\ref{eq:scaleheight}) concurrently.
We then use the so obtained values for $\alpha$, $\widehat{p}_{0}$ and $\widehat{\nu}_{0,\text{loc}}$ in subsequent iterations adopting increasing values of $\sigma_{0}$.
% , where we only need to solve (\ref{eq:ebalance}) and (\ref{eq:scaleheight}) by specifying initial values for $H_{0}$ and $c_{0}$.

% \begin{figure}
% \centering
% \includegraphics[width = 0.4\textwidth]{figs/gam.jpg}
% \caption{The ratio of the shear and the bulk viscosity $\gamma$ as given by (\ref{eq:gamma}) is drawn as solid curve. The dashed curve is the prescription used in our hydrodynamic simulations (Section \ref{sec:numerics}).}
% \label{fig:gamma}
% \end{figure}
% %\FloatBarrier
%
%\noindent

% For the value of $\gamma$, denoting the ratio of bulk and shear viscosity, we adopt a prescription which is very similar to the result obtained in \citet{chapman1970} (their Chapter 16) for a dense gas of 
% elastic spheres:
% \begin{equation}\label{eq:gamma}
%  \gamma = \frac{1.326 \mbox{ff}^2 \left(0.4+\mbox{ff}\right)^2}{0.013 + 0.042 \mbox{ff} +0.265 \mbox{ff}^2 + 0.8 \mbox{ff}^3 + \mbox{ff}^4},
% \end{equation}
%

\subsection{Results}\label{sec:equil_res}

\begin{figure*}
\centering
\includegraphics[width = \textwidth]{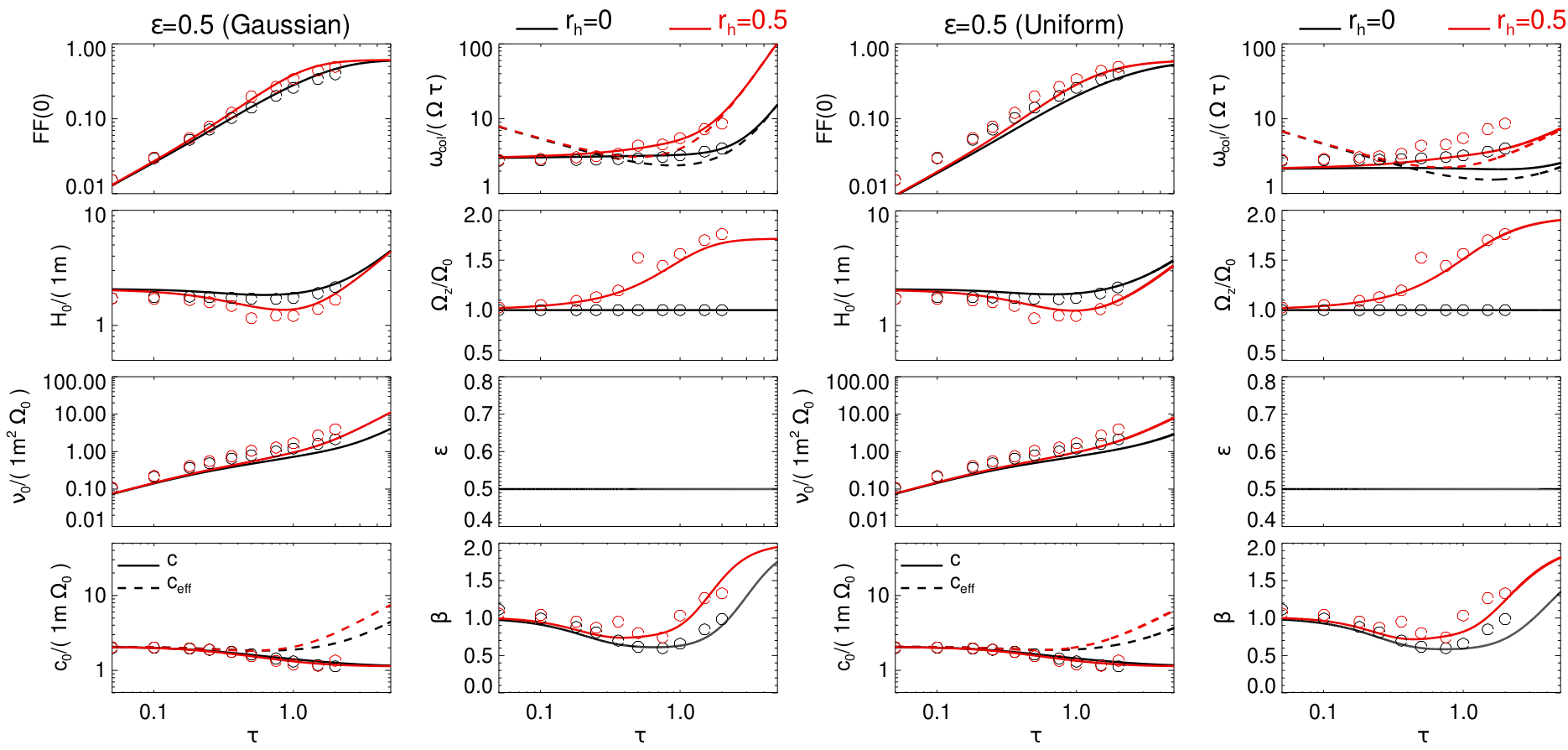}
\caption{Ring equilibrium states following from our hydrodynamic model (curves) using a coefficient of restitution $\epsilon=0.5$ are compared with corresponding results from N-body simulations (open circles), described in \S \ref{sec:nbody}. Shown are from top to bottom panels the filling factor, the scaleheight, the kinematic viscosity and the velocity dispersion (first and third columns), as well as the collision frequency, the vertical self-gravity factor, the coefficient of restitution, as well as the viscous slope (second and fourth columns).
The left panels correspond to a Gaussian vertical distribution (\ref{eq:rho_gauss}), whereas the right panels assume a uniform density distribution  (\ref{eq:rho_slab}). We compare ring states without ($r_{h}=0$) and with ($r_{h}=0.5$) vertical self-gravity. In the panel displaying the collision frequency $\omega_{c}$ the dashed curves represent Eq. (\ref{eq:omcol_dens}).}
\label{fig:gstate_eps05}
\end{figure*}

\begin{figure*}
\centering
\includegraphics[width = \textwidth]{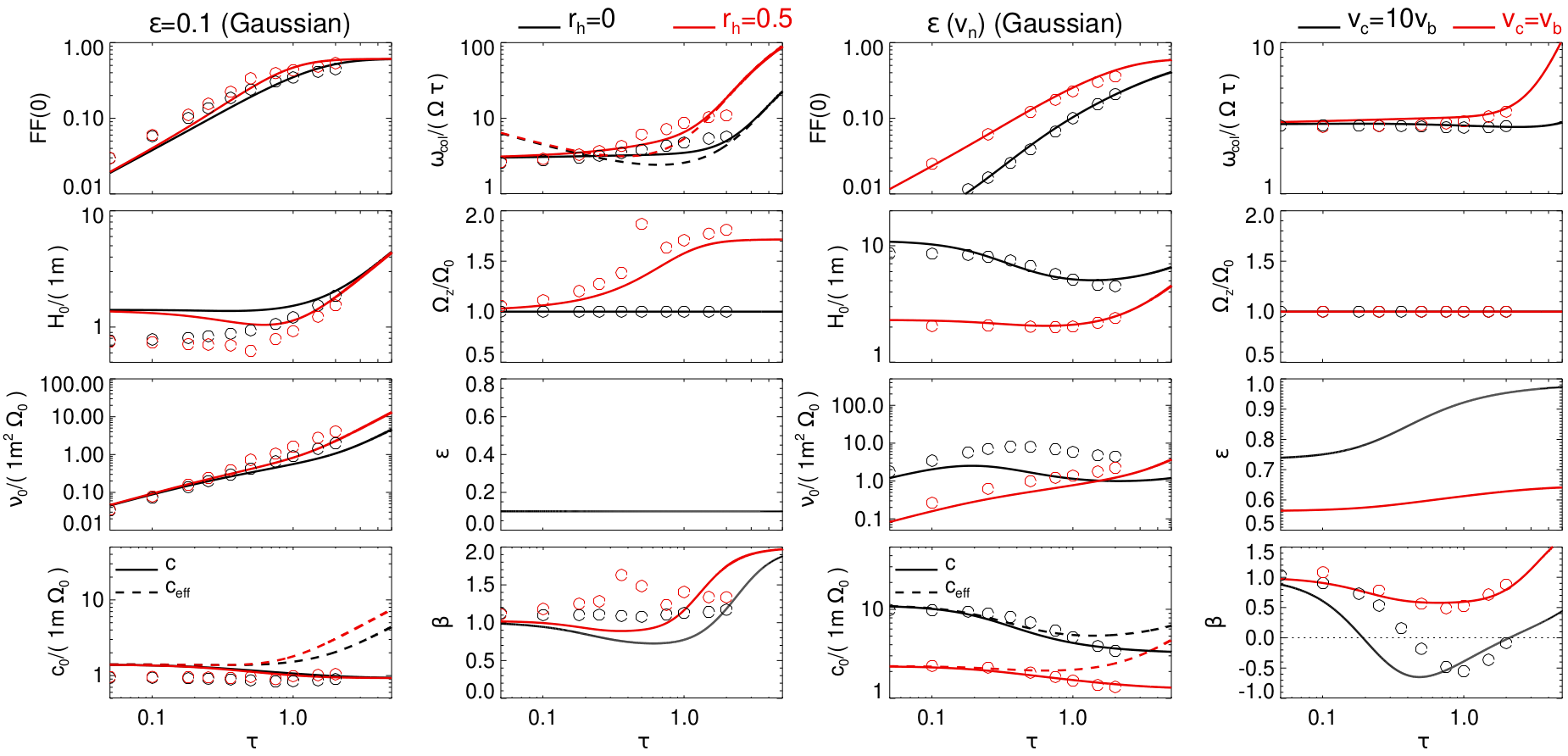}
\caption{Same as Figure \ref{fig:gstate_eps05}, but now assuming $\epsilon=0.1$ (left), and assuming a velocity dependent coefficient of restitution $\epsilon(v_{n})$ given by (\ref{eq:bridges}) (right). The N-body simulations adopted the scale parameters $v_{c}= v_{b}$ (red symbols) and$v_{c}=10 v_{b}$ (black symbols), respectively. For the computation of the hydrodynamic equilibra we used  $v_{c}=0.5 v_{b}$ (red curves) and$v_{c}=15 v_{b}$ (black curves), respectively.}
\label{fig:gstate_combo}
\end{figure*}

 % \begin{figure}
 % \centering
 % \includegraphics[width = 0.5\textwidth]{figs/equilibrium_bridges_new.png}
 % \caption{Same as Figure \ref{fig:gstate_eps05}, but now assuming a velocity dependent coefficient of restitution $\epsilon(v_{n})$ given by (\ref{eq:bridges}). The N-body simulations adopted the scale parameters $v_{c}= v_{b}$ (red symbols) and$v_{c}=10 v_{b}$ (black symbols), respectively. For the computation of the hydrodynamic equilibra we used  $v_{c}=0.5 v_{b}$ (red curves) and$v_{c}=15 v_{b}$ (black curves), respectively.}
 % \label{fig:gstate_epsbrid}
 % \end{figure}
 % %

In Figure \ref{fig:gstate_eps05} we present various equilibrium quantities as functions of the optical depth $\tau$, for non-gravitating ($r_{h}=0$) and self-gravitating ($r_{h}=0.5$) rings, where we consider a Gaussian vertical distribution (left panels) as well as a uniform vertical distribution (right panels). We use a constant $\epsilon=0.5$ and adopt a distance $r_{0}=10^5 \text{km}$ from the central planet. The particle size is $d=2\text{m}$ and we use $\widehat{\nu}_{0}=\widehat{\nu}_{1}=0.053$. The latter values are chosen as they yield good overall agreement with equilibrium properties inferred from N-body simulations, as can clearly be seen in the left panels. Note that the significant departure of quantities $\Omega_{z}$ and $\beta$ at $\tau\sim 0.5$ is related to the formation of a layer structure in the simulations.  That is, at $\tau=0.5$ the system is has a central peak, whereas at $\tau=0.75$ it starts to form two layers.
% The only quantity that shows notable departures from the simulation results is the local viscosity. This might indicate that Eq. (\ref{eq:nuloc_gt}) is not accurate enough to yield quantitative agreement with N-body simulations.

The displayed values of $H_{0}$ in N-body simulations represent the root mean squared distance of all particles from the mid-plane: $\sqrt{\langle z^2 \rangle}$. Likewise, the hydrodynamic results correspond to the value $\sqrt{\overline{z^2}}$ (cf. \S \ref{sec:vstruct}).
The resulting values correspond to a ring which is a few particle radii thick, and in the non-gravitating case the scaleheight slightly increases with increasing optical depth. This increase is a consequence of the increasing non-local pressure (indicated by $c_{\text{eff}}$) with increasing $\tau$. In the presence of vertical self-gravity the ring is subject to a significant flattening for optical depths $\tau\sim 1$, but eventually thickens, again on account of increasing non-local pressure.
On the other hand, the velocity dispersion $c$ is nearly independent of vertical self-gravity and in all cases steadily decreases, approaching the value $\Omega r_{p}$ in the optically thick limit $\tau\gg 1$. Furthermore, the ``viscous slope'' $\beta$ is
defined (as in \citealt{salo2001}) via
\begin{equation}\label{eq:beta2}
    \beta \equiv \frac{\partial \ln \overline{\nu_{0}}}{\partial \ln \sigma_{0}}, 
\end{equation}
where the subscript "0" is used to indicate that the derivative is computed (numerically) using the values of $\overline{\nu}$ corresponding to ring equilibrium states with increasing values of $\sigma_{0}$. As such, $\beta \neq  (\partial \ln \overline{\nu}/\partial \ln \sigma)_{0}$, since the latter derivative is understood to be taken using fixed values of $H_{0}$ and $c_{0}$, whereas the latter two parameters also vary in between ground states with different $\sigma_{0}$, as considered in (\ref{eq:beta2}) \footnote{Note that in the case of Eq. (\ref{eq:beta}) we have equality $\beta =  (\partial \ln \overline{\nu}/\partial \ln \sigma)_{0}$, since in this case $\sigma$ is the only independent variable characterising the viscosity.}.
That being said, we find that $\beta$ exceeds unity for values of the optical depth $\tau \sim 1-2$, depending on the magnitude of self-gravity. A value $\beta \gtrsim 1$ is expected to trigger the onset of the viscous overstability (\S \ref{sec:dense} and \S \ref{sec:linpert}). 
The flattening by vertical self-gravity results in an increased filling factor. Thus, non-local effects are augmented, yielding a larger viscosity and pressure. Consequently, $\beta$ approaches unity at a significantly smaller optical depth $\tau \sim 1$ than in the non-gravitating case where $\tau\sim 2$,  such that viscous overstability should be substantially promoted.
% In the plot for $\nu_{0}$ (left panels) the dashed curves represent the local viscosity (\ref{eq:visc_loc}). In addition, the circles correspond to Eq. (\ref{eq:nuloc_gt}) with a pre-factor of $\widehat{\nu}_{0,\text{GT}}=0.18$, such that the latter two curves are nearly identical for all $\tau$. 

Comparing the results for the uniform and Gaussian distributions, we see that a Gaussian results in an overall more self-consistent equilibrium state in that all quantities agree well with the simulation results. Interestingly, the viscosity, which is the key quantity for our linear analysis below takes very similar values for the two vertical distributions. It should be noted that strictly, the dilute expression for the collision frequency (\ref{eq:omcol_dil}) assumes a Gaussian vertical distribution. However, even if we correct this expression for a uniform distribution the overall agreement with N-body simulations is still significantly worse than for a Gaussian distribution.
% only marginal differences. However, the behavior of a planetary ring at large $\tau$-values is expected to be better described by the slab model. One indication for this is the behavior of the local viscosity for $\tau \gtrsim 1$. 
% Further note that the values of $\mbox{FF}$ and $\omega_{c}$ are the mid-plane values in the case of a Gaussian distribution.}

% Furthermore, at large values of $\rho_{b}$ the Gaussian prescription yields a ring thickness of less than a particle diameter $d=2\text{m}$, which is inconsistent, and indicates that the assumption of a Gaussian vertical structure is inadequate in this case.
In addition, Figure \ref{fig:gstate_combo} shows the same quantities as in Figure \ref{fig:gstate_eps05} but for a constant $\epsilon=0.1$ (left panels), as well as a velocity dependent coefficient of restitution (\ref{eq:bridges}) (right panels).  
In the case with $\epsilon=0.1$ we used $\widehat{\nu}_{0}=\widehat{\nu}_{0}=0.047$. Here the agreement between model and simulation is not as good as in the case with $\epsilon=0.5$. The equilibrium following from the simulations is significantly cooler at small optical depths. 

In the right panels we consider two values of the scale parameter $v_{c}=v_{b}$ (black curves) and $v_{c}=10 v_{b}$ (red curves). The former system is very similar to one with a constant $\epsilon=0.5$, which can seen in the panel displaying $\epsilon(\tau)$. On the other hand, a system with $v_{c}=10 v_{b}$ results in a substantially hotter system, such that the velocity dispersion $c\gg 1$. Note that we computed the hydrodynamic curves using $v_{c}=0.5 v_{b}$ and $v_{c}=15 v_{b}$, as these values yield better agreement with the simulations. Furthermore, we used $\widehat{\nu}_{0}=\widehat{\nu}_{1}=0.053$ for the case  $v_{c}=0.5 v_{b}$ and $\widehat{\nu}_{0}=0.019$, $\widehat{\nu}_{1}=0.094$ for the case $v_{c}=15 v_{b}$ (i.e. a significantly enhanced local viscosity). We can clearly see that the agreement between model and simulations is significantly reduced for the hotter equilibrium state. We again suspect that the main reason for this disagreement is the inaccuracy of the classical expression (\ref{eq:nuloc_gt}) for the local viscosity. Nevertheless, this is issue not of great concern in the present study as below we will focus on dense cold equilibrium states.

\section{Linear Analysis}\label{sec:linpert}

Now that we have established the equilibrium state of the planetary ring, we turn our attention to the linear perturbation modes it supports.
We analyze the stability of short-scale axisymmetric perturbations that we add to the equilibrium state, such that the perturbed variables read 
\begin{equation}
\begin{split}\label{eq:perturbations}
\{\sigma,u_x , u_y , T , H , \dot{H} \}& = \{\sigma_{0},0 , -\frac{3}{2}x , T_{0} , H_{0} , 0\} \\
 & \quad + \{\delta \sigma,\delta u_x,\delta u_y,\delta T,\delta H,\delta \dot{H} \} \cdot \exp\left(i k x + \omega t \right),
\end{split}
\end{equation}
with real positive radial wavenumber $k$ and complex eigenfrequency $\omega=\omega_{R} + i \omega_{I}$, 
such that $\omega_{R}$ and $\omega_{I}$ denote the growth rate and frequency, respectively, and where the subscript "$0$" denotes values at equilibrium.

\subsection{Linearised equations}

Inserting (\ref{eq:perturbations}) into (\ref{eq:contsig})---(\ref{eq:conth2}) and linearising with respect to the perturbations yields the dimensionless linear equations
\begin{align}
       %%%%%%%%%%%%%%%%%%%     
        \omega \delta \sigma  = &  -i k \delta u_{x} \label{eq:linsig}, \\
        %%%%%%%%%%%%%%%%%%%
    \begin{split}
        \omega \delta u_{x}  =& \, 2 \delta u_{y} + 2 i \zeta(k) g \delta \sigma -i k \delta p, \\
        \quad & +  i k (-\frac{2}{3} \overline{\eta_{0}} +  \overline{\xi_{0}})  \frac{\delta \dot{H}}{H_{0}} -k^2(\frac{4}{3} \overline{\eta_{0}} + \overline{\xi_{0}})  \delta u_{x}  \label{eq:linu} , 
    \end{split}\\
    %%%%%%%%%%%%%%%%%%%
    \begin{split}
        \omega \delta u_{y}  =& -\frac{1}{2}\delta u_{x} -\frac{3}{2} i k  \delta \eta  -k^2 \overline{\eta_{0}}  \delta u_{y}\label{eq:linv}  ,
    \end{split}\\
    %%%%%%%%%%%%%%%%%%%
    \begin{split}
        \omega \delta T  =& -\frac{2}{3}\overline{p_{0}}\left(i k \delta u_{x} + \frac{\delta \dot{H}}{H_{0}}\right) -\frac{2}{3} k^2 \overline{\kappa_{0}}\delta T \\
        \quad & +\frac{1}{6}\left(9\delta \eta  -4(\delta \Gamma -\overline{\Gamma_{0}}\delta \sigma) \right) - \overline{\eta_{0}}\left( \frac{3}{2} \delta\sigma + 2 i k \delta u_{y}\right) \label{eq:linT} ,
    \end{split}\\
    \begin{split}
    \omega \delta H   = & \,\delta \dot{H}\label{eq:linH} ,
    \end{split}\\
    \begin{split}
        \omega \delta \dot{H}   = & -\delta H + f_{\text{sg},z} \delta \sigma -\frac{\overline{p_{0}}}{\gamma H_{0}^2}\delta H + \frac{3}{\gamma H_{0}}\left(\delta p -\overline{p_{0}}\delta \sigma\right) \\
        \quad & -\frac{1}{3 \gamma H_{0}^2}\left(4 \overline{\eta_{0}} + 3 \overline{\xi_{0}} + 3 k^2 \overline{z^2 \eta_{0}}\right)\delta \dot{H} -\frac{i k (-2 \overline{\eta_{0}} + 3 \overline{\xi_{0}})  }{\gamma H_{0}}\delta u_{x}\label{eq:linHdot} ,
    \end{split}
\end{align}
where we use the non-dimensionalisation as stated in \S \ref{sec:equilibrium} and where
\begin{align}
    f_{\text{sg},z} & =   2 g \hspace{0.55cm}   \text{if}   \, \rho=\rho_{\text{S}},\label{eq:fsgz_slab}\\
  f_{\text{sg},z} & =  \frac{2 g}{\sqrt{\pi}}   \hspace{0.357cm} \text{if} \, \rho=\rho_{\text{G}}\label{eq:fsgz_Gauss},
\end{align}
describing the effect of vertical self-gravity on perturbations of the ring thickness. The thickness correction factor $\zeta(k)$ in the radial self-gravity term is provided in Appendix \ref{sec:rsg}.
% The equilibrium heat conductivity $\overline{\kappa_{0}}$ and bulk viscosity $\overline{\xi_{0}}$ will be specified in \S \ref{sec:linpars}.
Furthermore,
\begin{align}
    \delta p & \equiv \overline{p_{\sigma}}\delta \sigma + \overline{p_{H}}\delta H + \overline{p_{T}} \delta T\label{eq:dp},\\
       \delta \eta & \equiv \overline{\eta_{\sigma}}\delta \sigma + \overline{\eta_{H}}\delta H + \overline{\eta_{T}} \delta T\label{eq:deta},\\
        \delta \Gamma & \equiv \overline{\Gamma_{\sigma}}\delta \sigma + \overline{\Gamma_{H}}\delta H + \overline{\Gamma_{T}} \delta T\label{eq:dgamma},
\end{align}
with the derivatives $\overline{p_{\sigma}} \equiv \overline{\partial p/\partial \sigma}$, etc. .
Here we used the fact that the spatial dependence of the transport coefficients (\ref{eq:pressz})---(\ref{eq:gamz}) can in general be written as $p=p\left(\sigma(x),H(x),T(x),z\right)$, and similar expressions for $\eta$, $\kappa$ and $\Gamma$.
% In what follows, for clarity we drop the bar-symbols, with the understanding that all quantities are either vertically averaged or $z$-independent.

\begin{figure}
 \centering
 \includegraphics[width = 0.5\textwidth]{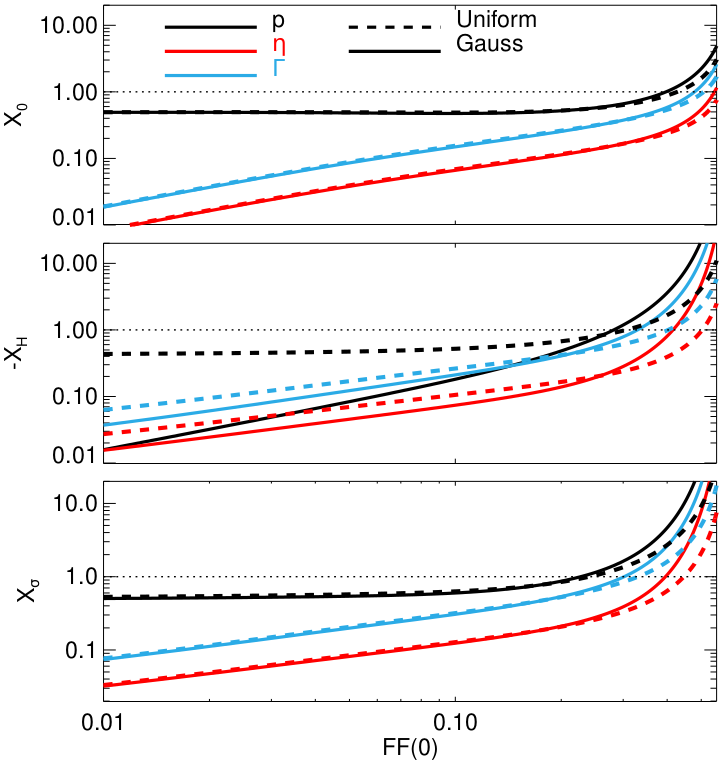}
 \caption{equilibrium values of  pressure $p$, viscosity $\eta$, and cooling rate $\Gamma$, as well as their derivatives with respect to scaleheight $H$ and surface mass density $\sigma$ for increasing filling factor $\mbox{ff}$. The plots correspond to the equilibrium shown in Figure \ref{fig:gstate_combo} with $\epsilon=0.1$. Note that the $H$-derivatives generally yield negative values.}
 \label{fig:derivs}
 \end{figure}

Explicit expressions for the derivatives appearing in (\ref{eq:dp})-(\ref{eq:deta}) are displayed together with the equilibrium quantities $\overline{p_{0}}$, $\eta_{0}$ and $\overline{\Gamma_{0}}$ for illustration in Figure \ref{fig:derivs} for the equilibrium shown in Figure \ref{fig:gstate_eps05}.
The plots show that deviations between the two density models (\ref{eq:rho_slab}) and (\ref{eq:rho_gauss}) are most pronounced for variations in the scaleheight $H$ in both limits of small and large filling factors.
% As noted in \S \ref{sec:vstruct}, for $\mbox{ff}\to 0$ ($\mbox{FF}\lesssim  0.6)$ the Gaussian (uniform) prescription is expected to be more adequate.

 Eqs. (\ref{eq:linsig})---(\ref{eq:linHdot}) can be written as 
\begin{align}\label{eq:eigenproblem}
    M\vec{b} = \omega \vec{b},
\end{align}
where $M$ is a $6\times 6$ complex matrix and $\vec{b}=\left\{\delta \sigma, \delta u_{x}, \delta u_y, \delta T, \delta H, \delta \dot{H} \right\}^T$ is a corresponding eigenvector. In the general case, we solve this eigenvalue problem with standard packages in IDL.

\subsection{Isothermal, vertically  static limit}\label{sec:isohstat}

% In contrast to \S \ref{sec:dilute} in this section we do not require $\sigma_{0}\ll 1$. 
In this section we neglect variations of the temperature and the disc scale height from their corresponding equilibrium values and thus also discard (\ref{eq:linT})---(\ref{eq:linHdot}). We then arrive at the standard isothermal equations that have been analyzed in numerous previous studies (e.g. \citealt{ward1981,schmit1995,latter2009}).
The dispersion relation, which is given by $\text{Det}\left[M\right]=0$ (cf. \ref{eq:eigenproblem}) in this limit reads
\begin{equation}
\begin{split}\label{eq:deteqiso}
  0 & = \omega^3 + \frac{1}{3} (7 \overline{\eta_{0}} + 3 \overline{\xi_{0}}) k^2 \omega^2 \\
  \quad & + \left(1 -2 g \zeta\left(k\right) k +\overline{p_{\sigma}} k^2 + \frac{1}{3} (4 \overline{\eta_{0}} + 3 \overline{\xi_{0}})^2 k^4  \right)\omega \\
  \quad & + k^2 \left[3 \overline{\eta_{\sigma}} + \overline{\eta_{0}} k (\overline{p_{\sigma}} k -2 g \zeta\left(k\right))\right],
  \end{split}
\end{equation}
giving rise to three eigenmodes, which are the viscous instability (VI) mode and a pair of overstable modes (OS).
Since the radial wavenumber $k\ll 1$ we can adopt a perturbation series 
\begin{equation*}
    \omega = \omega_{0} + \omega_{1} k + \omega_{2} k^2 + \ldots
\end{equation*}
where the $\omega_i$ ($i=0,1,2$) are complex. Inserting this in (\ref{eq:deteqiso}) and expanding the latter in orders of $k$ yields the approximate eigenfrequencies
\begin{align}
\omega_{\text{VI,iso$\&$hstat}} & = -3 \overline{\eta_{\sigma}} k^2 -2 g \zeta\left(k\right) \left(3 \overline{\eta_{\sigma}} -\overline{\eta_{0}}\right)k^3 + \mathcal{O}\left(k^4\right)\label{eq:vi_iso_hstat},\\
\begin{split}
\omega_{\text{OS,iso$\&$hstat}} & = i\bigg( \pm 1 \mp  g \zeta\left(k\right) k \pm \frac{1}{2}   \left[\overline{p_{\sigma}} -g^2 \zeta\left(k\right)^2\right] k^2 \\
\quad & + \frac{1}{2} \zeta(k) g \left[-\zeta(k)^2 g^2 + \overline{p_{\sigma}}\right]k^3 \bigg)\\
\quad & + \frac{1}{6}k^2 \left(9\overline{\eta_{\sigma}} -\overline{\eta_{0}}\left[7  + 3 \widetilde{\xi_{0}}\right]\right) -k^3 \left(\overline{\eta_{0}}-3 \overline{\eta_{\sigma}}\right) \zeta(k) g   \\
\quad & + \mathcal{O}\left(k^4\right).
\end{split}
\end{align}
Thus, the VI is triggered if $\overline{\eta_{\sigma}}<0$. As discussed in \S \ref{sec:equil_res}, this is not expected to occur in a dense planetary ring.
Nevertheless, radial self-gravity amplifies the instability at shorter wavelengths, as described by the term $\propto k^3$. 
On the other hand, for the onset of viscous overstability we find the condition
\begin{equation}\label{eq:os_iso}
\frac{\overline{\eta_{\sigma}}}{\overline{\eta_{0}}} > \frac{1}{3}\left(\frac{7}{3} + \widetilde{\xi_{0}}\right)
\end{equation}
on sufficiently long wavelengths. 
For a viscosity of the form (\ref{eq:beta}) Eq. (\ref{eq:os_iso}) yields
 the condition $\beta>\beta_{c}$, where $\beta_{c}$ is given by the expression (\ref{eq:betc}) derived by \citet{schmit1995}.
Inspection of Figure \ref{fig:derivs} reveals that radial self-gravity is expected to increase the growth rates of the visous overstability via the $k^3$-term, at least for sufficiently large $\mbox{ff}$, since then $3\overline{\eta_{\sigma}}>\overline{\eta_{0}}$.
It can also readily be seen that self-gravity and pressure decrease and increase the modes oscillation frequency, respectively, at sufficiently short wavelengths. This gives rise to a frequency minimum in $k$-space, which has been found to play a role in the nonlinear saturation process of the viscous overstability \citep{lehmann2017}.

\subsection{Vertically static limit}\label{sec:hstat}

Next, we consider the vertically static limit, such that thickness variations are neglected (but temperature may depart from its equilibrium value). This limit has been investigated in several studies \citep{spahn2000a,schmidt2001b,lehmann2017}.

To leading order, the VI mode is now described by
\begin{align}
\begin{split}
    \omega_{\text{VI,hstat}} & = \left(-3 \frac{\overline{\eta_{\sigma}}\, \overline{\Gamma_{T}}}{ 4 \overline{\Gamma_{T}}-9 \overline{\eta_{T}}} - 3 \overline{\eta_{T}} \frac{4 (\overline{\Gamma_{\sigma}}-4 \overline{\Gamma_{0}} + 9 \overline{\eta_{0}})}{9 \overline{\eta_{T}} -4 \overline{\Gamma_{T}}}\right)k^2 +  \mathcal{O}\left(k^3\right) \\
    \quad & = \left(-\frac{9}{2}\overline{\eta_{\sigma}} + \frac{2}{3} \overline{\Gamma_{\sigma}}\right)k^2 +  \mathcal{O}\left(k^3\right)  \label{eq:vi_hstat},
    \end{split}
\end{align}
where for the second equality we used (\ref{eq:ebalance}), as well as
\begin{equation}\label{eq:etat}
    \overline{\eta_{T}} = \frac{1}{2}\frac{\overline{\eta_{0}}}{T_{0}},  \overline{\Gamma_{T}}  = \frac{3}{2}\frac{\overline{\Gamma_{0}}}{T_{0}}
\end{equation}
In addition, we find the thermal mode
\begin{equation}
    \omega_{\text{T,hstat}} = \frac{3}{2}\overline{\eta_{T}}-\frac{2}{3}\overline{\Gamma_{T}} +\left(F_{1} -\frac{2}{3}\kappa_{0}\right) k^2,
\end{equation}
where 
\begin{equation}
\begin{split}
    F_{1} & = \overline{\eta_{T}}\frac{9 \overline{\eta_{\sigma}} -4 \overline{\Gamma_{\sigma}} + 6 \overline{E_{T}}^2 \overline{\eta_{0}} -4 \overline{E_{T}} \overline{p_{0}}}{2 \overline{E_{T}}(1+\overline{E_{T}}^2)}\\
    \quad & + \overline{p_{T}}\frac{-9 \overline{\eta_{\sigma}} + 4 \overline{\Gamma_{\sigma}} + 6 \overline{\eta_{0}} + 4 \overline{E_{T}} \overline{p_{0}}}{6(1+\overline{E_{T}}^2)},
    \end{split}
\end{equation}
with
\begin{equation}
    \overline{E_{T}} \equiv \frac{2}{3}\left(\overline{\Gamma_{T}} - \frac{9}{4}\overline{\eta_{T}}\right),
\end{equation}
describing the temperature dependence of the thermal equilibrium (\ref{eq:ebalance}), as defined in \citet{schmidt2001b}.
The isothermal limit can be recovered for $\overline{\Gamma_{T}}\to \infty$, implying that any temperature variation is instantly extinguished. In this limit (\ref{eq:vi_hstat}) becomes (\ref{eq:vi_iso_hstat}).
Furthermore, the viscous overstability is described by the pair
\begin{equation}\label{eq:os_hstat}
    \begin{split}
\omega_{\text{OS,hstat}} & = \pm i \left(  1 - g \zeta\left(k\right) k +\frac{1}{2}   (\overline{p_{\sigma}} -g^2 \zeta\left(k\right)^2 + F_{2}) k^2\right)\\
\quad & + \frac{1}{6}k^2 \left(9\overline{\eta_{\sigma}} - \overline{\eta_{0}}(7 + 3 \widetilde{\xi_{0}}) + 3 F_{3} \right) + \mathcal{O}\left(k^3\right).
\end{split}
\end{equation}
with \citep{schmidt2001b}
\begin{align}
\begin{split}\label{eq:F2}
    F_{2} & = \overline{p_{T}}\frac{9 \overline{\eta_{\sigma}} \overline{E_{T}} -4 \overline{\Gamma_{\sigma}} \, \overline{E_{T}} -6 \overline{E_{T}} \overline{\eta_{0}} + 4 \overline{p_{0}}}{6(1+\overline{E_{T}}^2)}\\
    \quad & + \overline{\eta_{T}} \frac{-9 \overline{\eta_{\sigma}} + 4 \overline{\Gamma_{\sigma}} + 6 \overline{\eta_{0}} + 4 \overline{E_{T}} \overline{p_{0}}}{2(1+ \overline{E_{T}}^2)},
    \end{split}\\
    \begin{split}\label{eq:F3}
    F_{3} & = \overline{\eta_{T}} \frac{ 9 \overline{\eta_{\sigma}} \overline{E_{T}} -4 \overline{\Gamma_{\sigma}} \, \overline{E_{T}} -6 \overline{E_{T}}\overline{\eta_{0}} + 4 \overline{p_{0}} }{2(1+ \overline{E_{T}}^2)}\\
    \quad & + \overline{p_{T}}\frac{9 \overline{\eta_{\sigma}} -4 \overline{\Gamma_{\sigma}} -6 \overline{\eta_{0}} -4 \overline{E_{T}} \overline{p_{0}}}{6(1+ \overline{E_{T}}^2)}.
    \end{split}
\end{align}
Thus, viscous overstability now occurs (in the limit of long wavelengths) if
\begin{equation}\label{eq:os_noniso}
\frac{\overline{\eta_{\sigma}}}{\overline{\eta_{0}}} > \frac{1}{3}\left(\frac{7}{3} +\widetilde{\xi_{0}} - \frac{F_{3}}{\overline{\eta_{0}}}\right).
\end{equation}
For the equilibrium shown in Figure \ref{fig:gstate_eps05} we find that $F_{3}<0$ decreases for all $\mbox{ff}$ such that, in agreement with previous studies, thermal effects reduce growth rates of the viscous overstability (cf. (\ref{eq:os_iso})), and in addition increase the critical value of $\mbox{ff}$ for its onset. On the other hand, $F_{2}>0$ for $\mbox{ff}\lesssim 0.6$, but turns negative for larger $\mbox{ff}$. Thus, for most feasible values of $\mbox{ff}$ thermal effects enhance the oscillation frequency of overstable modes (see also Figure \ref{fig:Fs} below).

\subsection{Isothermal limit}\label{sec:os_iso}

Finally, we include variations of the ring thickness but neglect temperature variations, such that the dispersion relation is of fifth order in the complex frequency $\omega$. The corresponding expression is lengthy and will not be displayed here.
To leading order in $k$, the VI mode is now described by
\begin{align}
\begin{split}
    \omega_{\text{VI,iso}} & = \left(-3 \overline{\eta_{\sigma}}  + \frac{3 \overline{\eta_{H}} H_{0}\left( \overline{p_{0}} -\overline{p_{\sigma}} + \gamma f_{\text{sg},z} H_{0}\right)}{\gamma H_{0}^2 +\overline{p_{0}}-H_{0} \overline{p_{H}}}\right) k^2 + \mathcal{O}\left(k^3\right) ,
    \end{split}
\end{align}
where the second term in the bracket encompasses effects of ring thickness variations and vertical self-gravity. Based on Figure \ref{fig:derivs} we find that vertical self-gravity has an amplifying effect on the VI, whereas the other term $\propto (\overline{p_{\sigma}}-\overline{p_{0}}$), describing non-local pressure effects, has a damping effect.

In addition, we find the pair of vertical `breathing' modes
\begin{align}
\begin{split}\label{eq:hmodes_iso}
    \omega_{\text{H,iso}} = &  \pm  \sqrt{ \overline{\eta_{0}}^2 \left(\frac{4 + 3 \widetilde{\xi_{0}}}{6 \gamma H_{0}^2}\right)^2 - \frac{\overline{p_{0}} - H_{0} \overline{p_{H}}}{\gamma H_{0}^2} -1 } -\frac{\overline{\eta_{0}}(4 + 3 \widetilde{\xi_{0}})}{6 \gamma  H_{0}^2} \\
    \quad & 
     + \mathcal{O}\left(k^2\right).
    \end{split}
\end{align}
It can readily be seen that in the here considered limit of long wavelengths viscosity (as expected) damps these modes, and in addition, decreases their frequency. In contrast, pressure forces increase the modes' frequency, since $\overline{p_{H}}<0$ (Figure \ref{fig:derivs}). In the dilute limit ($\overline{p_{H}}\to 0$, as well as vanishing viscosity)
we find $\omega_{\text{H,iso}}\to \pm i \sqrt{2}$,
% which is slightly smaller than  (\ref{eq:hmodes_dilute}), indicating that temperature variations slightly increase the modes' frequencies. 
consistent with Eq. (31) in \citet{lubow1981}.

\begin{figure}
\centering
\includegraphics[width = 0.4\textwidth]{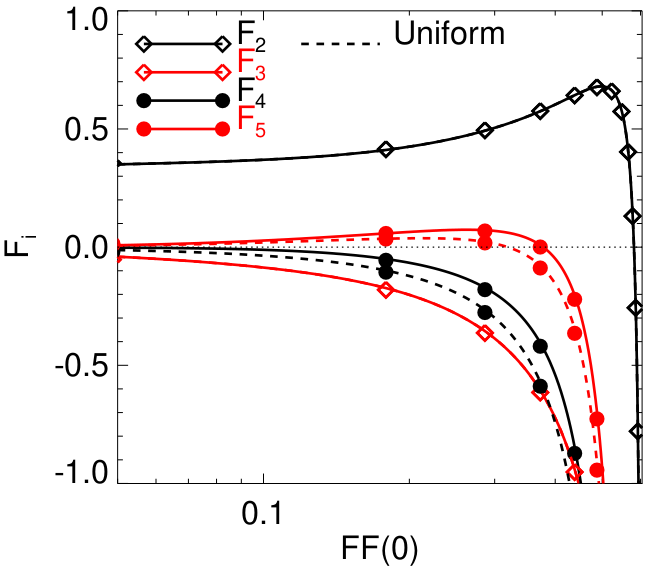}
\caption{The "thermal" coefficients $F_{2}$ and $F_{3}$, as given by (\ref{eq:F2}) and (\ref{eq:F3}), represented by diamond symbols. Also plotted are the coefficients $F_{4}$ and $F_{5}$ related to vertical motions in absence of self-gravity, as described in the text. The solid (dashed) curves apply to a Gaussian (uniform) distribution (\ref{eq:rho_gauss}). The corresponding profiles of $F_2$ and $F_3$ for a uniform distribution (\ref{eq:rho_slab}) are identical to the Gaussian case.}
\label{fig:Fs}
\end{figure}

The corresponding eigenvalues describing the viscous overstability modes in the isothermal limit $\omega_{\text{OS,iso}}$ can be written in the same form as (\ref{eq:os_hstat}), but with $F_{2}$ and $F_{3}$ replaced by corresponding expressions $F_{4}$ and $F_{5}$, respectively.
Compared to the "thermal" expressions (\ref{eq:F2})-(\ref{eq:F3}), the expressions $F_{4}$ and $F_{5}$ are rather unwieldy and will not be displayed here.
Nevertheless, the functions $F_{4},F_{5}$ are plotted for the non-gravitating case $g= 0$, together with their "thermal" counterparts $F_{2},F_{3}$ in Figure \ref{fig:Fs} . 
We find that compared to the effect of temperature variations, the effect of vertical disc motions (i.e. scaleheight variations) is somewhat more complicated.
That is, based on the behavior of $F_{5}$ the latter can in principle either reduce or increase the critical value of $\overline{\eta_{\sigma}}$ (or equivalently $\mbox{FF}$) for the onset of viscous overstability, depending on the critical $\mbox{FF}$ resulting from (\ref{eq:os_iso}).
Furthermore, scaleheight variations are likely to reduce the overstable modes' oscillation frequencies, unless $\mbox{FF}_{cr}$ resulting from (\ref{eq:os_iso}) is very small. This is opposite to the effect of temperature variations. However, at larger values of $\mbox{FF}$, scale height variations are expected to have a damping effect on overstable modes, as the function $F_{5}$ becomes positive at a given $\mbox{ff}$.
We find that it is the bulk viscosity which is responsible for $F_{5}$ taking positive values. For $\widetilde{\xi_{0}}\lesssim 7/3$ we find that $F_{5}\lesssim 0$ for all $\mbox{FF}$. It should be noted though that overall a bulk viscosity has a damping effect (cf. Eq. (\ref{eq:os_noniso})).
Moreover, vertical self-gravity mildly increases both $F_{4}$ and $F_{5}$ (not shown), such that $F_{4}$ can attain positive values for a larger range of $\mbox{FF}$.

\section{Viscous Overstability}\label{sec:OS}

 In this section we focus on the viscous overstability. We assume the dense ring equilibrium as shown in Figure \ref{fig:gstate_combo} with constant $\epsilon=0.1$. For these parameters (see also \S \ref{sec:linpert}), the viscous overstability is the only instability that can occur. If not stated otherwise, presented results correspond to a Gaussian vertical density distribution.
Moreover, in addition to the solution of the full system of equations (\ref{eq:linsig})---(\ref{eq:linHdot}), we will consider various limiting cases as in \S \ref{sec:isohstat}-\S \ref{sec:hstat}, which will allow us to clarify the role of  different physical elements in the viscous overstability mechanism. In this section we assume a constant bulk viscosity $\widetilde{\xi}_{0}=4$.

\subsection{Driving and damping agents}

  \begin{figure}
  \centering
  \includegraphics[width = 0.41\textwidth]{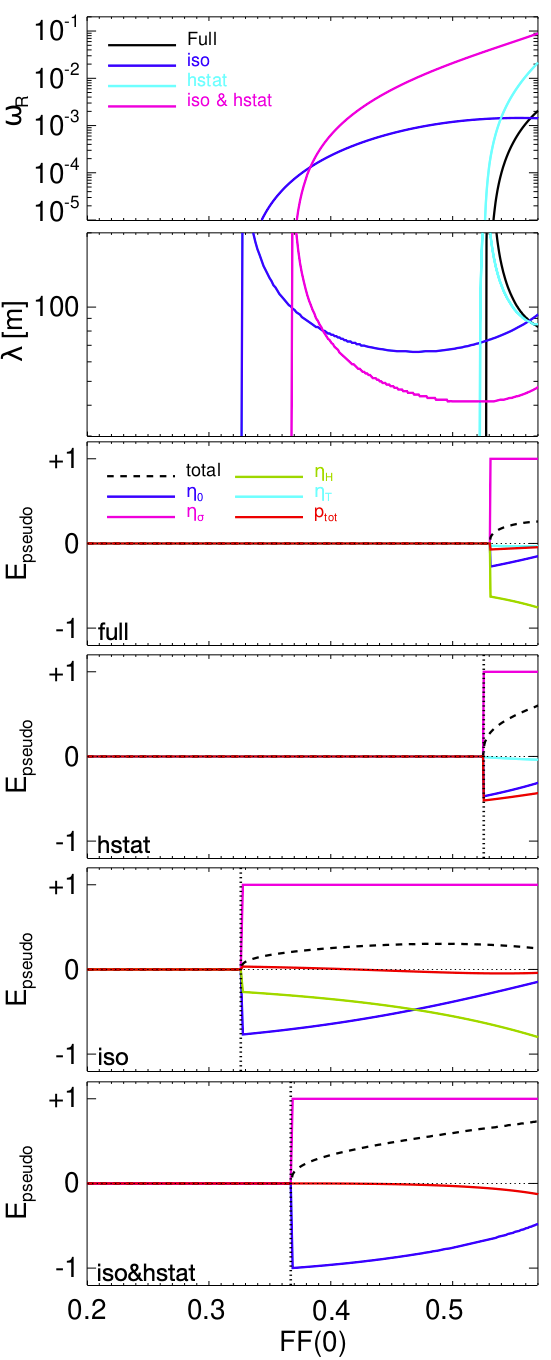}
  \caption{Maximum linear growth rates and corresponding wavelengths of the viscous overstability in absence of self-gravity, for increasing filling factor (top panels). The remaining panels show the pseudo-energy decomposition (Appendix \ref{sec:pseudo}) of the modes for the various limiting cases as explained in the text. The pseudo-energies are normalised such that the dominating component (in all presented cases $E_{\eta_{\sigma}}$) yields unity. Furthermore, we plot the cube root of the total energy for increased visibility.}
  \label{fig:lam_gr}
  \end{figure}
 
In this section we neglect self-gravity, but briefly consider its effect in \S \ref{sec:os_sg}.
In absence of self-gravity the equilibrium and linear stability properties of the planetary ring can be parameterised through a single parameter, e.g. the filling factor $\mbox{FF}$ (its mid-plane value) or the optical depth $\tau$ for identical particles with a given constant $\epsilon$.
This can be seen from Eqs. (\ref{eq:linsig})---(\ref{eq:linHdot}). Using $g=0$ the above equations depend on $\mbox{ff}$, $H_{0}$, and $T_{0}$. In absence of self-gravity the latter two quantities are fully determined by the value of $\tau$, or, interchangeably, $\mbox{FF}$.
The reason is that in this case the equilibrium filling factor (\ref{eq:ff_true}) and optical (\ref{eq:tau}) depth differ by a factor of the scaleheight $H_{0}$ (\ref{eq:scaleheight_omz}) which, in turn, only depends on either $\tau$ or $\mbox{FF}$. This is not true in the self-gravitating case, where the equilibrium scaleheight $H_{0}$ (and also the temperature $T_{0}$) explicitly depends on the particle bulk density $\rho_{b}$ or equivalentlty $r_{h}$ (cf. Eqs. (\ref{eq:scaleheight}), (\ref{eq:scaleheight_gauss}) as well as Figure \ref{fig:gstate_eps05}).

Figure \ref{fig:lam_gr} (upper panel) shows the maximal growth rate of the viscous overstability (across all wavenumbers $k$) for increasing midplane filling factor $\mbox{FF}(0)$, assuming a Gaussian density distribution, and for the different approximations as discussed above. These are "hstat" (vertically static), "iso" (isothermal), and their combination. The label "full" corresponds to the solution of the full set of Eqs. (\ref{eq:linsig})---(\ref{eq:linHdot}).
 The second panel shows the corresponding wavelengths of the fastest growing modes. 
These plots readily reveal the values of the critical filling factor $\mbox{FF}_{\text{cr}}$ for the onset of instability in the different cases, indicated by the sharp ``rise'' of the wavelengths of growing modes as functions of $\mbox{FF}$. The remaining panels show the pseudo-energy decomposition (Appendix \ref{sec:pseudo}) for the different cases. 
% Table \ref{tab:os_crit} compares critical values of different quantities for a Gaussian and a slab distribution.

\begin{figure}
 \centering
 \includegraphics[width = 0.45\textwidth]{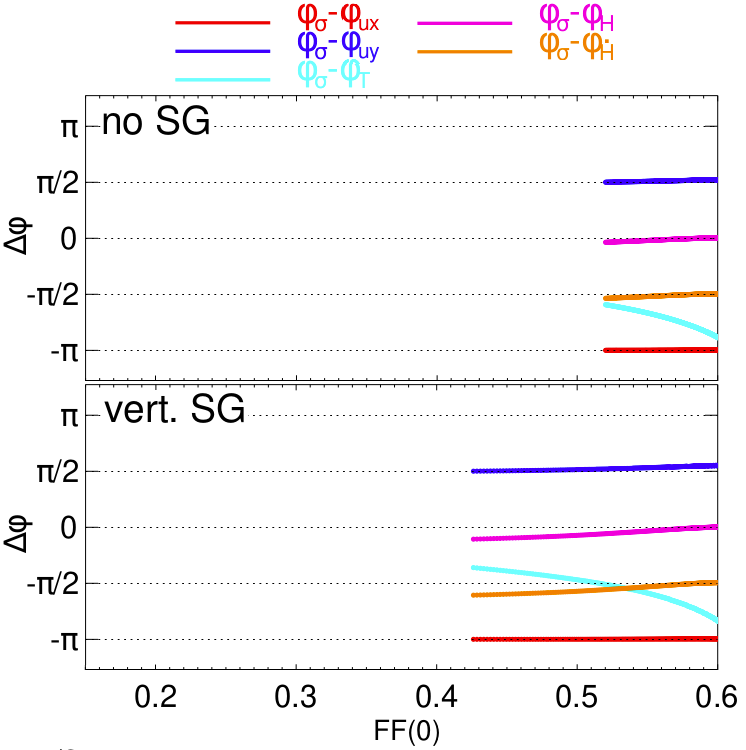}
 \caption{Phase shifts between the different hydrodynamic quantities of fastest growing overstable modes for a left traveling wave, resulting from numerical solution of the full linear eigenvalue problem (\ref{eq:eigenproblem}) for non-gravitating (upper panel) and vertically self-gravitating rings (lower panel) with $r_{h}=0.5$ ($\rho_{b}=204 \text{kg} \text{m}^{-3}$).}
 \label{fig:phas}
 \end{figure}

The energy decomposition confirms that the driving force of the viscous overstability is in general the increase of the viscous stress with increasing surface mass density, described by $\overline{\eta_{\sigma}}$, in agreement with Eqs. (\ref{eq:os_iso}) and (\ref{eq:os_noniso}). This quantity is positive, increases with increasing $\mbox{FF}$ (Figure \ref{fig:derivs}), and is closely related to the viscous slope $\beta$ defined by (\ref{eq:beta2}). From  Eq. (\ref{eq:eetasig}) we see that instability then requires a phase shift $\pi < \varphi_{\sigma} - \varphi_{u_{y}}<0$ between the surface mass density perturbation and the azimuthal velocity perturbation. According to Figure \ref{fig:phas} (upper panel), where we display phase shifts between individual hydrodynamic quantities corresponding to the overstable modes resulting from the full solution as shown in Figure \ref{fig:lam_gr} (for a left traveling wave), we find $\varphi_{\sigma} - \varphi_{u_{y}}\approx \pi/2$ in all cases, which is optimal for instability.

As explained above, the critical value of $\overline{\eta_{\sigma}}$  corresponds to a unique critical filling factor $\mbox{ff}_{\text{cr}}$ in our model. 
The requirement of a critical filling factor to be exceeded for the onset of instability is in agreement with the kinetic model of \hyperlinkcite{latter2008}{LO08} and the N-Body simulations of \hyperlinkcite{mondino2023}{MS23}, and is therefore considered to be the most fundamental criterion for the viscous overstability.

 The largest damping effect on overstable modes in the full model we find to be related to variations of the viscous stress with ring thickness $H$, described by the curve labeled $\overline{\eta_{H}}$. 
   % This finding suggests that  variations of the disc scaleheight to investigate the viscous overstability.
 This damping can easily be understood when comparing (\ref{eq:eetasig}) and (\ref{eq:eetah}), since variations of the surface density $\delta \sigma$ and the scaleheight $\delta H$ are practically in phase in overstable waves (See figure \ref{fig:phas}). Thus, the increase of the viscous stress with increasing $\sigma$ is partially compensated by a corresponding decrease due to an increasing scaleheight. Intuitively, a damping related to ring thickness variations is expected, since a vertical expansion of ring material reduces the local volume filling factor, such that  viscous overstability is expected to be weakened. This is a novel finding, as previous studies on the viscous overstability  did not include the effect of vertical motions. 
% The pressure-related damping effect can be understood by inspecting the phase shift between pressure perturbations $\delta P$ and density perturbations $\delta \sigma$ (Figure \ref{fig:phas}, upper panel, upper panel). It is a well known result in fluid dynamics that when pressure and density perturbations are out of phase, work is done by pressure (e.g. \citealt{cox1967}). If pressure leads (lags) the density, this work is negative (positive), such that oscillations are being damped (amplified). 
% (see for instance \citealt{lin2015} for an application of this principle to dusty gas in proto-planetary discs). 
% We note that we find the thermal energy associated with overatable modes $\propto \delta \left(\sigma T\right)$ to be negligible compared to the kinetic energy considered here. this is expected since the viscous overstability is a dynamical rather than a thermal instability.

In agreement with our calculations in the limit of long wavelengths (\S \ref{sec:linpert})
we find that the inclusion of temperature variations \emph{increases} $\mbox{FF}_{\text{cr}}$, which can be seen by comparing the panels labeled "iso$\&$hstat" and "hstat". A closer inspection of the energy-decomposition (not shown) reveals that the largest difference between the two cases lies in the term related to $\overline{p_{T}}$, which results in a larger pressure-related damping effect compared to the isothermal limit. On the other hand, the contribution due to $\overline{\eta_{T}}$ is found to be subdominant.
Interestingly, the inclusion of variations of the scaleheight (i.e. vertical motions) slightly decreases $\mbox{FF}_{\text{cr}}$ (seen by comparing the plots labeled "iso$\&$hstat" and "iso"). However, this reduction is not seen when comparing the cases "hstat" and "full", where we find only a very small difference. Thus, it appears that thermal effects and the effects of vertical motions are not independent.
% We suspect that this reduction is only small because of a reduced damping effect of pressure, which can clearly be seen in the energy decomposition.  
% We note that this reduction of $\mbox{ff}_{\text{cr}}$ is not found if we assume the slab vertical density distribution (\ref{eq:rho_slab}). We further suspect that this difference is related to the different magnitudes of the derivatives $\overline{p_{H}}$ and $\overline{\eta_{H}}$ for the Gaussian and slab distributions. Figure \ref{fig:derivs} shows that for $\mbox{ff}\gtrsim 0.2$ these derivatives for the slab distribution become increasingly smaller than those for a Gaussian one, which might explain the absence of a shift in $\mbox{ff}_{\text{cr}}$. 

% Moreover, as indicated by the behavior of the function $F_{5}$ displayed in Figure \ref{fig:Fs}, for larger $\mbox{ff}$ the growth rates resulting from the vertically static solution exceed those of the full solution (Figure \ref{fig:lam_gr}, top panel). This is due to the aforementioned damping effect on the latter associated with variations of the viscous stress with varying ring thickness, indicated by $E_{\eta_{H}}$, which becomes increasingly negative with increasing $\mbox{ff}$.
%
The vertical dashed lines in the three lowest panels of Figure \ref{fig:lam_gr} indicate the critical values for $\mbox{ff}_{\text{cr}}$, obtained from (\ref{eq:os_iso}) for the case "iso$\&$hstat", (\ref{eq:os_noniso}) for the case "hstat", and (\ref{eq:os_noniso}) with $F_{3}$ replaced by $F_{5}$ (see Figure \ref{fig:Fs}) for the case "iso". These are in excellent agreement with the numerical results. In Table \ref{tab:os_crit} we summarise critical values of different quantities for both Gaussian and slab distributions and for the different limiting cases. We find in general quite notable differences between the two density distributions, which we will, however, not further discuss here. Note also that each of the individual approximations strictly requires a different prescription of the bulk viscosity $\widetilde{\xi}_{0}$, which is ignored here.

In accordance with previous studies, the longest wavelengths become unstable first, albeit with negligible growth rates.
Overall, the typical wavelengths of modes (with meaningful growth rates $\omega_{R}\gtrsim 10^{-4}$) are $\lambda \sim 100 \text{m}$, a property shared with previous hydrodynamic models \citep{schmit1995,schmidt2001b,latter2009,lehmann2017}, kinetic modeling \hyperlinkcite{latter2008}{LO08} and N-body simulations (\citealt{salo2001};~\citealt{lehmann2017};~\hyperlinkcite{mondino2023}{MS23}). On the other hand, the behaviour of the maximal growth rates with increasing filling factor is less obvious and does not seem to correlate with the corresponding wavelengths of the modes.

% One aspect that is noteworthy is the positive contribution of the viscous stress term $E_{\eta_{0}}$ to the mode energies in the full solution and also the isothermal approximation, for sufficiently large $\mbox{ff}$,  represented by the dark blue curve. The corresponding energy expression is given by (\ref{eq:e_eta0}). Considering the different terms in this expression, the only term that can in principle give rise to amplification is the last term $\propto \text{Im}\left[\delta u_{x} \delta \dot{H}\right]$, which describes the viscous coupling of radial and vertical motions. By again resorting to the phase relations in Fig. xx, we find that this term is amplifying if $\widetilde{\xi_{0}} > 2/3$. For the value $\widetilde{\xi_{0}}=4$ adopted here this term starts to dominate the remaining damping terms for sufficiently large $\mbox{ff}$. This effect is also responsible for the reduction of $\mbox{ff}_{\text{cr}}$ when shifting from the isothermal, vertically static approximation to the isothermal approximation, since essentially all other additional contributions are damping. Nevertheless, as already noted in \S \ref{sec:os_iso}, the net effect of a bulk viscosity is a damping effect.

\begin{table}
 \caption{Critical values of important model parameters for the onset of viscous overstability in linear calculations excluding self-gravity, using a constant $\epsilon$, and for different limiting cases, as explained in the text.}
 \label{tab:os_crit}
 \begin{tabular*}{\columnwidth}{@{}l@{\hspace*{20pt}}l@{\hspace*{20pt}}l@{\hspace*{20pt}}l@{}}
  \hline
\text{Slab / Gauss} & $\beta_{\text{cr}}$  &   $\tau_{\text{cr}}$    &  $\mbox{FF}_{\text{cr}}$  \\ \hline
full &   1.34 / 1.23 & 3.70 / 2.11  & 0.52 / 0.53 \\ 
hstat &  1.19 / 1.21 & 3.07 / 2.06 & 0.49 / 0.52  \\ 
iso &  1.11 / 0.76   & 2.73 / 0.92 & 0.47 / 0.33\\
iso$\&$hstat &   0.77 / 0.79 & 1.40 / 1.08 & 0.33 / 0.37  \\ 
  \hline
 \end{tabular*}
\end{table}

\subsection{Impact of self-gravity}\label{sec:os_sg}

\subsubsection{Vertical self-gravity}\label{sec:os_vsg}
With the inclusion of vertical self-gravity, the characterisation of the equilibrium and, in principle also the perturbation quantities requires one additional parameter. Here we choose to work with the optical depth $\tau$ (replacing $\mbox{FF}$) as well as $r_{h}$, the latter quantifying the strength of self-gravity.
Therefore, in Figure \ref{fig:gr_sg} we display similar quantities as in Figure \ref{fig:lam_gr}, except using $\tau$ as free parameter (the bottom frame assumes a fixed $r_{h}=0.5$). Note that $\sigma_{0}$ (or $\tau$) enters the linear problem via the self-gravity parameter $g$. The upper frame displays the growth rates as a contour plot of both $\tau_{0}$ and $r_{h}$.  
% In such a plot, constant filling factors are described by super-linear curves in the $\rho_{b}-\sigma_{0}$ plane. In contrast, in absence of vertical self-gravity, these are described by straight lines (see Eqs. (\ref{eq:ff_slab}),(\ref{eq:ff_gauss})). Super-linearity arises since with increasing $\rho_{b}$ the scaleheight $H_{0}$ drops (Figure \ref{fig:gstate_eps05}).

The \emph{critical} filling factor $\mbox{FF}_{cr}$ for the onset of overstability significantly reduces as $0.52\gtrsim  \mbox{ff}_{\text{cr}}\gtrsim 0.37$ with increasing $0<r_{h} \lesssim 0.6$,
such that vertical self-gravity appears to have a significant effect on overstable perturbations. The pseudo-energy decomposition reveals that vertical self-gravity affects the $E_{\eta_{0}}$ and $E_{\eta_{T}}$-terms. The former becomes less damping, whereas the latter - in contrast to the non-selfgravitating case - becomes slightly amplifying for a range of sufficiently small filling factors. Considering Eq. (\ref{eq:eetat}) and using the fact that $\overline{\eta_{T}}>0$, this implies that the relative phases between the perturbations $\delta T$ and $\delta u_{y} $ are affected by vertical self-gravity, which is indeed confirmed in Figure \ref{fig:phas} (by comparing the upper and lower panels).
Furthermore, according to Eq. (\ref{eq:esgz}) the ability of vertical self-gravity to directly
amplify a particular eigenmode depends on the relative phase shifts of the perturbations  $\delta \sigma$ and $\delta \dot{H}$,
% In the case of the VI we find a phase shift of $\pm \pi$ between these two quantities in numerical solutions to the eigenvalue problem. This is also confirmed in corresponding hydrodynamic test-simulations of the VI. 
such that a phase shift $\pm \pi$ is optimal, whereas a phase shift of $\pm \pi/2$ erases the direct effect of vertical self-gravity.
Figure \ref{fig:phas} shows that for $\mbox{FF}\gtrsim \mbox{FF}_{\text{cr}}$ the phase shift $\varphi_{\sigma}-\varphi_{\dot{H}}\lesssim -\pi/2$, but it approaches $-\pi/2$ for increasing $\mbox{FF}$, such that the effect of vertical self-gravity leads to a mild reduction of $\mbox{FF}_{\text{cr}}$, but its effect eventually vanishes at larger $\mbox{FF}$. We find (not shown) that the aforementioned indirect effect of vertical self-gravity through the  $E_{\eta_{T}}$-term dominates the direct effect.
It should be noted though that the main effect of vertical self-gravity on the region in ($\tau-r_{h}$)-space that is overstable enters via the altered equilibrium state (i.e. via a reduction of the equilibrium $\mbox{FF}$).
% for vertical self-gravity to amplify the perturbation since then the perturbation of $\mbox{ff}$ and hence that of $\Omega_{z}$ [Equation (\ref{eq:omz})] is exactly in phase with those of $\hat{\sigma}$ and $\hat{H}$ such that increases/decreases of $\hat{\sigma}$/$\hat{H}$ result in an increase/decrease of $\Omega_{z}$.  
% It is therefore plausible that vertical self-gravity is essential for the occurrence of the VI in our disk model. For the the viscous overstability we find relative absolute phase shifts $\lesssim \pi/2$ between $\hat{\sigma}$ and $\hat{H}$, depending on the model parameters. This means that the oscillation of $\Omega_{z}$ is not exactly in phase with the former two quantities. Therefore the amplifying effect of vertical self-gravity is much weaker in the case of the viscous overtability.

\subsubsection{Radial self-gravity}\label{sec:os_rsg}

\begin{figure}
 \centering
 \includegraphics[width = 0.45\textwidth]{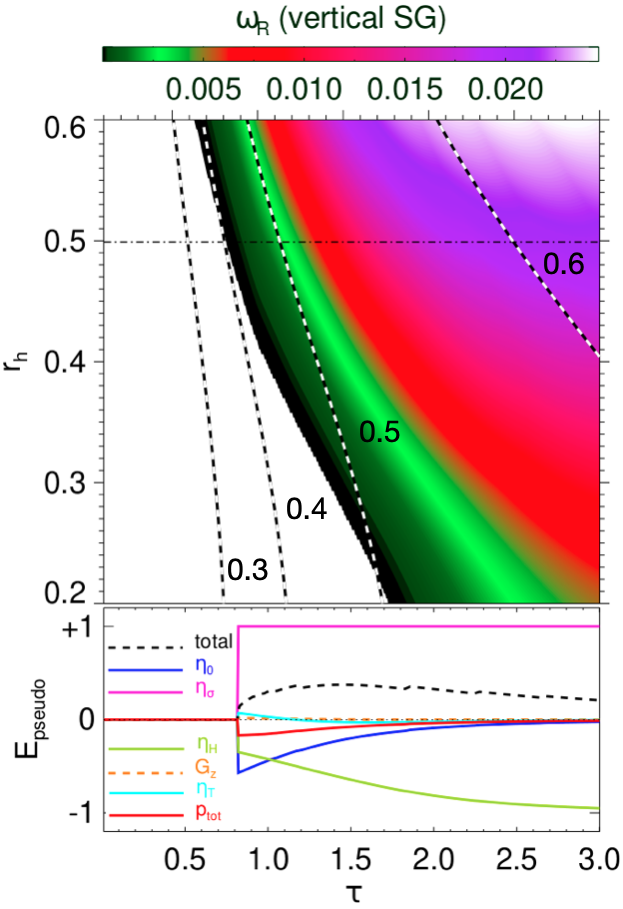}
 \caption{Contours of maximum (across all wavelengths) linear growth rates of the viscous overstability in presence of vertical self-gravity for different equilibrium optical depths $\tau$ and values of $r_{h}$, quantifying the strength of self-gravity. In the upper panels dashed curves describe a constant filling factor $\mbox{FF}$. The lower panel shows the pseudo-energy decomposition for  fixed $r_{h}=0.5$, as indicated by the black dash-dotted line in the upper panel.}
 \label{fig:gr_sg}
 \end{figure}

Compared to the impact of vertical self-gravity, we find that radial self-gravity has a rather strong effect on overstable perturbations, with overall substantially larger growth rates up to an order of magnitude higher.
For $r_{h}\gtrsim 0.5$ we actually find overstability for all $\mbox{FF} \gtrsim 0$, which is related to the "spurious branch" of overstable modes at small filling factor (\S \ref{sec:constit}), which starts to merge with the "physical branch" of overstable modes (cf. Figure \ref{fig:gr_sg}).
%
% (Figure \ref{fig:gr_sg}, right panels).
% For particle bulk densities $\rho_{b}\gtrsim 100 \text{kg}\text{m}^{-3}$ we find a strongly reduced $\mbox{ff}_{\text{cr}}$ compared to computations involving only vertical self-gravity. The pseudo-energy decomposition suggests that radial self-gravity indirectly affects the values of $\mbox{ff}_{\text{cr}}$ via an increased $E_{\eta_{T}}$-term as well as  a reduced $E_{\eta_{0}}$-term , as the direct contribution due to radial self-gravity vanishes at the critical $\sigma_{0}$. For completion, we also present phase shifts of individual hydrodynamic quantities of the fastest growing overstable modes in the presence of vertical and radial self-gravity in Figure \ref{fig:phas} (bottom panel).
We will come back to the effect of radial self-gravity on overstability in \S \ref{sec:nbody_hydro}, where we compare results from our model with those from our N-body simulations.

\section{N-Body simulations}\label{sec:nbody}

In addition to our hydrodynamic analysis, we perform N-body simulations of a small patch of a dense planetary ring, using the local simulation method \citep{wisdom1988}.
The simulation method is the same as described in \citet{lehmann2017} (see also \citealt{salo1992a,salo1995}, with slight modifications to the computation of vertical and radial self-gravity forces, as outlined below. Note that the azimuthal component of self-gravity is omitted. This is done so as to faci\textit{}litate comparison with our axisymmetric linear hydrodynamic model for the viscous overstability without complications from non-axisymmetric self-gravity wakes.

We perform a series of simulations with constant $\epsilon=0.1$ (for some cases we use $\epsilon=0.5$), at $r_{0}=100,000 \text{km}$ in order to determine the threshold (or critical) values of the optical depth
\begin{equation}\label{eq:tau_sim}
    \tau = \frac{\pi r_{p}^2}{L_{x} L_{y}}
\end{equation}
and the volume filling factor \citep{mondino2023}
\begin{equation}\label{eq:ff_sim}
    \mbox{FF}(z) = \frac{1}{L_{x}L_{y}}\sum\limits_{i} a_{i}(z)
\end{equation}
for the onset of viscous overstability, where $a_{i}(z)$ describes the area of particle $i$ cut by the plane at height $z$. The filling factor (\ref{eq:ff_sim}) represents an average value over several orbital periods and the sum is taken over all particles. Note that prior to the onset of instability the system is uniform in planar directions. In figures below we present the mid-plane value of (\ref{eq:ff_sim}) at $z=0$.
We consider different strengths of self-gravity, measured by the $r_{h}$ parameter (\ref{eq:rh}). We adopt the values $r_{h} = 0.61,0.5,0.4,0.3,0.2$, for a range of optical depths around the threshold. For the applied distance $r_{0}$ these correspond to particle bulk densities $13\text{kg}\text{m}^{-3} \lesssim \rho_{b}\lesssim 350\text{kg}\text{m}^{-3}$.
In all simulations the radial size of the box and the number of particles are the same, i.e. $N_{p} = 4776$ and $L_{x} = 300 r_{p}$, where $r_{p}=1\text{m}$. Also the number of radial bins is $m_{\text{max}}=64$ in all cases. These fixed values assure that the accuracy of the statistical sampling of forces, vertical profiles, etc., is similar in all cases. The simulations are evolved for 300 orbital periods.
In each of the $r_{h}$-series different optical depths corresponds to a different tangential size $L_{y} = 25 r_{p}/\tau$ particle radii. Furthermore, the surface mass density $\sigma_{0}$, is determined by the prescribed values of $\tau$ and $r_{h}$ via (\ref{eq:tau}) and (\ref{eq:rh}):
\begin{equation}
\frac{\sigma_{0}}{(1000\,\text{kg}/\text{m}^2)} = 2.18 \tau\, \left(\frac{r_{p}}{1\text{m}}\right)  \left(\frac{r_{h}}{(r_{0}/100,000\,\text{km})}\right)^3.
\end{equation}

\subsection{Calculation of self-gravity forces}\label{sec:nbody_sg}

The radial self-gravity calculation is made with the FFT-method as in \citet{salo2010} and \citet{lehmann2017}, except that here we correct radial self-gravity for the finite thickness of the ring in a similar fashion as in our hydrodynamic model (see \S \ref{sec:linpert}). The simulation region of radial and tangential size $L_{x}$ and $L_{y}$ is divided into $m_{\text{max}}$ radial bins.
Then, using a radial Fourier decomposition of the tangentially averaged surface density
\begin{equation}
\sigma(x)=\sigma_0\left[1+\sum_{m=1}^{m_{\text{max}}} A_m \cos \left(m \frac{2 \pi}{L_{\text {x }}}\left(x-\phi_m\right)\right)\right],
\end{equation}
the corresponding radial force is
\begin{equation}
F_{\text{sg},x}(x)=-2 \pi G \sigma_0 \sum_{m=1}^{m_{\max }} \zeta(k_{m}) A_m \sin \left[m \frac{2 \pi}{L_{\text {x }}}\left(x-\phi_m\right)\right],
\end{equation}
where we now add the thickness correction factor $\zeta(k_{m})$, as given by (\ref{eq:zeta}) and with
\begin{align}
k_{m} & = m \frac{2 \pi}{L_{\text{x}}}.
\end{align}
We note that it is essential that this thickness correction factor is used in our simulations. That is, without the suppression of very short wavelength modes, overstability would grow much more vigorous.
In above expressions $\sigma_0$ is the mean surface density of the simulation region, $A_m$ and $\phi_m$ are the fractional amplitude and phase of different $m$-components with wavenumber $k_{m}$ (i.e. wavelength $\lambda=L_{\text {x }}/m$). In practice the Fourier decomposition is performed by tabulating the surface density in the radial bins and applying a FFT to obtain $F_{\text{sg},x}$ at these bins. The force at particle locations is obtained by interpolation.
The thickness correction factor is computed by approximating the particle distribution as a homogeneous slab (Eq. \ref{eq:rho_slab}) with semi-thickness $H \equiv \frac{1}{2}\sqrt{12\langle z^2\rangle} \approx 1.73 \sqrt{\langle z^2\rangle}$, where  $ \sqrt{\langle z^2\rangle}$ is the root-mean-squared distance of all simulation particles from the midplane.

Furthermore, 
the vertical (self)-gravity force at a given radial location is assumed to be of the form (see Eq. (\ref{eq:navz2}) with (\ref{eq:fz_central}) and (\ref{eq:fsgz_slab}))
\begin{equation}\label{eq:fsgz_nbody}
F_{\text{sg},z}=-\left(\Omega^2+\frac{2 \pi G \sigma(x)}{H(x)}\right) z \equiv -\Omega_{z}^2 z,
\end{equation}
which is consistent with the vertical self-gravity (\ref{eq:omz}) appearing in our hydrodynamic model. The local value $\sigma(x) / H(x)$ is calculated via a FFT, in similar manner as used for radial forces.
Note that this approximation for the vertical self-gravity force is significantly more accurate than the assumption a constant frequency $\Omega_{z}\geq \Omega_{0}$, as it accounts for local variations of the particle configuration which arise if the system is perturbed or unstable.
Below we will find that our N-body simulations using (\ref{eq:fsgz_nbody}) are largely consistent with the fully self-consistent self-gravitating simulations of \hyperlinkcite{mondino2023}{MS23}, as long as self-gravitational wakes are expected to be weak.
% This follows from Poisson equation written for $z=0$ and ignoring the radial terms. 

% \subsection{Calculation of the particle mean free path length}

% \subsection{Calculation of phase shifts}

% ...?

\subsection{Results}

\begin{figure*}
\centering
\includegraphics[width = 0.85\textwidth]{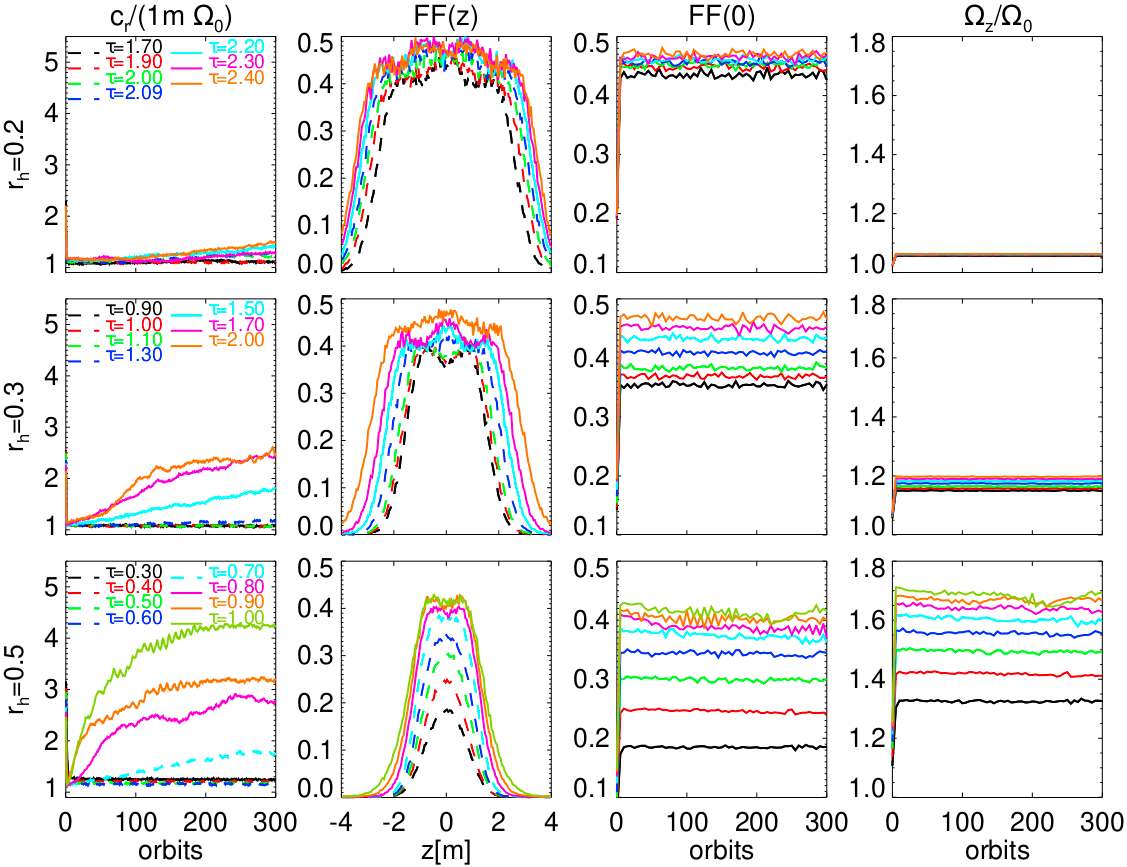}
\caption{Summary of a number of conducted N-body simulations wit $\epsilon=0.1$ including radial and vertical self-gravity. Different rows compare different values of $r_{h}$ as indicated. Different colors correspond to different optical depths $\tau$ (\ref{eq:tau_sim}). From left to right columns we plot the time evolution of the radial component of the root-mean-squared radial velocity $c_{r}$, the time-average of the vertical distribution of the filling factor $\mbox{FF(z)}$ (\ref{eq:ff_sim}), the time evolution of the central (midplane) filling factor $\mbox{FF}(0)$, and the vertical frequency $\Omega_{z}$ (cf. (\ref{eq:fsgz_nbody})). In the first and second columns runs which did (did not) develop overstability are drawn as solid (dashed) curves.}
\label{fig:nbody}
\end{figure*}

Figure \ref{fig:nbody} collects various quantities extracted from simulations with $r_{h}=0.2,0.3,0.6, \text{and} \, 0.7$, where $r_{h}$ increases from top to bottom rows. Different optical depths $\tau$ as defined by (\ref{eq:tau_sim}) are represented by different colors.
From left to right columns, we plot the root-mean-squared radial velocity $c_{r}$, the time-averaged vertical profile of the volume filling factor $\mbox{ff}(z)$ as defined by (\ref{eq:ff_sim}), the time evolution of the midplane filling factor $\mbox{ff}(0)$, and the vertical self-gravity factor $\Omega_{z}/\Omega_{0}$ as defined by (\ref{eq:fsgz_nbody}).
Actually the occurrence of overstability has been determined from the evolution of radial Fourier modes of the velocity field.
Simulations that develop viscous overstability are characterised by a saturated value $c_{r}\gtrsim 1 \text{m}\, \Omega_{0}$. The plots in the second column confirm that the vertical density distribution indeed transitions from a near-Gaussian to a near-uniform distribution for increasing $\tau$ (cf. \S \ref{sec:vstruct}), such that for $\tau\gtrsim 1$ the midplane region attains a fairly flat density profile.

% \begin{figure*}
% \centering
% \includegraphics[width = 0.6\textwidth]{figs/nbody_spec.png}
% \caption{Time evolution of Fourier power spectra of radial profile of the optical depth.}
% \label{fig:nbody_spec}
% \end{figure*}

%  \begin{figure}
%  \centering
%  \includegraphics[width = 0.4\textwidth]{figs/Epseudo.png}
%  \caption{...}
%  \label{fig:epseudo}
%  \end{figure}

\section{Comparison of simulations and model}\label{sec:nbody_hydro}

\subsection{Threshold for viscous overstability - constant vertical self-gravity}

We first consider the approximation of a constant vertical self-gravity force, described by a constant $\Omega_{z}$, as given by (\ref{eq:omz}), in absence of radial self-gravity. This assumption, while artificial, provides a simple means to shift smoothly between the non-selfgravitating and strongly self-gravitating limits, as far as vertical self-gravity is concerned. 
% Moreover, it simplifies a comparison of our results with those from N-body simulations adopting the same approximation.
%
Table \ref{tab:os_ng} summarises critical values of the optical depth $\tau_{cr}$, the filling factor $\mbox{FF}_{cr}$, and the viscous slope $\beta_{cr}$,
 for the onset of viscous overstability, resulting from the solution of (\ref{eq:eigenproblem}). We adopt the values $\Omega_{z}/\Omega_{0}=1, 2$ for the constant vertical self-gravity, as well as the constant coefficients of restitution $\epsilon=0.1,0.3,0.5$. Results for both the slab and Gaussian vertical distribution are shown, and are compared to corresponding results reported in \hyperlinkcite{mondino2023}{MS23} for the same values of $\epsilon$  and $\Omega_{z}$.
 For $\Omega_{z}/\Omega_{0}=1$, corresponding to vanishing vertical self-gravity, we find our values of $\mbox{FF}_{cr}$ to agree quite well with those from N-body simulations. There is only little difference between the two density distributions regarding the values of $\mbox{FF}_{cr}$, which also holds for larger values of $\Omega_{z}$. On the other hand, the critical values for the  optical depth $\tau_{cr}$ found in N-body simulations are much more closely reproduced with the Gaussian distribution.
 However, deviations from the N-body results increase with increasing value of $\Omega_{z}$. This is not surprising, and is most likely related to the circumstance that the vertical particle distribution in N-body simulations using a constant $\Omega_{z}/\Omega_{0}\gtrsim 2$ - in contrast to the vertical distribution assumed in our model - is subject to a layering, such that the number of vertical particle layers increases with increasing optical depth in a discontinuous fashion.
 % Indeed, for $\Omega_{z}/\Omega_{0}=3.6$ we find values of the ring thickness below a particle diameter, which is unphysical. In the true particulate system this is avoided through the aforementioned formation of layers.
 This results in a oscillatory behaviour of the central filling factor with increasing optical depth, as the formation of each new layer abruptly changes the midplane filling factor.

% %   \begin{figure}
% %   \centering
% %   \includegraphics[width = 0.4\textwidth]{figs/taucrit.png}
% %   \caption{}
% %   \label{fig:os_crit_ng}
% %   \end{figure}
% %   %

\begin{table}
 \caption{Critical values of important model parameters for the onset of viscous overstability using a constant $\epsilon$ and $\Omega_{z}$. The three values of $\beta_{\text{cr}}$, $\tau_{\text{cr}}$ and $\mbox{ff}_{\text{cr}}$ correspond to model calculations using a slab profile (S) and a Gaussian profile (G), as well as N-body simulations (NB; values from \citet{mondino2023}).}
 \label{tab:os_ng}
 \begin{tabular*}{\columnwidth}{@{}l@{\hspace*{15pt}}l@{\hspace*{15pt}}l@{\hspace*{15pt}}l@{\hspace*{15pt}}l@{}}
  \hline
 $\epsilon$ &  $\Omega_{z}/\Omega_{0}$ & $\beta_{\text{cr,(S/G\,/NB)}}$  &   $\tau_{\text{cr,(S/G\,/NB)}}$    &  $\mbox{FF}_{\text{cr,(S/G\,/NB)}}$  \\ \hline %%%%OMZ=1
  0.1 & 1.0  &  1.29/1.18/1.20 & 3.50/1.99/2.1  & 0.51/0.52/0.47  \\ 
  0.3 & 1.0  & 1.30/1.18/1.20 & 3.91/2.23/2.3 & 0.51/0.52/0.46  \\ 
  0.5 & 1.0  & 1.33/1.20/1.15 & 4.89/2.77/2.9 & 0.52/0.53/0.45 \\
  \hline %%%OMZ=2
    0.1 & 2.0  & 1.16/1.01/1.35 & 1.46/0.82/0.95  & 0.48/0.48/0.41  \\ 
  0.3 & 2.0  & 1.16/1.00/1.30 & 1.63/0.92/1.05 & 0.48/0.48/0.36 \\ 
  0.5 & 2.0  & 1.18/1.00/1.10 & 2.02/1.13/1.1 & 0.49/0.49/0.50 \\
    \hline %%%%OMZ=3.6
    % 0.1 & 3.6  & 1.14/0.99/1.95 & 0.78/0.44/0.42 & 0.48/0.47/0.67 \\ 
  % 0.3 & 3.6  & 1.14/0.99/1.40 & 0.87/0.49/0.65 & 0.48/0.48/0.69 \\ 
  % 0.5 & 3.6  & 1.16/0.99/0.95 & 1.09/0.61/0.85 & 0.49/0.49/0.51 \\
  % \hline
 \end{tabular*}
\end{table}

\subsection{Threshold for viscous overstability - vertical and radial self-gravity}

\begin{figure}
  \centering
  \includegraphics[width = 0.4\textwidth]{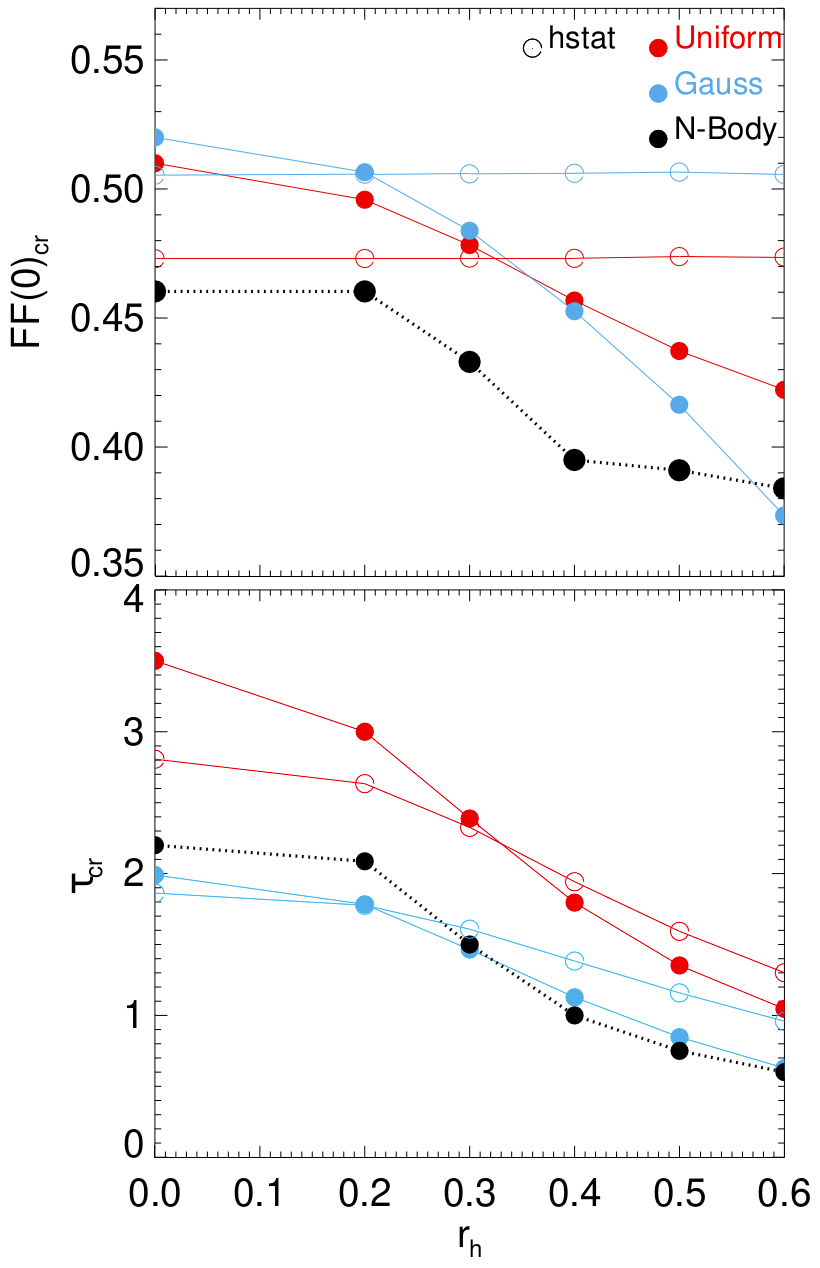}
  \caption{Critical values of the filling factor (top) and optical depth (bottom) for the onset of viscous overstability resulting from hydrodynamic model calculations using a uniform distribution (red solid curves) and a Gaussian distribution (blue solid curves) are compared with corresponding values from N-body simulations (black dotted curves) for increasing $r_{h}$ in the presence of vertical self-gravity. The simulations used $\epsilon=0.1$ and the model calculations assumed the equilibrium shown in Figure \ref{fig:gstate_combo} (left panels). Open circles show results using the vertically static approximation (see \S \ref{sec:hstat}).}
  \label{fig:os_crit_sg}
  \end{figure}

We now compare critical quantities for the onset of overstability in presence of (radially varying) vertical and radial self-gravity. Here we consider a fixed $\epsilon=0.1$ in simulations and model calculations. In Figure \ref{fig:os_crit_sg} we first plot $\mbox{FF}_{\text{cr}}$ and $\tau_{\text{cr}}$ for increasing $r_{h}$ (i.e. increasing strength of self-gravity), resulting from our hydrodynamic model and our N-body simulations adopting  only vertical self-gravity forces. Again, we find overall only mild differences between $\mbox{FF}_{\text{cr}}$ for the slab (red) and Gaussian (blue) distributions, which match reasonably well the values found from our simulations (black).
On the other hand, the values of $\tau_{\text{cr}}$ resulting with the Gaussian distribution significantly better match those from N-body simulations. The open circles are results obtained using the vertically static approximation "hstat" (see \S \ref{sec:hstat}), where vertical motions are ignored. These results clearly demonstrate that it is the vertical self-gravity acting on overstable waves (via Eq. \ref{eq:linHdot}) which lowers the critical filling factor for overstability.
% We compare model calculations and simulations with radial and vertical self-gravity (solid curves), as well as with only vertical self-gravity (dashed curves). 
% We find that in absence of radial self-gravity the results of model (using a Gaussian vertical distribution) and simulations compare reasonably well. 

If we in addition include radial self-gravity (without otherwise changing any of the model parameters), a discrepancy develops with increasing $r_{h}\gtrsim 0.4$, such that the values of $\mbox{FF}_{\text{cr}}$ resulting from the model become increasingly smaller than those resulting from the simulations. As mentioned in \S \ref{sec:os_rsg} for $r_{h}\gtrsim 0.5$ we find overstability for all $\mbox{FF}\gtrsim 0$ when radial self-gravity is included.
Thus, the effect of radial self-gravity seen in our model (cf. \S \ref{sec:os_sg}) is much stronger than in our N-body simulations, where it results only in a mild reduction of $\mbox{FF}_{cr}$.
Nevertheless, we find that agreement between model and simulations in presence of radial self-gravity can be attained by using values of the bulk viscosity $\widetilde{\xi_{0}}$ larger than (\ref{eq:nub_omc}).
 This is discussed in Appendix \ref{app:rsg}.
There we also show that the measured values of the bulk viscosity in \citet{salo2001} are likely to be altered if radial self-gravity is included in the simulations.

Furthermore, we find that critical values for $\mbox{FF}$ and $\tau$ resulting from our N-body simulations including radial and vertical self-gravity are very similar to those resulting from the fully self-gravitating simulations of \hyperlinkcite{mondino2023}{MS23} as long as $r_{h}\lesssim 0.65$ (Their Figures 10 and C1). 
% This is shown in Figure \ref{fig:nbody_vs_ms23}. 
For larger values of $r_{h}$ overstability is being suppressed by strong self-gravity wake structures, not captured by our axisymmetric self-gravity implementaton (\hyperlinkcite{mondino2023}{MS23}).

  %

% \begin{figure}
%   \centering
%   \includegraphics[width = 0.5\textwidth]{figs/nbody_vs_ms23.png}
%   \caption{Critical values of the filling factor (top) and optical depth (bottom) for the onset of viscous overstability resulting from our N-body simulations are compared with corresponding values from fully self-consistent self-gravitating N-body simulations of \citet{mondino2023} (red solid curves) for increasing $r_{h}$. Open (solid) circles describe results including vertical (vertical and radial) self-gravity. Note that the apparent scatter of the red curve mainly stems from the $\tau$-spacing between subsequent simulations to measure the critical optical depth for overstability.}
%   \label{fig:nbody_vs_ms23}
% \end{figure}
  %

\subsection{Linear growth rates}
\begin{figure}
  \centering
  \includegraphics[width = 0.5\textwidth]{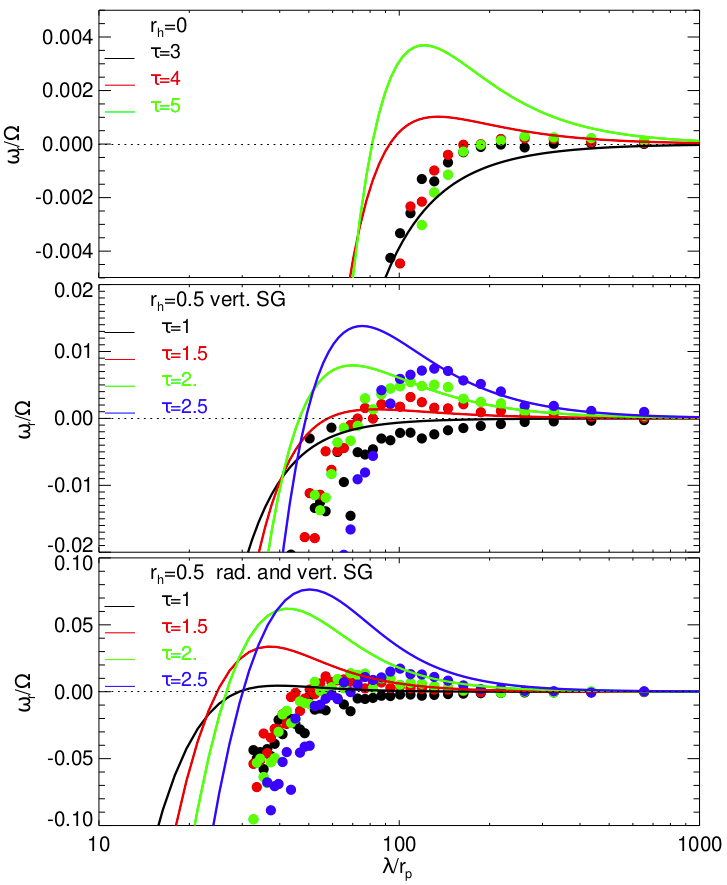}
  \caption{Comparison of linear growth rates of the viscous overstability as measured from N-body simulations (cirles) and as predicted by our fluid model (curves). From top to bottom panels we consider a non-gravitating ring, a ring subject to vertical self-gravity, and a ring subject to radial and vertical self-gravity. The results assume $\epsilon=0.5$.}
  \label{fig:grates}
\end{figure}

In the previous section we compared the threshold values of important equilibrium  quantities that lead to the onset of viscous overstability in our model and simulations.
Here we compare the linear growth rates of resulting overstable modes, as predicted by our model and as measured from our simulations. Figure \ref{fig:grates} shows growth rates measured in simulations (represented by circles) without self-gravity (top panel), with only vertical self-gravity (middle panel), and with both radial and vertical self-gravity (bottom panel). All simulations adopt a constant $\epsilon=0.5$. The self-gravitating cases adopt $r_{h}=0.5$. Compared in each panel are different optical depths.
The curves represent hydrodynamic growth rates, based on the equilibrium state shown in Figure \ref{fig:gstate_eps05} (left panel) using a Gaussian distribution.

The method to measure growth rates from our simulations is largely similar to that used in \citet{lehmann2017}. That is, we seeded sinusoidal radial velocity perturbations of all modes (with identical amplitudes $r_{p}\Omega_{0}$) simultaneously and tracked the amplitudes of individual modes in time by applying a Fourier transform on the total radially binned velocity profile.

In the case with $r_{h}=0$ we find that the hydrodynamic growth rates are substantially larger than those measured from simulations. In contrast, the wavelengths of maximal growth are substantially smaller. Similar discrepancies were observed in previous hydrodynamic studies (e.g. \citealt{salo2010} (their Figure 5) and \citealt{lehmann2017} (their Figure 4)). On the other hand, the threshold for overstability as found in the simulations is quite well reproduced by our model (see Figure \ref{fig:os_crit_sg}), which is reflected by the reasonable agreement for the case with $\tau=3$. The reason for the large hydrodynamic growth rates for the cases with $\tau\geq 4$ is that the model is very close to the densest packing with $\mbox{FF}\sim 0.61$. The value of $\mbox{FF}$ in the simulations is slightly smaller, which might partially explain the large difference in the growth rates.

Appreciably better agreement is seen in the growth rates resulting with the inclusion of vertical self-gravity. One possible reason for the better agreement is that the vertical layering of particles (not captured by the hydrodynamic model) is reduced when vertical self-gravity is included (\citet{latter2008}).
However, when including radial self-gravity, deviations are again larger, which is not surprising based on our findings described in the previous section. Note that for the hydrodynamic calculations in the bottom panel we used the bulk viscosity as displayed in Figure \ref{fig:nub}. Despite the threshold values for the onset of overstability being in good agreement (Figure \ref{fig:nub}, right panels), the linear growth rates show a similar deviation from the measured values as for the non-gravitating case.

\section{Discussion}\label{sec:summary}

\subsection{summary}

We developed a fluid model to describe the dynamics of dense planetary rings, such as the Saturnian A and B rings, or the Uranian $\epsilon$ ring, inspired by the granular flow model of \citet{hutter1995}. In contrast to existing hydrodynamic models, and more in line with kinetic models, our model is largely based on parameters that can be determined from observations and experiments, such as the particle size $d$, the coefficient of restitution $\epsilon$, and the optical depth $\tau$. 
% Few additional parameters are required, values of which can be motivated based on results from N-body simulations and existing kinetic models. 
Our model generalises previous hydrodynamic models of planetary rings by incorporating the dynamical evolution of the disc thickness, where motions are restricted to be symmetric with respect to the midplane. 
This generalisation enables us to study the effect of vertical motions, different vertical density stratifications, and the effect of variations in the volume filling factor on ring dynamics.
Furthermore, while our model is able to adequately capture the behavior of densely packed particulate systems, it exhibits a substantially decreased mathematical complexity as compared to dense ring kinetic models.

  We applied our model to compute axisymmetric planetary ring equilibrium states, where we considered a vertically uniform distribution, as well as a Gaussian vertical distribution for the volume mass density. The computed equilibrium states compare (at least qualitatively) well with those resulting from N-body simulations for small and large optical depths. 
  % We find that the precise form of the vertical density distribution only mildly affects the equilibrium quantities.

Next, we performed an axisymmetric linear stability analysis of our model, where we reproduced the viscous instability, the thermal relaxation mode, and the pair of overstable density waves, analysed in previous studies. In addition, we find a pair of vertical breathing modes, which describe oscillations of the ring thickness. We showed that in the appropriate limits the behaviour of all these modes agree with previous studies.

We then focused on the viscous overstability. 
In absence of self-gravity we find the occurrence of the viscous overstability as soon as a certain critical volume filling factor of the ring material is surpassed. This finding is in agreement with the kinetic study of \hyperlinkcite{latter2008}{LO08} and the N-body simulations of \hyperlinkcite{mondino2023}{MS23}. 
% Interestingly, we find that the inclusion of vertical scaleheight variations slightly reduces the critical filling factor for onset of instability. 
In our model we find that the largest damping effect on overstable modes results from variations in the disc thickness. 
Furthermore, in agreement with the hydrodynamic studies of \citet{spahn2000a} and \citet{schmidt2001b}, our model predicts that temperature variations have a mitigating effect on the viscous overstability.

In addition, we performed local N-body simulations including vertical and radial self-gravity. In absence of radial self-gravity the critical values for the filling factor and the optical depth for the onset of overstability agree well between our simulations and our hydrodynamic model using a Gaussian density distribution. In particular, the critical filling factor mildly reduces with increasing particle bulk density. As shown, this reduction is caused by the vertical self-gravity acting on overstable modes, and is absent if vertical motions are neglected.
 On the other hand, when radial self-gravity is included, the hydrodynamic model predicts for increasing $r_{h} \gtrsim 0.4$ increasingly smaller values of the critical filling factor and optical depth than our simulations. We speculate that the strong effect of radial self-gravity in our hydrodynamic model results from a reduction of the fluid's compressibility, which in turn mitigates the damping effect of the bulk viscous stress on overstable modes. We find that agreement between model and simulations can be attained by using increased values of the bulk viscosity, depending on the value of $r_h$.
We also showed that the method used to measure the bulk viscosity in \citet{salo2001} needs to be modified when radial self-gravity is included, thus providing a possible explanation for the need of an increased bulk viscosity.
Moreover, we find that our N-body simulations with vertical and radial self-gravity correctly describe the threshold for viscous overstability as found in fully self-gravitating simulations of \hyperlinkcite{mondino2023}{MS23} for $r_{h}\lesssim 0.65$, beyond which self-gravity wakes suppress the instability in the latter simulations.

\subsection{Caveats and outlook}

Even though our hydrodynamic model is able to reproduce well (in some parameter regimes even quantitatively) the equilibrium state of a dense planetary ring as resulting from N-body simulations, agreement is substantially reduced when perturbations to the equilibrium state are considered. This can be seen for instance when comparing linear growth rates of overstable modes.  

Furthermore, despite being largely framed in kinetic parameters, such as the optical depth $\tau$ and coefficient of restitution $\epsilon$, our model still needs to rely on a number of additional parameters, such as the "epicycle-parameter" $\delta$, and probably most importantly, the bulk viscosity $\xi$. These quantities are strongly related to the failure of a fluid description in the dilute regime, characterised by low particle collision frequencies.
Perhaps the only way of getting around the need of such parameters is the employment of a kinetic model. 
However, even sophisticated dense ring kinetic models (\citealt{latter2008}) struggle to achieve good agreement with N-body simulations when perturbations to the equilibrium are considered.

In addition, limitations of our description of disc thickness variations likely occur when vertical motions become strong, as for instance in the vicinity of a Lindblad resonance, or in strongly nonlinear density wave trains, which should lead to a "splashing" of the ring material \citep{borderies1985,salo2001} . A correct analytical description in such situations might require a fully three-dimensional analysis of the fluid equations.

Regarding the description of the viscous overstability, our hydrodynamic model may potentially be improved by the construction of better suited transport coefficients, in particular the viscosity. On the other hand, kinetic models for the viscous overstability might be further improved by adopting a more accurate description of the vertical ring structure (\citealt{araki1991}), rather than adopting a pre-specified form. Even though this would further increase the mathematical complexity of such models, it could help to pin down possible sources for disagreement between models and N-body simulations. In this regard, it may also be useful to compare hydrodynamic and kinetic models directly.

% While the transport coefficients derived in this paper capture well the optically thick regime with sufficiently small values of the coefficient of restitution $\epsilon\lesssim 0.5$, relevant to the viscous overstability, 
% they disagree with those obtained from N-body simulations and kinetic models at small optical depths. More generally speaking, our set of transport coefficients is not applicable to describe dilute rings, characterised by small collision frequencies and larger values of $\epsilon$. For instance, Eq. (\ref{eq:veldisp}) fails to predict a critical $\epsilon$ above which no equilibrium is possible.
% While we do not believe that this qualitatively affects the dynamics in the optically thick regime, it should be attempted to construct improved transport coefficients which are valid also for small optical depths and more elastic systems.

The strong impact of radial self-gravity exhibited in our linear hydrodynamic model at large particle bulk densities is puzzling, and should be addressed in future work. This may require a modification to the computation of the bulk viscosity in N-body simulations as utilised in \citet{salo2001}.
Also regarding the effect of radial self-gravity a direct comparison with dense ring kinetic models could be useful.

In a future study we plan to perform nonlinear hydrodynamic simulations based on our fluid model. These simulations can be used to investigate the effect of vertical thickness variations on the long-term nonlinear saturation of the viscous overstability. 
% Even though their effect resulting from our model is found to be small for linear perturbations, this need not be the case in the nonlinear regime.
 Moreover, once the axisymmetric dynamics are understood, 2D non-axisymmetric shearing sheet  simulations can be performed to study the interaction of the viscous overstability with self-gravity wakes, which cannot directly be accounted for in a linear model.

\section*{Acknowledgements}

ML is indebted to J\"urgen Schmidt
for valuable contributions, especially regarding Appendix A.
We acknowledge support from the National Science and Technology Council through grants 110-2112-M-001-034-, 111-2112-M-001-062-, 111-2124-M-002-013-, 112-2124-M-002-003- and an Academia Sinica Career Development Award (AS-CDA-110-M06)".

%%%%%%%%%%%%%%%%%%%%%%%%%%%%%%%%%%%%%%%%%%%%%%%%%%
\section*{Data Availability}

The data underlying this article will be shared on reasonable
request to the corresponding author.
%\FloatBarrier
%\clearpage


\bibliographystyle{mnras}
\bibliography{lit} % if your bibtex file is called example.bib

%\begin{thebibliography}{99}
%\end{thebibliography}

%%%%%%%%%%%%%%%%%%%%%%%%%%%%%%%%%%%%%%%%%%%%%%%%%%

%%%%%%%%%%%%%%%%% APPENDICES %%%%%%%%%%%%%%%%%%%%%
% \clearpage

\appendix

 \section{Vertical Integration of hydrodynamic Equations}\label{app:zint}
 
 \subsection{Basic equations and assumptions}
 In the following we perform the vertical integration of the three-dimensional hydrodynamic equations of a planetary ring, resulting in the dynamical equations (\ref{eq:contsig})-(\ref{eq:conth2}) used in this paper.
%Basic hydrodynamic equations and assumptions on parity
For simplicity, we adopt a local description from the outset. That is, we neglect any possible global gradients of ring properties, such as pressure and self-gravity, as these are not expected to affect the local instabilities investigated here, which has also been assumed in numerous previous analytical studies and N-body simulations of planetary rings. 
Thus, we directly focus on local dynamics with scales $\ll r$ and adopt Cartesian coordinates $(x,y,z)$ with unit vectors $\vec{\mathbf{e}}_x,\, \vec{\mathbf{e}}_y, \, \vec{\mathbf{e}}_z$ in a (for the time being) non-rotating frame at constant distance $r_{0}$ from the central planet.
The three-dimensional continuity and the Navier-Stokes equations then read
\begin{align}
\mathcal{D}_{t} \rho &=-\rho\left(\partial_{x} u_{x}+\partial_{y} u_{y}+\partial_{z} u_{z}\right),\label{eq:conteq_3D} \\
\rho \mathcal{D}_{t} u_{x} &=\rho F_{x}-\left(\partial_{x} P_{x x}+\partial_{y} P_{y x}+\partial_{z} P_{z x}\right),\label{eq:navx} \\
\rho \mathcal{D}_{t} u_{y} &=\rho F_{y}-\left(\partial_{x} P_{x y}+\partial_{y} P_{y y}+\partial_{z} P_{z y}\right),\label{eq:navy} \\
\rho \mathcal{D}_{t} u_{z} &=\rho F_{z}-\left(\partial_{x} P_{x z}+\partial_{y} P_{y z}+\partial_{z} P_{z z}\right),\label{eq:navz}
\end{align}
where $\mathcal{D}_{t} =\partial_{t}+u_{x} \partial_{x}+u_{y} \partial_{y} + u_{z} \partial_{z} $ and
where $F_{i}$ are the components of the sum of central gravity and self-gravity forces. In the general case $\rho, u_{x}, u_{y}, u_{z}, F_{x}, F_{y}, F_{z}$ and the components of the pressure tensor ($P_{ij}$ with $i,j=x,y,z$) are depending on $x, y$, and $z$. For the balance of energy we take the three-dimensional heat-flow equation
\begin{equation}
\begin{split}\label{eq:heateq}
\frac{3}{2} \rho \mathcal{D}_{t} T=&-\left(P_{x x} \partial_{x} u_{x}+P_{y y} \partial_{y} u_{y}+P_{z z} \partial_{z} u_{z}\right.\\
\quad &+P_{x y} \partial_{y} u_{x}+P_{x y} \partial_{x} u_{y}+P_{x z} \partial_{x} u_{z} \\
\quad &\left.+P_{x z} \partial_{z} u_{x}+P_{y z} \partial_{y} u_{z}+P_{y z} \partial_{z} u_{y}\right) \\
\quad &-\left(\partial_{x} q_{x}+\partial_{y} q_{y}+\partial_{z} q_{z}\right)-\Gamma,
\end{split}
\end{equation}
where $T$ is the kinetic temperature and $q_{i}$ with $i=x,y,z$ are the components of the heat flux. The function $\Gamma$ stands for the steady energy loss in the system due to inelastic collisions.

Now we make a few (for a thin disk reasonable) assumptions on the parity of some of the employed functions
We assume that the disk is symmetric with respect to the mid-plane 
such that
\begin{align}
\rho(x, y, z) &=\rho(x, y,-z) ,\\
u_{z}(x, y, z) &=-u_{z}(x, y,-z), \\
F_{z}(x, y, z) &=-F_{z}(x, y,-z).
\end{align}
As in \citet{stehle1999} we assume the ring to be sufficiently thin such that the variation of the planar components of the central gravity and self-gravity forces over the height of the ring is negligible. Also the planar components of the velocity shall not vary with $z$. Thus,
\begin{align}
u_{x} &=u_{x}(x, y) \label{eq:uxpar},\,u_{y} =u_{y}(x, y),\\
F_{x} &=F_{x}(x, y) \label{eq:Fxpar},\, F_{y}  =F_{y}(x, y).
\end{align}
Furthermore, we require that the density shall drop off exponentially as $z \rightarrow \pm \infty$. Hence, the pressure as well as the transport coefficients will also vanish at large $z$.

\subsection{Pressure tensor}

The pressure tensor is given by Eq. (\ref{eq:pten})
% To treat the pressure terms in the Navier-Stokes equation we now have to specify the pressure tensor. We take the usual Newtonian form
% %
% \begin{equation}
% \begin{split}
%     P_{i j}(x, y, z) &=p(x, y, z) \delta_{i j}-\eta(x, y, z)\left(\partial_{i} u_{j}+\partial_{j} u_{i}-\frac{2}{3} \delta_{i j} \mathbf{\nabla} \cdot \mathbf{u}\right)\\
%     \quad & -\xi(x, y, z) \delta_{i j} \mathbf{\nabla} \cdot \mathbf{u}\\
% \quad &=p \delta_{i j}-\eta\left(\partial_{i} u_{j}+\partial_{j} u_{i}\right)+\left[\frac{2}{3} \eta-\xi\right] \delta_{i j} \mathbf{\nabla} \cdot \mathbf{u},
% \end{split}
% \end{equation}
% with the scalar pressure $p$ and the dynamic shear and bulk viscosities $\eta$ and $\xi$. 
and the disk is assumed to be vertically isothermal, such that its temperature $T$ is independent of $z$. Therefore
\begin{align}
p(x, y, z) &=p(x, y,-z), \\
\eta(x, y, z) &=\eta(x, y,-z), \\
\xi(x, y, z) &=\xi(x, y,-z),
\end{align}
since these quantities depend only on position through $\rho$ (which is symmetrical with respect to the midplane) and $T$. From the definition of the (symmetric) pressure tensor we then read the parity of its components
\begin{align}\label{eq:Pii}
&P_{i i}(z)=p-2 \eta \partial_{i} u_{i}+\left(\frac{2}{3} \eta-\xi\right) \sum \limits_{k=x,y,z}^{}\partial_{k} u_{k}  =P_{i i}(-z) ,\\
&P_{x y}(z)=-\eta\left(\partial_{x} u_{y}+\partial_{y} u_{x}\right)=P_{x y}(-z) ,\\
&P_{x z}(z)=-\eta \partial_{x} u_{z}=-P_{x z}(-z)\label{eq:Pxz}, \\
&P_{y z}(z)=-\eta \partial_{y} u_{z}=-P_{y z}(-z)\label{eq:Pyz}.
\end{align}

\subsection{Continuity equation}
%The continuity equation
Now we $z$-integrate, starting with the continuity equation (\ref{eq:conteq_3D}). With the assumption (\ref{eq:uxpar}) we have
\begin{align}
\int_{-\infty}^{\infty} d z u_{x} \partial_{x} \rho &=u_{x} \partial_{x} \int_{-\infty}^{\infty} d z \rho ,\\
\int_{-\infty}^{\infty} d z \rho \partial_{x} u_{x} &=\partial_{x} u_{x} \int_{-\infty}^{\infty} d z \rho,
\end{align}
and similar relations for the terms containing $u_{y}$ in the equation. Thus, the z-integrated continuity equation reads
\begin{equation}\label{eq:conteq}
\begin{split}
\left(\partial_{t}+u_{x} \partial_{x}+u_{y} \partial_{y}\right) \sigma& =-\sigma\left(\partial_{x} u_{x}+\partial_{y} u_{y}\right)\\
\quad & -\int_{-\infty}^{\infty} d z\left(\rho \partial_{z} u_{z}+u_{z} \partial_{z} \rho\right).
\end{split}
\end{equation}
% where the surface density $\sigma$ is defined through
%  \begin{equation}
%      \sigma = \int \limits_{-\infty}^{+\infty} \rho \mathrm{d}z.
%  \end{equation}
The last term in $(\ref{eq:conteq})$ yields a boundary term which vanishes if $\rho$ decays to zero more rapidly for $z \rightarrow \pm \infty$ than $u_{z}$ grows. This is a reasonable assumption, since the density should drop off exponentially while the velocity should grow at most algebraically. Thus, we are left with the z-integrated continuity equation 
\begin{equation}
D_{t} \sigma=-\sigma\left(\partial_{x} u_{x}+\partial_{y} u_{y}\right),
\end{equation}
where $D_{t}=\left(\partial_{t}+u_{x} \partial_{x}+u_{y} \partial_{y}\right)$, which is identical to (\ref{eq:contsig}).
%The Navier-Stokes equation

\subsection{Navier-Stokes equations}
Next we consider the x-component (\ref{eq:navx}) of the Navier-Stokes equation and $\mathrm{z}$-integrate. Again by the assumptions $(\ref{eq:uxpar},\ref{eq:Fxpar})$ we have
\begin{align}
\int_{-\infty}^{\infty} d z \rho \partial_{t} u_{x} &=\partial_{t} u_{x} \int_{-\infty}^{\infty} d z \rho ,\\
\int_{-\infty}^{\infty} d z \rho u_{x} \partial_{x} u_{x} &=u_{x} \partial_{x} u_{x} \int_{-\infty}^{\infty} d z \rho, \\
\int_{-\infty}^{\infty} d z \rho u_{y} \partial_{y} u_{x} &=u_{y} \partial_{y} u_{x} \int_{-\infty}^{\infty} d z \rho, \\
\int_{-\infty}^{\infty} d z \rho F_{x} &=F_{x} \int_{-\infty}^{\infty} d z \rho,
\end{align}
and similar relations for the terms containing $u_{y}$. The terms proportional to $\partial_{z} u_{x}$ and $\partial_{z} u_{y}$ vanish. Therefore we get for the derivatives of the pressure tensor in the Navier-Stokes equation
\begin{align}
\int_{-\infty}^{\infty} d z \partial_{x} P_{x x} &=\overline{\partial_{x} P_{x x}}+ \int_{-\infty}^{\infty} d z \partial_{x}\left(\left[\frac{2}{3} \eta-\xi\right] \partial_{z} u_{z}\right) ,\\
\int_{-\infty}^{\infty} d z \partial_{y} P_{y y} &=\overline{\partial_{y} P_{y y}}+\int_{-\infty}^{\infty} d z \partial_{y}  \left(\left[\frac{2}{3} \eta-\xi\right] \partial_{z} u_{z}\right), \\
\int_{-\infty}^{\infty} d z \partial_{x} P_{x y} &=\overline{\partial_{x} P_{x y}} ,\\
\int_{-\infty}^{\infty} d z \partial_{y} P_{x y} &=\overline{\partial_{y} P_{x y}},
\end{align}
%
% and
% \begin{equation}
%     \{\overline{p}, \overline{\eta}, \overline{\xi}\} \equiv \int_{-\infty}^{\infty} d z\{p, \eta, \xi\}
% \end{equation}
% %
% are $z$-integrated scalar quantities. 
Furthermore, if we assume that the components of the pressure tensor vanish at infinity (i.e. that $p, \eta$, and $\xi$ should vanish with $\rho$), we have
\begin{equation}
\int_{-\infty}^{\infty} d z \partial_{z} P_{i j}=\lim _{z \rightarrow \infty}\left(P_{i j}(z)-P_{i j}(-z)\right)=0.
\end{equation}
Consequently the planar components of the Navier-Stokes equations reduce to
\begin{align}
\sigma D_{t} u_{x} &=\sigma F_{x}-\left(\overline{\partial_{x} P_{x x}}+\overline{\partial_{y} P_{y x}}+\overline{\partial_{x} f\left(u_{z}\right)}\right), \\
\sigma D_{t} u_{y} &=\sigma F_{y}-\left(\overline{\partial_{x} P_{x y}}+\overline{\partial_{y} P_{y y}}+\overline{\partial_{y} f\left(u_{z}\right)}\right),
\end{align}
where $f\left(u_{z}\right)$, given by (\ref{eq:fz}), 
% \begin{equation}
% f\left(u_{z}\right) \equiv \left(\frac{2}{3} \eta-\xi\right) \partial_{z} u_{z}
% \end{equation}
is the only $u_{z}$-dependent term in the Navier-Stokes equations that does not vanish after z-integration. Equations (\ref{eq:contux},\ref{eq:contuy}) then follow upon changing to a frame rotating with $\Omega_{0}$.

\subsection{Scale height equation}
The z-component of the Navier-Stokes equations vanishes identically after z-integration, only due to the assumptions made on the parity. However, from it we can derive an equation describing the evolution of the disk scale height as follows.
With the assumption (\ref{eq:velz})
the vertical Navier-Stokes Equation (\ref{eq:navz}) reads
\begin{equation}\label{eq:navz2}
 \rho \frac{z}{H}D^2_{t} H = \rho \left( F_{\text{c},z} + F_{\text{sg},z} \right) - \partial_{x} P_{xz}- \partial_{y} P_{yz} - \partial_{z} P_{zz}
\end{equation}
where $F_{\text{c},z}$ and $F_{\text{sg},z}$ denote the vertical components of the planetary gravity and the self-gravity force, respectively.
In the current approximation the former is given by
\begin{equation}\label{eq:fz_central}
F^{c}_{z}(z)  = -\Omega_{0}^2 z .
\end{equation}
% Now we multiply (\ref{eq:navz2}) by $z$ and integrate. Using (\ref{eq:rho_slab}), (\ref{eq:Pii}), (\ref{eq:Pxz}) and (\ref{eq:Pyz}) we obtain Eq. (\ref{eq:conth2}).
% \begin{equation}
%     \begin{split}\label{eq:Keq1}
% D_{t} K  & =  -\Omega_{0}^2 H - \frac{3}{ H \sigma}  \int\limits_{-\infty}^{\infty} \mathrm{d}z \, \rho z \left(F^{\text{sg}}_{z}\right)  
% +  \frac{3 \overline{\overline{P_{zz}}} }{\delta H \sigma} \\
% &  \quad +\frac{H}{\sigma} \left[ \partial_{x} \left(\overline{\eta} \, \partial_{x} \frac{K}{H} \right) +\partial_{y} \big(\overline{\eta} \, \partial_{y} \frac{K}{H} \big)\right].
% \end{split}
% \end{equation}
% In order to arrive at (\ref{eq:Keq1}) we made use of the fact that $\eta$ is independent of $z$ with the density (\ref{eq:rho_slab}) and (\ref{eq:shearvis}) when evaluating the terms containing $P_{xz}$ and $P_{yz}$.
% In addition, as explained in Section \ref{sec:transportcoef} we introduce an additional factor $\delta$ in the vertical stress term to account for the effect of vertical non-locality of collisions.

In order to evaluate the self-gravity term we integrate the Poisson equation (\ref{eq:poissoneq}) with respect to $z$ and assume that the variation of $\phi_{\text{sg}}$ in vertical direction is much larger
than in planar directions, i.e. 
\begin{equation}
F_{\text{sg},z}  \equiv -\partial_{z} \phi_{\text{sg}}\approx  -4 \pi G \int\limits_{0}^{z} \mathrm{d}\widehat{z}\, \rho\left(\widehat{z}\right)
\end{equation}
where we also used the symmetry of $\phi_{\text{sg}}$ with respect to the ring midplane. This yields
  \begin{align}
    F_{\text{sg},z}  &  = - \frac{2 \pi G \sigma}{H} z \hspace{2.1 cm}   \text{if}   \, \rho=\rho_{\text{S}},\label{eq:fsgz_slab_z}\\
       F_{\text{sg},z}  &  = - 2 \pi G \sigma \,\text{Erf}\left(\frac{z}{\sqrt{2}H}\right) \hspace{0.55cm}   \text{if}   \, \rho=\rho_{\text{G}},\label{eq:fsgz_Gauss_z}
\end{align}
where "Erf" denotes the error function\footnote{\url{https://mathworld.wolfram.com/Erf.html}}.
We now multiply (\ref{eq:navz2}) by $z$ and perform a $z$-integration. Using (\ref{eq:gamma}) and an integration by parts on the last term on the right hand side yields (\ref{eq:conth2}) and (\ref{eq:conth}).

\subsection{Heat flow equation}
%The heat-flow equation
Finally, we consider the heat-flow equation (\ref{eq:heateq}).
Assuming $T=T(x, y)$ by vertical isothermality we have
\begin{align}
\int_{-\infty}^{\infty} d z\left(\frac{3}{2} \rho \mathcal{D}_{t} T\right) &=\frac{3}{2} \sigma D_{t} T, \\
\int_{-\infty}^{\infty} d z\left(\partial_{x} q_{x}+\partial_{y} q_{y}+\partial_{z} q_{z}\right) &=\overline{\partial_{x} q_{x}}+\overline{\partial_{y} q_{y}} ,\\
& \equiv-\overline{\partial_{x} (\kappa \partial_{x} T)}-\overline{ \partial_{y} (\kappa \partial_{y} T)}.
\end{align}
The pressure-tensor terms including $\partial_{z} u_{x / y}$ vanish by assumption (\ref{eq:uxpar}. Other components read
z-integrated

\begin{equation}
\begin{split}
\int_{-\infty}^{\infty} d z P_{z z} \partial_{z} u_{z} &=\overline{p \partial_{z} u_{z}}+\left(\partial_{x} u_{x}+\partial_{y} u_{y}\right)\overline{\left(\frac{2}{3} \eta-\xi\right) \partial_{z} u_{z}}\\
\quad & -\overline{\left(\frac{4}{3} \eta+\xi\right)\left(\partial_{z} u_{z}\right)^{2}}
\end{split}
\end{equation}
\begin{align}
\int_{-\infty}^{\infty} d z P_{x x} \partial_{x} u_{x} &=\overline{P_{x x}} \partial_{x} u_{x}+\overline{f\left(u_{z}\right)} \partial_{x} u_{x} ,\\
\int_{-\infty}^{\infty} d z P_{y y} \partial_{y} u_{y} &=\overline{P_{y y}} \partial_{y} u_{y}+\overline{f\left(u_{z}\right) }\partial_{y} u_{y} ,\\
\int_{-\infty}^{\infty} d z P_{x y} \partial_{y} u_{x} &=\overline{P_{x y}} \partial_{y} u_{x}, \\
\int_{-\infty}^{\infty} d z P_{x y} \partial_{x} u_{y} &=\overline{P_{x y}} \partial_{x} u_{y} ,\\
\int_{-\infty}^{\infty} d z P_{x z} \partial_{x} u_{z} &=-\overline{\eta\left(\partial_{x} u_{z}\right)^{2}} ,\\
\int_{-\infty}^{\infty} d z P_{y z} \partial_{y} u_{z} &=-\overline{\eta\left(\partial_{y} u_{z}\right)^{2}}.
\end{align}
where
\begin{align}
&\overline{P_{x x}}=\overline{p}+\left(\frac{2}{3} \overline{\eta}-\overline{\xi}\right) \partial_{y} u_{y}-\left(\frac{4}{3} \overline{\eta}+\overline{\xi}\right) \partial_{x} u_{x}, \\
&\overline{P_{y y}}=\overline{p}+\left(\frac{2}{3} \overline{\eta}-\overline{\xi}\right) \partial_{x} u_{x}-\left(\frac{4}{3} \overline{\eta}+\overline{\xi}\right) \partial_{y} u_{y}, \\
&\overline{P_{x y}}=-\overline{\eta}\left(\partial_{x} u_{y}+\partial_{y} u_{x}\right),
\end{align}
The z-integrated heat-flow equation is then readily given by (\ref{eq:conttemp}).

% \begin{equation}\label{eq:heatflow}
% \begin{split}
% \frac{3}{2} \sigma D_{t} T=&-\sum_{i, j=x, y} \overline{P_{i j}} \partial_{i} u_{j}-\left(\partial_{x} \overline{q_{x}}+\partial_{y} \overline{q_{y}}\right)-\overline{\Gamma} \\
% \quad &-g\left(u_{z}\right)+\overline{\eta\left(\partial_{x} u_{z}\right)^{2}}+\overline{\eta\left(\partial_{y} u_{z}\right)^{2}}\\
% \quad & -f\left(u_{z}\right)\left(\partial_{x} u_{x}+\partial_{y} u_{y}\right).
%\end{split}
%\end{equation}
% In order to arrive at (\ref{eq:conttemp}) we use (\ref{eq:velz}) to replace factors $\partial_{z} u_{z}$ appearing in (\ref{eq:heatflow}). Then, with (\ref{eq:rho_slab}) and (\ref{eq:shearvis}) we obtain
% \begin{equation}\label{eq:zvis}
%  \overline{\eta\left(\partial_{x/y} u_{z}\right)^{2}}=   \frac{1}{3} H^2 \overline{\eta} \left[ \partial_{x/y} \left(\frac{K}{H}\right)\right]^2.
% \end{equation}

% From 

%%%%%%%%%%%%%%%%%%%%%%%%%%%%%%%%%%%%%%%%%%%%%%%%%%%%%%%%%%%%%%%%%%%%%%%%%%%%%%%%%%%%%%%%%%%%%%%%%%%%%%%%%%%%%%%%%%%%%%%%%%%%%%%%%%%%%%%%%%%%%%%%%%%%%%%%%%%%%%%%%%%%%%%%%%%%%%%%%%%%%%%%%%%%%%%%%%%%%%%%

\section{Pseudo-energy decomposition}\label{sec:pseudo}

Following \citet{ishitsu09}, we perform an energy decomposition of the linearised Eqs. (\ref{eq:linsig})-(\ref{eq:linHdot}). This technique has been proven to be a useful tool to better understand linear instabilities in terms of their driving forces and damping agents (see for instance  \citealt{lin2021} for an application to instabilities in proto-planetary discs). We define the total `pseudo'-energy\footnote{The factor 4 has by convention been chosen as in \citet{ishitsu09} to eliminate the contributions due to the epicyclic terms.}
\begin{equation}
\begin{split}\label{eq:epseudo}
    E_{\text{pseudo}} & \equiv |\delta u_{x}|^2 + 4 |\delta u_{y}|^2 + \frac{1}{4}\delta\dot{H}^2 \\
    \quad &  \equiv E_{\text{sg},x}+E_{\text{sg},z}+E_{p_{0}}+E_{\eta_{0}}+E_{p_{\sigma}}+E_{\eta_{\sigma}}\\
    \quad & +E_{p_{H}}+E_{\eta_{H}}+E_{p_{T}}+E_{\eta_{T}}.
    \end{split}
    \end{equation}
For the vertical contribution to $E_{\text{pseudo}}$ (last term in the first equality of (\ref{eq:epseudo})) we used
\begin{equation}
  \left|\,\,\int\limits_{z=0}^{H} \delta u_{z}\right|^2 = \frac{1}{4}\delta\dot{H}^2,
      \end{equation}
 where we used (\ref{eq:velz}).
 The energy decomposition as constructed here measures the energies that determine the growth rates of the linear modes.
The individual components of $E_{\text{pseudo}}$ are given by
\begin{align}
\label{eq:esgx}
    E_{\text{sg},x}  & = -2 g \zeta\left(k\right)  \text{Im}\left[\delta \sigma \delta u_{x}^{*}\right],\\
\label{eq:esgz}
    E_{\text{g},z}  & = -\frac{1}{4} f_{\text{sg},z} \text{Re}\left[\delta\dot{H} \delta \sigma^{*} \right] -\frac{1}{4} \text{Re}\left[\delta\dot{H} \delta H^{*}\right],\\
\label{eq:e_eta0}
 \begin{split}
    E_{\eta_{0}}  &= - \overline{z^2 \eta_{0}} \frac{k^2}{4 \gamma H_{0}^2} |\delta \dot{H}|^2 - \overline{\eta_{0}}\Bigg( \frac{4 + 3 \widetilde{\xi_{0}}   }{12 \gamma H_{0}^2}|\delta \dot{H}|^2  \\
    \quad & +\frac{1}{3} k^2 \left(\left(4 + 3 \widetilde{\xi_{0}} \right) |\delta u_{x}|^2 + 12 |\delta u_{y}|^2 \right) \\
    \quad & -\left(\frac{1}{12 \gamma} +\frac{1}{3}\right) \frac{-2 + 3 \widetilde{\xi_{0}}}{H_{0}} k \text{Im}\left[\delta u_{x} \delta \dot{H}^{*}\right]
    \Bigg),
    \end{split}\\
     E_{\eta_{\sigma}}  & = \overline{\eta_{\sigma}}6 k \text{Im}\left[\delta \sigma \delta u_{y}^{*}\right] ,\\[0.2cm]\label{eq:eetasig}
     E_{\eta_{H}}  & = \overline{\eta_{H}} 6 k \text{Im}\left[\delta H \delta u_{y}^{*}\right],\\[0.2cm]\label{eq:eetah}
      E_{\eta_{T}}  & = \overline{\eta_{T}} 6 k \text{Im}\left[ \delta T \delta u_{y}^{*}\right],\\[0.2cm]\label{eq:eetat}
    E_{p_{0}} &  = - \overline{p_{0}}\frac{1}{4 \gamma H_{0}^2} \left(\text{Re}\left[\delta\dot{H} \delta H^{*}\right] + H_{0} \text{Re}\left[\delta \dot{H} \delta \sigma^{*}\right]\right) ,\\
    E_{p_{\sigma}}  &  = \overline{p_{\sigma}}\left( \frac{1}{4 \gamma H_{0}}\text{Re}\left[\delta \dot{H} \delta \sigma^{*}\right] + k \text{Im}\left[\delta \sigma \delta u_{x}^{*}\right] \right),\\
    E_{p_{H}}  & = \overline{p_{H}} \left(\frac{1}{4 \gamma H_{0}} \text{Re}\left[\delta\dot{H} \delta H^{*}\right] + k \text{Im}\left[\delta H \delta u_{x}^{*}\right] \right),\\[0.2cm]
    E_{p_{T}}  & = \overline{p_{T}} k \text{Im}\left[ \delta T \delta u_{x}^{*}\right],
 \end{align}
 %where "$\text{Re}$" stands for the real part 
 where "$^*$" indicates the complex conjugate of a quantity.
In turn, (\ref{eq:esgx})---(\ref{eq:eetat}) describe, as the notation suggests, contributions from radial self-gravity, vertical self-gravity, viscosity, and pressure.

%\newpage
% \subsection{Analysis}

\section{Linearised radial self-gravity force}\label{sec:rsg}
As outlined in \S \ref{sec:sg} we assume that radial self-gravity forces do not depend on the distance $z$ away from the disc midplane, which in most situations is a reasonable assumption for thin discs.
The dimensionless perturbed self-gravity potential is given by \citep{shu1984}
\begin{equation}\label{eq:wkbsgos}
 \delta \phi_{\text{sg}} = -\frac{g}{k} \delta \sigma,
\end{equation}
yielding the perturbed radial self-gravity force
\begin{equation}
   \delta F_{\text{sg},x} = -i g \delta \sigma.
\end{equation}
However, \citet{shu1984} also derived a corrected radial self-gravity force 
\begin{equation}
    \delta F_{\text{sg},x} = -i g \zeta(k)\, \delta \sigma,
\end{equation}
which takes into account the finite thickness of the disc via the factor
\begin{equation}\label{eq:zeta_int}
\zeta\left(k\right) = \int \limits_{-\infty}^{+\infty} \int \limits_{-\infty}^{+\infty} h(x) h(y) \exp\left(-k H_{0} |x-y|\right) \mathrm{d}x\mathrm{d}y   
\end{equation}
with
\begin{align}
   h(z) & = \frac{1}{2} \hspace{0.1cm} \text{for} \hspace{0.1cm}  |z|\leq \frac{1}{2}  \hspace{0.6cm} \text{if} \, \rho=\rho_{\text{S}}\label{eq:g_slab},\\
  h(z) & =  \frac{1}{2} \exp \left(-\frac{\pi}{4}z^2\right)  \hspace{0.35cm}   \text{if}   \, \rho=\rho_{\text{G}}\label{eq:g_gauss}.
\end{align}
Analytical treatment of the integral (\ref{eq:zeta_int}) for the case $\rho=\rho_{s}$ given by (\ref{eq:rho_slab}) yields the approximate solution \citep{shu1984}
\begin{equation}\label{eq:zeta_2}
\zeta\left(k\right) = \frac{1}{k H_{0}} \left(1-\frac{1}{k H_{0}} \exp \left(-k H_{0}\right) \sinh \left(kH_{0}\right) \right).   
\end{equation}
Alternatively, \citet{vdvoort1970} found using a similar method:
\begin{equation}\label{eq:zeta}
\zeta\left(k\right) = \frac{1}{1+H_{0} k}   . 
\end{equation}
The function $\zeta\left(k\right)$ is a correction factor accounting for the finite thickness of the disc, which becomes significant for perturbations with wavelengths approaching the disc thickness $H_{0}$, i.e. when $k\sim 1/H_{0}$.
In our analysis below we find a small but not negligible effect of this factor on the stability boundary for viscous overstability, when radial self-gravity is included. Moreover, we include a similar correction to the computation of radial self-gravity forces in our N-body simulations, where we find that it can  have a significant effect.

We find (not shown) that (\ref{eq:zeta_2}) is indeed very accurate if $\rho=\rho_{S}$, whereas (\ref{eq:zeta}) is a better approximation if $\rho=\rho_{G}$. However, the differences are small and the choice of $\zeta\left(k\right)$ should not significantly affect the results.
For simplicity, we use (\ref{eq:zeta}) in al computations involving radial self-gravity.

\section{Effect of radial self-gravity in the hydrodynamic model}\label{app:rsg}

We speculate that the strong effect of radial self-gravity in our hydrodynamic model at large particle bulk densities (\S \ref{sec:nbody_hydro}) is related to the fluid's compressibility.
% Figure \ref{fig:pdv_rsg} (lower panel) shows the velocity divergence $|\vec{\nabla} \cdot \vec{u}|=|\partial_{x} u_{x} + \partial_{z} u_{z}|$ (a measure for compressibility) of fastest growing overstable modes obtained from numerical solution of the linear eigenvalue problem (\ref{eq:eigenproblem}), for calculations including vertical and radial self-gravity (solid curve) and calculations including only vertical self-gravity (dashed curve).
% The calculations are performed for increasing $r_{h}$, and for values of $\mbox{FF}$ at the threshold for viscous overstability in the presence of only vertical self-gravity (the latter values are displayed in the upper panel). 
% Therefore, the difference between the two curves in the lower panel corresponds to the effect of radial self-gravity on the compressibility of overstable waves, which appears to be strongly reducing.
We find (not shown) that radial self-gravity for increasing $r_{h}\gtrsim 0.3$ strongly reduces the compressibility (i.e. the velocity divergence $|\vec{\nabla} \cdot \vec{u}|=|\partial_{x} u_{x} + \partial_{z} u_{z}|$) of overstable waves.
A reduction of compressibility implies a reduced effect of the bulk viscosity on fluid motions, which might provide an explanation for the much more vigorous viscous overstability in the presence of radial self-gravity.
In this regard we deem it possible that the values of the bulk viscosity obtained in \citet{salo2001}, and which we used in our model, are inadequate in the current situation.
Indeed, it can be shown that the equation used to compute the bulk viscosity in \citet{salo2001}:
\begin{equation}\label{eq:bulk_equation}
    \overline{p(t)} = \overline{p_{0}} - \overline{\xi_{0}} \partial_{x} u_{x},
\end{equation}
is - strictly speaking - inadequate for our model.
In this equation the instantaneous isotropic (vertically averaged) pressure $p(t)$ is related to the trace of the pressure tensor:
\begin{equation}
\overline{p(t)}=\overline{ \frac{1}{3} Tr\left[\widehat{P}\right]},
\end{equation}
and is measured at an instance of time and at a specific radial location (averaged over some radial width of the simulation region).

Strictly speaking, in order to be suitable for our model, Equation (\ref{eq:bulk_equation}) requires correction in two aspects.
First of all, we need to replace
\begin{equation}
\begin{split}
\partial_{x}u_{x}\to \partial_{x}u_{x} + \overline{\partial_{z} u_{z}} = \partial_{x} u_{x} + \frac{\dot{H}}{H}, 
\end{split}
\end{equation}
due to the effective inclusion of vertical motions in this study, in contrast to the hydrodynamic model considered in \citet{salo2001}. 
Furthermore, in the presence of self-gravity, further modifications occur. This can be seen as follows.
The portion of the radial momentum equation relevant to this discussion reads
\begin{equation}
    \partial_{t} u_{x} = \cdots -\frac{1}{\sigma}\Vec{\nabla} \widehat{P} -\Vec{\nabla} \phi_{\text{sg}}.
\end{equation}
The self-gravity force can be written as \citep{lynden1972}
\begin{equation}
    -\Vec{\nabla} \phi_{\text{sg}} = -\frac{1}{\sigma}\Vec{\nabla} \widehat{T}
\end{equation}
with the self-gravity tensor
\begin{equation}
\widehat{T} \equiv \int \limits_{-\infty}^{\infty} \mathrm{d}z \left[ \frac{\Vec{\nabla} \phi_{\text{sg}} \Vec{\nabla} \phi_{\text{sg}}}{4 \pi G}  - \frac{\left(\Vec{\nabla} \phi_{\text{sg}}\right)^2}{8 \pi G} \widehat{U}\right],
\end{equation}
where $\widehat{U}$ denotes the unity tensor.
According to \citet{lynden1972} the second term in the bracket amounts to an isotropic tension, i.e. a negative pressure.
If we now define the ``generalised'' instantaneous isotropic pressure as
\begin{equation}
\overline{p(t)}=\overline{ \frac{1}{3} Tr\left[\widehat{P}+ \widehat{T}\right]},
\end{equation}
we find (assuming axisymmetry) for the slab vertical distribution (\ref{eq:rho_slab}):
\begin{equation}\label{eq:bulk_measure}
\overline{p(t)}=\overline{p_{0}} -\overline{\xi_{0}} \left( \partial_{x} u_{x} + \frac{\dot{H}}{H} \right) -\frac{1}{24 \pi G}\left( F_{\text{sg},x}^2 + \frac{8 \left( \pi G \sigma\right)^2 H}{3}  \right),
\end{equation}
where the terms in the second bracket arise from radial and vertical self-gravity, respectively.

\begin{figure}
 \centering
 \includegraphics[width = 0.45\textwidth]{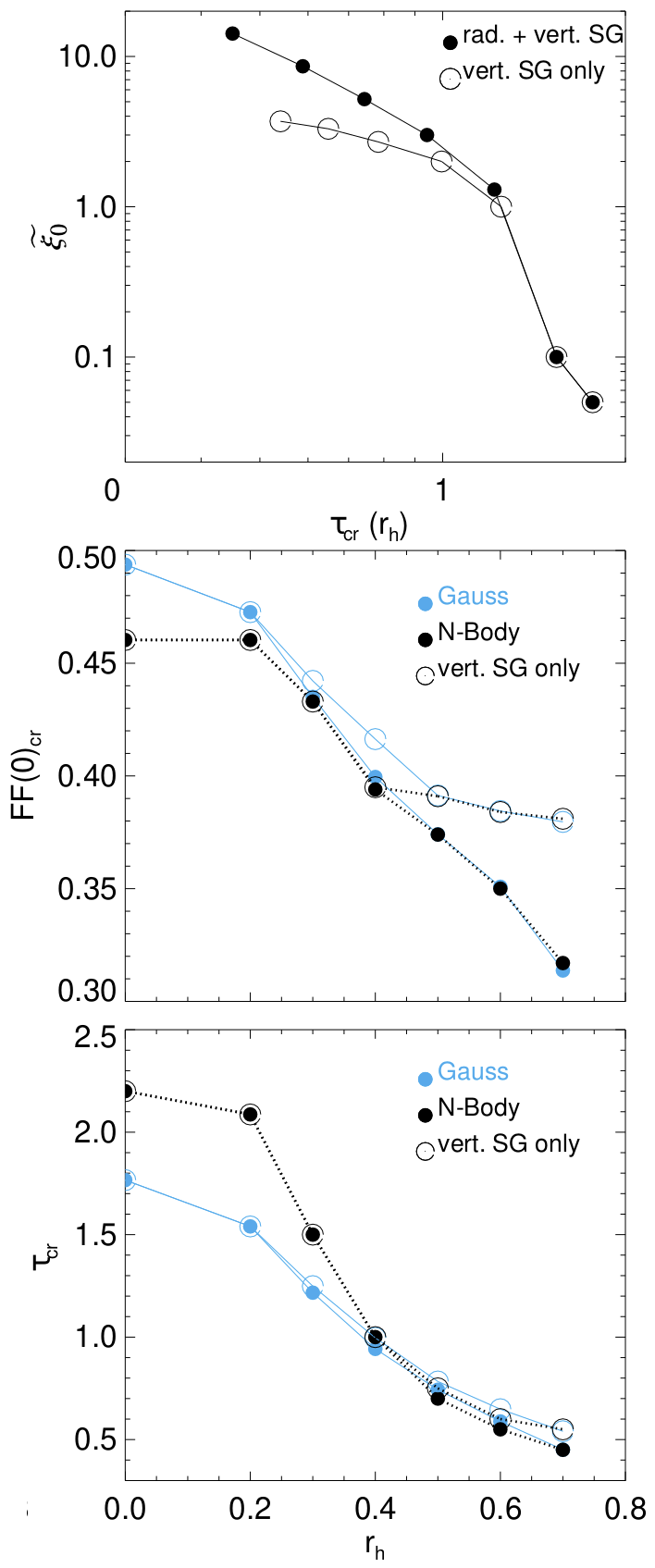}
 \caption{Values of the bulk viscosity (upper panel) that result in good agreement between our hydrodynamic model and our N-body simulations for the critical values of $\mbox{FF}$ (middle panel) and $\tau$ (bottom panel).
 The bulk viscosity is drawn against the critical values of $\tau$ for which overstability occurs in the model. Note that the latter decrease with increasing $r_{h}$ (cf. Figure \ref{fig:os_crit_sg}).  Open (solid) circles correspond to computations without (with) radial self-gravity.}
 \label{fig:nub}
 \end{figure}

It is a priori not clear how the above modifications affect the values of the bulk viscosity measured in self-gravitating N-body simulations.
Nevertheless, we find that we can indeed adjust the value of $\widetilde{\xi_{0}}$ in our model for each value of $r_{h}$, such that the values of $\mbox{FF}_{cr}$ nearly match those from our N-body simulations in the presence of radial self-gravity.
These values of $\widetilde{\xi_{0}}$ are shown in Figure  \ref{fig:nub} (upermost panel), and are plotted against the corresponding optical depth values for which overstability occurs in the model (using a Gaussian density distribution) for the cases of radial and vertical self-gravity, and only vertical self-gravity. The corresponding values of $\mbox{FF}_{cr}$ and also $\tau_{cr}$ are indeed nearly identical to those from our N-body simulations, as displayed in the middle and right panels, respectively.
In absence of radial self-gravity  we find optimal values $0.05\lesssim \widetilde{\xi_{0}} \lesssim 4$. The upper values are very similar to those of relation (\ref{eq:nub_omc}) which we used throughout the paper. Interestingly, for large optical depths our linear model requires only a very small bulk viscosity to obtain the best possible agreement with the simulation results. That is, at large optical depths the effect of the bulk viscosity becomes very small, such that the use of Eq. (\ref{eq:nub_omc}) yields very similar results for $\tau\gtrsim 1$.
Substantially larger values of $\widetilde{\xi_{0}}$  are required to attain agreement when radial self-gravity is included. This suggests that the effect of radial self-gravity in (\ref{eq:bulk_measure}) dominates over the other additional terms, but this remains to be verified in future N-body simulations.

 %  \begin{figure}
 % \centering
 % \includegraphics[width = 0.5\textwidth]{figs/pdv_rsg.png}
 % \caption{Explanation of the strong impact of radial self-gravity on the quantity $\mbox{ff}_{\text{cr}}$. The top panel shows values of $\mbox{ff}_{\text{cr}}$ for calculations with vertical (dashed) and (vertical and radial) self-gravity. The remaining panels describe corresponding values of terms that lead to the amplification of overstable modes due to radial self-gravity, as explained in the text.}
 % \label{fig:pdv_rsg}
 % \end{figure}
 % %

 %  \begin{figure}
 % \centering
 % \includegraphics[width = 0.5\textwidth]{figs/pdv_rsg_2.png}
 % \caption{Values of the fluid compressibility $|\vec{\nabla} \vec{u}| = |\partial_{x} u_{x} + \partial_{z} u_{z}|$ (lower panel) of the fastest growing overstable mode at the threshold for the onset of instability for increasing $r_{h}$ (i.e. for increasing particle bulk density). The dashed curve corresponds to computations with only vertical self-gravity. The solid curve is computed for the same values of $r_{h}$ and $\mbox{ff}$ as the dashed curve, but including vertical and radial self-gravity. The latter values of $\mbox{ff}$ are drawn in the top panel. }
 % \label{fig:pdv_rsg}
 % \end{figure}
 % %

 %

%%%%%%%%%%%%%%%%%%%%%%%%%%%%%%%%%%%%%%%%%%%%%%%%%%

% Don't change these lines
\bsp	% typesetting comment
\label{lastpage}
\end{document}